\documentclass[aps,prb,superscriptaddress,twocolumn]{revtex4-1}
\usepackage{amsmath,amssymb}
\usepackage{natbib}
\usepackage{braket}
\usepackage{hyperref}
\usepackage{graphicx}
\usepackage{color}
\usepackage[caption=false]{subfig}
\usepackage{epstopdf}

\newcommand*{\rom}[1]{\expandafter\@slowromancap\romannumeral #1@}

\begin{document}
\title{Lifetime renormalization of weakly anharmonic superconducting qubits: \\ I. Role of number non-conserving terms}
\author{Moein Malekakhlagh}
\affiliation{Department of Electrical Engineering, Princeton University, Princeton, New Jersey, 08544}
\author{Alexandru Petrescu}
\affiliation{Department of Electrical Engineering, Princeton University, Princeton, New Jersey, 08544}
\author{Hakan E. T\"ureci}
\affiliation{Department of Electrical Engineering, Princeton University, Princeton, New Jersey, 08544}
\affiliation{Rigetti Computing, 2919 Seventh Street, Berkeley, CA 94710}
\date{\today}
\begin{abstract}
The dynamics of a weakly anharmonic superconducting qubit in a complex electromagnetic environment is generally well-described by an effective multimode Kerr Hamiltonian at sufficiently weak excitation. This Hamiltonian can be embedded in a master equation with losses determined by the details of the electromagnetic environment. Recent experiments indicate, however, that when a superconducting circuit is driven with microwave signals populating the system with sufficiently high excitations the observed relaxation rates appear to be substantially different from expectations based on the electromagnetic environment of the qubit alone. This issue is a limiting factor in the optimization of superconducting qubit readout schemes. We claim here that an effective master equation with drive-power dependent parameters is an efficient approach to model such quantum dynamics. In this sequence of papers, we derive effective master equations, whose parameters exhibit nonlinear dependence on the excitation level of the circuit as well as the electromagnetic environment of the qubit. We show that the number non-conserving terms in the qubit nonlinearity generally lead to a renormalization of dissipative parameters of the effective master equation, while the number conserving terms give rise to a renormalization of the system frequencies. Here, in Part~I, we consider the excitation-relaxation dynamics of a transmon qubit that is prepared in a certain initial state, but is not driven otherwise. A unitary transformation technique is introduced to study the renormalization of i) qubit relaxation due to coupling to a generic bath and ii) Purcell decay. Analytic expressions are provided for the dependence of the non-linear dissipative terms on the details of the electromagnetic environment of the qubit. The perturbation technique based on unitary transformations developed here is generalized to the continuously driven case in Part~II.
\end{abstract}
\maketitle

\section{Introduction}
\label{Sec:Intro}

Radiative corrections to the properties of a discrete-level system have an important bearing on any quantum technology relying on such systems. It is well understood that the radiative lifetime of an atom, whether natural or artificial, sensitively depends on its electromagnetic environment \cite{Purcell_Resonance_1946, Purcell_Spontaneous_1995}. This fact is most transparently expressed by the dependence of the Purcell decay rate on the imaginary part of the classical electromagnetic Green's function computed at the source position and oscillation frequency \cite{Dung_Spontaneous_2000}. In circuit quantum electrodynamics, the equivalent view expresses the Purcell decay rate in terms of the admittance seen by the qubit as a classical oscillator \cite{Esteve_Effect_1986}. Though radiative corrections are inherently quantum in character, their computation at the linear-response level depends on classical electromagnetic properties most compactly expressed through the electromagnetic Green's function. From this point of view, it does not matter whether the object that is radiatively damped is a classical antenna or a quantum object. Here we will focus on the dependence of the qubit lifetime on (i) the detailed quantum mechanical structure of the emitter and (ii) the excitation level of the emitter and the electromagnetic environment.

\begin{figure}[t!]
\centering
\subfloat[\label{subfig:cQEDopenSybolic}]{%
\includegraphics[scale=0.225]{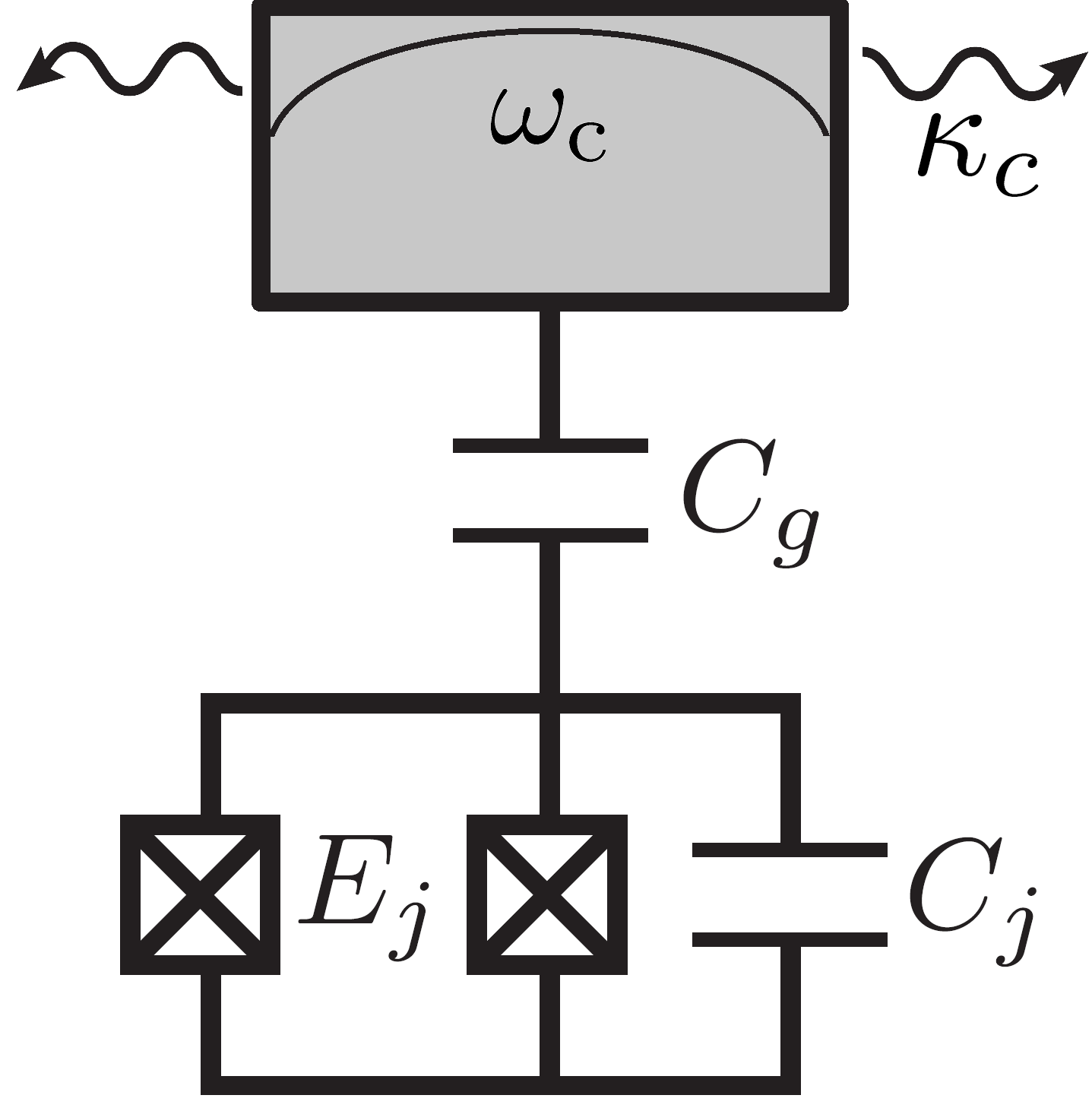}%
} \hfill
\subfloat[\label{subfig:cQEDopenSybolicSpider}]{%
\includegraphics[scale=0.225]{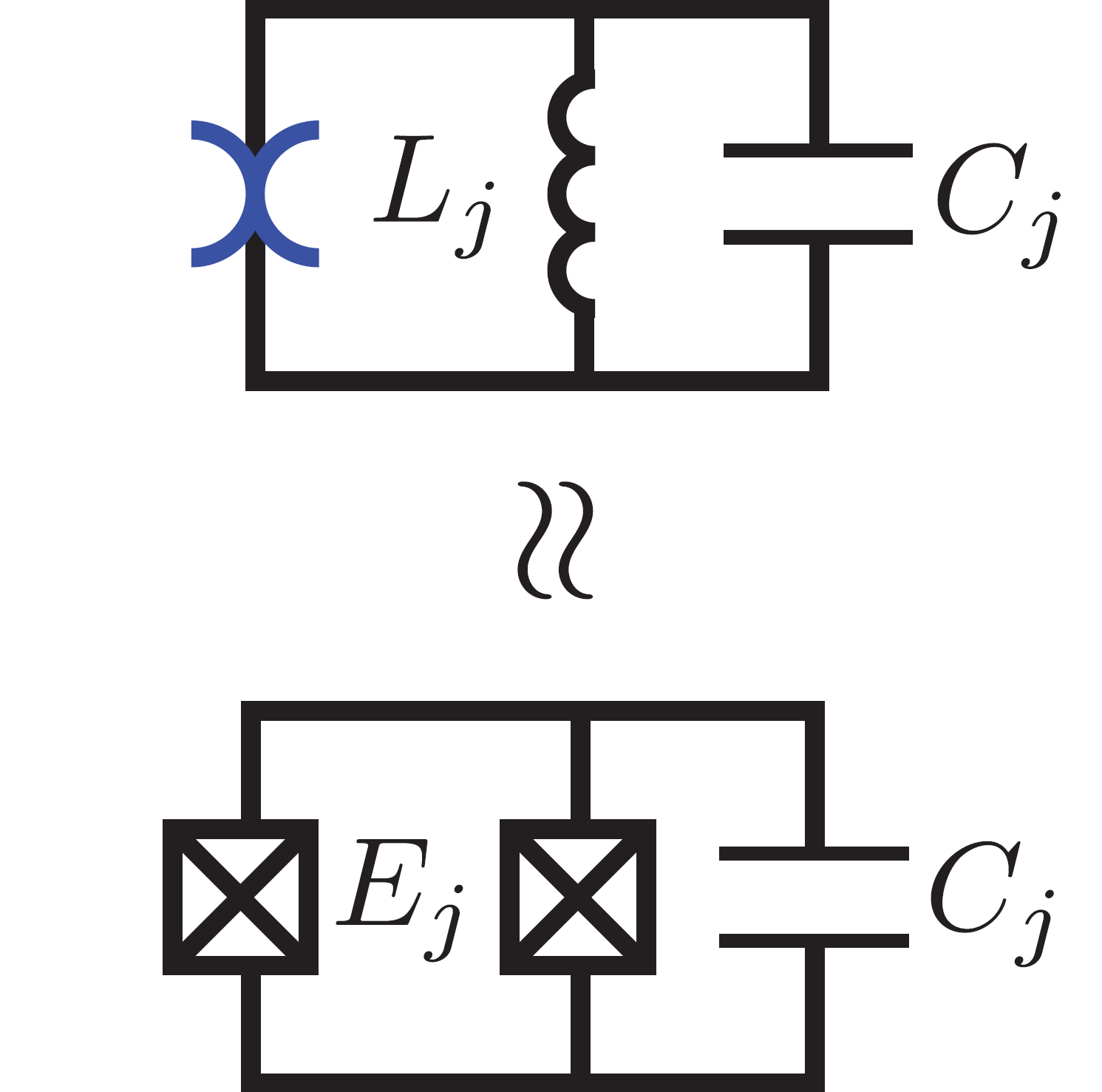}%
}\hfill
\subfloat[\label{subfig:TransmonSpec}]{%
\includegraphics[scale=0.265]{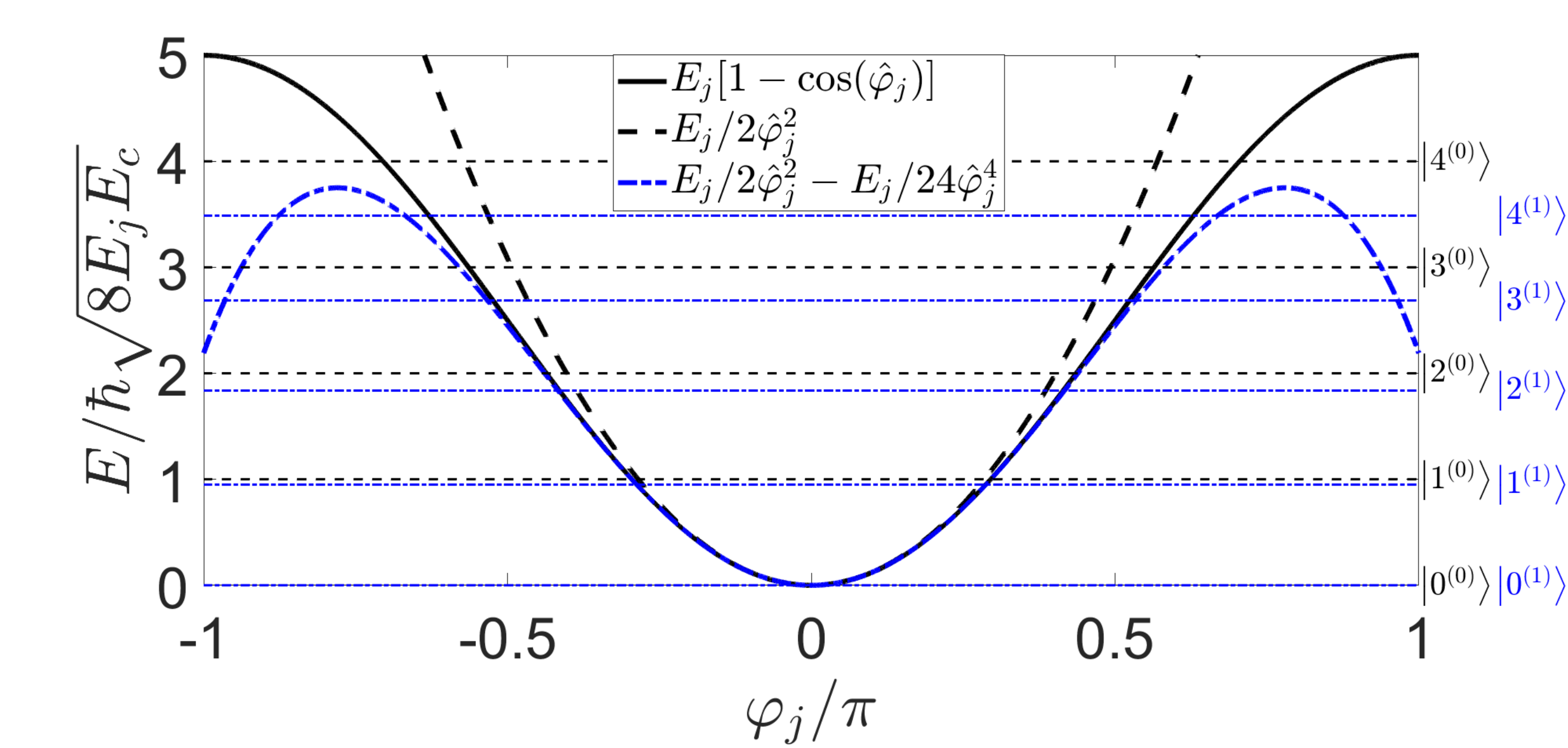}%
} 
\caption{a) A weakly anharmonic qubit ($E_{\text{j}}\gg E_{\text{c}}$) linearly (capacitively) coupled to an open cavity. b) Separation of linear and anharmonic parts of the Josephson potential. c) Josephson potential and its perturbative spectrum for a weak anharmonicity characterized by $E_{\text{j}}=50E_{\text{c}}$. The \textcolor{black}{solid black} curve shows the normalized Josephson potential, while the dashed black and dash-dotted blue are the resulting quadratic and quartic theories along with their corresponding energy levels.}
\label{Fig:cQEDopenSybolic}
\end{figure}

A detailed understanding of these aspects of radiative decay is becoming increasingly relevant in superconducting quantum computing. A number of the schemes devised for accurate and rapid readout of the quantum state of superconducting qubits rely on the understanding of the dissipative dynamics in the presence of a resonator excited beyond the linear response regime (sometimes referred to as the ``nonlinear dispersive regime'' \cite{Boissonneault_Nonlinear_2008}). In particular, several experiments \cite{Johnson_Heralded_2012, Slichter_Measurement-Induced_2012, Sank_Measurement-Induced_2016} have observed anomalous state transitions when the resonator photon occupancy is increased past a certain point. Other experiments inherently rely on the strong excitation regime for a rapid readout \cite{Reed_High-Fidelity_2010}. Even when the resonator is moderately excited a strong renormalization of qubit lifetimes is observed as a function of resonator occupancy \cite{Mundhada_Dependence_2016, Minev_Catch_2019}. While it is clear that a number of different mechanisms are at play in the renormalization of qubit lifetimes at finite excitation \cite{Boissonneault_Dispersive_2009, Boissonneault_Nonlinear_2008, Wilson_Dressed_2010, Slichter_Measurement-Induced_2012, Sete_Purcell_2014, Sank_Measurement-Induced_2016, Muller_Deriving_2017}, an improved understanding of the renormalization of the qubit lifetime due to purely radiative processes (i.e. due to the open nature of the electromagnetic environment) is vital to the implementation of rapid high-fidelity read-out protocols.

The goal of these two papers, referred to as Part I and Part II, is to develop a systematic perturbation theory based on unitary transformations to derive an effective master equation whose parameters depend on the nonlinearity of a weakly anharmonic Josephson artificial atom (e.g. a transmon\cite{Koch_Charge_2007}) and its electromagnetic environment. In particular we show how the effective relaxation rates are renormalized by the qubit nonlinearity. These results were made public in Refs.~[\onlinecite{Malekakhlagh_Systematic_2018}] and [\onlinecite{Tureci_T1_2018}]. Here, we provide the detailed discussion. 

In this first part, we discuss renormalization effects in the absence of a coherent microwave drive. This is the basic physics of spontaneous emission, but departing from the conventional approach, we analyze the impact of the specific nonlinearity of the qubit. The central result of this calculation is the delineation of the important role of number non-conserving terms in the renormalization of the qubit lifetime. The driven case requires a sufficiently different technical approach to warrant a separate discussion, which we undertake in Part~II. 
 
In what follows, we focus on a qubit coupled (i) to a flux bath (pure qubit relaxation), and (ii) to an open single-mode resonator (Purcell). However, the technique of unitary transformations presented here can be extended to other possible sources of decoherence as well as the multimode case, where hybridized modes can be found via a first-principles calculation \cite{Nigg_BlackBox_2012, Malekakhlagh_NonMarkovian_2016, Malekakhlagh_Cutoff-Free_2017}. Our results naturally complement and extend the black-box quantization technique \cite{Nigg_BlackBox_2012}, which drops the number non-conserving terms in the Josephson nonlinearity. 

The remainder of this article is organized as follows: In Sec.~\ref{Sec:Model}, we provide a brief summary of the main results and discuss our modeling of a circuit-QED setup involving a weakly anharmonic qubit. In Sec.~\ref{Sec:HierEqs}, we introduce a perturbation theory based on a unitary transformation to systematically compute the correction to frequency and radiative lifetimes of weakly anharmonic qubits. We apply this method to two specific environments to which the qubit can be coupled, in Secs.~\ref{SubSec:EffME-Qu}, \ref{SubSec:EffME-Qu1Mode}. Appendices~\ref{App:1stSWPT} and~\ref{App:2ndPT} contain the details of the first and second orders of our perturbation theory, respectively. Appendix~\ref{App:EMEDer} provides a derivation of the effective master equations used in the numerical simulation. In App.~\ref{App:ObsEOM}, we discuss the derivation of equations of motion for relevant physical observables based on our effective master equations.

\section{Model and Main results}
\label{Sec:Model}

The subgap dynamics of a superconducting circuit containing a weakly anharmonic artificial atom can generally be described by a multimode Kerr Hamiltonian \cite{Nigg_BlackBox_2012} at sufficiently weak excitation. Such a Hamiltonian has the virtue that the linear hybridization described by Maxwell's equations is fully accounted for in the effective parameters of the Kerr Hamiltonian. The Kerr parameters have a direct experimental relevance in the dispersive limit: the self-Kerr interaction terms give rise to a nonlinear dependence of the oscillator frequencies on the excitation level, while the cross-Kerr coupling between the qubit-like and resonator-like modes give rise to a qubit-state dependent shift in resonator normal mode frequencies. The latter forms the foundation of dispersive read-out schemes, discussed in Part~II.

Here, we focus on the impact of \textit{number non-conserving terms} in the original Josephson nonlinearity of the qubit, when the system is prepared in an initial state but is not driven otherwise. Results presented in this paper and summarized below suggest the use of an effective multimode master equation with renormalized dissipative parameters that, like the oscillator frequencies, depend nonlinearly on the excitation level in the initial state [See Eqs.~(\ref{eqn:EffME-DissQu-Lind Eq1}) and~(\ref{eqn:EffME-Qu1Mode-PT Lind Eq}) below].

Our starting point is a weakly anharmonic superconducting qubit coupled to an open resonator as depicted in Fig.~\ref{subfig:cQEDopenSybolic}. Though the results below can be generalized to a multimode cavity \cite{Malekakhlagh_NonMarkovian_2016, Malekakhlagh_Cutoff-Free_2017}, the basic mechanism of lifetime renormalization is already contained in the case of a single cavity mode, which we focus on here. The Hamiltonian describing the setup is \cite{Malekakhlagh_NonMarkovian_2016}
\begin{align}
  \hat{\mathcal{H}}&=\hat{\mathcal{H}}_{\text{s}} + \hat{\mathcal{H}}_{\text{b}}+ \hat{\mathcal{H}}_{\text{sb}},
\end{align}
where the system Hamiltonian
\begin{align}
\begin{split}
\hat{\mathcal{H}}_{\text{s}}&=\hat{\mathcal{H}}_{\text{a}}+\hat{\mathcal{H}}_{\text{c}}+\hat{\mathcal{H}}_{\text{ac}} \\
&\equiv \frac{\bar{\omega}_{\text{a}}}{4}\left[\hat{
\bar{Y}}_{\text{a}}^2-\frac{2}{\epsilon}\cos\left(\sqrt{\epsilon}\hat{\bar{X}}_{\text{a}}\right)\right]\\
&+\frac{\bar{\omega}_{\text{c}}}{4}\left(\hat{\bar{X}}_{\text{c}}^2+\hat{\bar{Y}}_{\text{c}}^2\right)+g\hat{\bar{Y}}_{\text{a}}\hat{\bar{Y}}_{\text{c}},
\end{split}
\label{eqn:Model-Hs 1}
\end{align}
and the resonator is in contact with a bath described by the bath Hamiltonian $\hat{\mathcal{H}}_\text{b}  = \sum_{k} \omega_{k} \hat{B}_{k}^\dagger \hat{B}_{k}$ and the resonator-bath coupling $\hat{\mathcal{H}}_\text{sb} = \hat{\bar{Y}}_{\text{c}}  \sum_{k} g_{k} \hat{Y}_k$. Here, we have used $\hat{\bar{X}}_{l}\equiv(\hat{l}+\hat{l}^{\dag})$, $\hat{\bar{Y}}_{l}\equiv -i (\hat{l}-\hat{l}^\dag)$ to denote the unitless phase and number operators of the qubit (cavity), such that $\hat{\varphi}_l \equiv \varphi_{l,\text{zpf}}\hat{X}_l$ and $\hat{n}_l\equiv n_{l,\text{zpf}}\hat{Y}_l$ for $l=a,c$. Moreover, the bare modes are denoted with a bar notation, in order to distinguish from the normal modes of the system, to be introduced shortly, which are more relevant to the development of our perturbation technique. In addition, $\bar{\omega}_{\text{a}}\equiv \sqrt{8E_{\text{j}}E_{\text{c}}}$ is the bare qubit frequency and $\epsilon\equiv\sqrt{2E_{\text{c}}/E_{\text{j}}}$ is a measure for the anharmonicity of the qubit, with $E_{\text{c}}$ and $E_{\text{j}}$ being the charging and Josephson energy scales. The qubit-cavity coupling strength is denoted by $g$, which can be found via the second quantization of the underlying circuit \cite{Malekakhlagh_Origin_2016}. 

Based on Hamiltonian~(\ref{eqn:Model-Hs 1}), there are two \textit{independent} mixing mechanisms between the bare qubit and cavity modes. First, there is a linear coupling of strength $g$, which is responsible for the mixing of the qubit/cavity-like degrees of freedom at the linear level. We refer to this as ``hybridization'', and to the resulting basis, in which the linear Hamiltonian is diagonal, as the ``normal mode basis'' \cite{Nigg_BlackBox_2012}. Second, because the qubit mode is intrinsically anharmonic, there will be a \textit{nonlinear} mixing of the modes on top of hybridization. 

To separate the two aforementioned sources of mode-mode mixing, it is helpful to first express Hamiltonian~(\ref{eqn:Model-Hs 1}) in the normal mode basis, where the effect of linear hybridization is exactly accounted for. In this basis, the Hamiltonian reads 
\begin{align}
\begin{split}
&\hat{\mathcal{H}}_{\text{s}}\equiv\omega_{\text{a}}\left(\hat{a}^{\dag}\hat{a}+\frac{1}{2}\right)+\omega_{\text{c}}\left(\hat{c}^{\dag}\hat{c}+\frac{1}{2}\right)\\
& + \, \frac{\bar{\omega}_{\text{a}}}{2} \sum_{n=2}^{\infty} (-\epsilon)^{n-1} \frac{\left[u_{\text{aa}}(\hat{a}+\hat{a}^{\dag})+u_{\text{ac}}(\hat{c}+\hat{c}^{\dag}) \right]^{2n}}{(2n)!}.
\end{split}
\label{eqn:Model-Hs in normpic}
\end{align}
Here, $u_{\text{aa}}$, $u_{\text{ac}}$, $u_{\text{ca}}$ and $u_{\text{cc}}$ are hybridization coefficients relating the bare and normal mode $X$-quadratures. The corresponding hybridization for the $Y$-quadratures are denoted by $v$. Together we write:
\begin{align}
\begin{bmatrix}
\hat{\bar{X}}_{\text{a}}\\
\hat{\bar{X}}_{\text{c}}
\end{bmatrix}=
\begin{bmatrix}
u_{\text{aa}} & u_{\text{ac}}\\
u_{\text{ca}} & u_{\text{cc}}
\end{bmatrix}
\begin{bmatrix}
\hat{X}_{\text{a}}\\
\hat{X}_{\text{c}}
\end{bmatrix}, \  
\begin{bmatrix}
\hat{\bar{Y}}_{\text{a}}\\
\hat{\bar{Y}}_{\text{c}}
\end{bmatrix}=
\begin{bmatrix}
v_{\text{qq}} & v_{\text{ac}}\\
v_{\text{ca}} & v_{\text{cc}}
\end{bmatrix}
\begin{bmatrix}
\hat{Y}_{\text{a}}\\
\hat{Y}_{\text{c}}
\end{bmatrix}.
\label{eqn:Model-HybrCoeffs}
\end{align}
The hybridization coefficients in Eq.~(\ref{eqn:Model-HybrCoeffs}) can be found in terms of bare parameters via successive application of scaling and rotational transformations, as discussed in App.~\ref{SubApp:1stSWPT-Qu1Mode}.
\color{black}

At this point, the bath modes can be integrated out and in the Born-Markov approximation a Lindblad master equation can be obtained for the system density matrix as
\begin{align}
\begin{split}
\dot{\hat{\rho}}_{\text{s}}(t)=-i\left[\hat{\mathcal{H}}_{\text{s}},\hat{\rho}_{\text{s}}(t)\right]&+2\kappa_{\text{c}}\mathcal{D}[v_{\text{cc}}\hat{c}]\hat{\rho}_{\text{s}}(t)\\
&+2\kappa_{\text{a}}\mathcal{D}[v_{\text{ca}}\hat{a}]\hat{\rho}_{\text{s}}(t),
\end{split}
\label{eqn:Model-Lind Eq.}
\end{align}
where $\mathcal{D}[\hat{C}](\bullet)=\hat{C}(\bullet)\hat{C}^{\dag}-1/2\{\hat{C}^{\dag}\hat{C},(\bullet)\}$ represents the dissipator for a collapse operator $\hat{C}$. The loss rates are given in terms of the bath spectral function as $2\kappa_{\text{a,c}} = S_{YY} (\omega_{\text{a,c}})$, where $S_{YY} (\omega) = \int_{-\infty}^\infty d\tau \, \text{e}^{-i\omega \tau} \text{tr}_{\text{b}} \left[\hat{\rho}_\text{b}(0)\hat{Y}_{\text{b}}(\tau) \, \hat{Y}_{\text{b}}(0) \right]$. $\hat{Y}_{\text{b}}= \sum_{k} g_{k} \hat{Y}_k$ is the noise operator that the cavity quadrature couples to and $\hat{\rho}_{\text{b}}(0)\equiv (1/Z_{\text{b}}) \text{e}^{- \hat{\mathcal{H}}_{\text{b}}/ k_\text{B} T} $ is the bath density matrix that is assumed to obey thermal distribution. We assume here the bath modes to be thermalized at $T=0$ and that the qubit sees no other loss channel than the radiative one through the resonator.

It is important to note that the loss rates calculated above account for what is typically referred to as the Purcell \cite{Purcell_Resonance_1946, Purcell_Spontaneous_1995} losses of the qubit (and similarly to resonator losses modified by the hybridization with the qubit). Here this loss rate is expressed through a properly secularized Markov approximation \cite{Breuer_Theory_2002} as also discussed in Refs.~[\onlinecite{Nigg_BlackBox_2012}] and~[\onlinecite{Malekakhlagh_Cutoff-Free_2017}]. This approximation is accurate for resonators with non-overlapping resonances (i.e. in the high-finesse regime). For low-finesse cases, the calculation of the exact linearized qubit dynamics beyond the Markov approximation can be implemented as well \cite{Malekakhlagh_NonMarkovian_2016}. We assume a high-finesse situation here to most transparently reveal the effects we are after, namely the role of the number non-conserving terms in the nonlinearity. 

An explanation of terminology to be used is in order before discussing our distinct treatment of the nonlinearity. In what follows, we employ the more generic terminology of secular and non-secular, commonly used in the theory of classical dynamical systems \cite{Bender_Advanced_1999, Strogatz_Nonlinear_2014}, to refer to number conserving and number non-conserving terms in the {\it hybridized} basis, respectively. The distinction between the two terminologies is as follows. The terms number conserving and non-conserving refer to the category of operators that commute with the number operator of a given set of bosonic modes. This definition do not necessarily require the modes to be the actual \textit{normal} modes of the system. Therefore, whether a non-linear term is number conserving or not depends on the choice of the modal expansion that is employed. However, ``secularity'' has a unique meaning in the sense that it is only defined with respect to the normal modes that are unique for a particular system up to the linear order. In the normal (hybridized) basis [used in Eq.~(\ref{eqn:Model-Hs in normpic})], such terms will appear also as number non-conserving terms. That would not be the case had we carried out the perturbation theory in the bare basis employed in Eq.~(\ref{eqn:Model-Hs 1}). The transformation to the hybridized basis is essential for the development of a systematic perturbation theory.
\color{black}

In pursuit of an accurate effective model in the low-excitation regime, we will devise an appropriate unitary transformation to remove the Josephson nonlinearity [last line of Eq.~(\ref{eqn:Model-Hs in normpic})] to successively higher orders in the parameter $\epsilon$. Such an approach has been implemented before in the context of superconducting circuits for the Jaynes-Cummings model \cite{Boissonneault_Dispersive_2009, Sete_Purcell_2014}, for the Rabi model \cite{Beaudoin_Dissipation_2011} and for lattice models \cite{Zhu_Dispersive_2013, Aron_Photon-Mediated_2016}. The treatment here is distinct, because it accounts for the Josephson nonlinearity perturbatively rather than making a two-level approximation. The existence of a small parameter $\epsilon$ for weakly anharmonic qubits, such as the transmon qubit, allows for a controlled expansion for the parameters of the effective master equation obtained. This limit is the opposite to that in which the qubit can be approximated as a two-level system, and which underlies the Rabi and Jaynes-Cummings models. 

Consider applying a unitary transformation to the {\it full} Hamiltonian including the system and bath as
\begin{align}
\hat{\mathcal{H}}_{\text{eff}}\equiv e^{-\hat{G}}\hat{\mathcal{H}}e^{+\hat{G}},
\end{align}
where $\hat{G}$ is an unknown anti-Hermitian operator that acts as the generator of this transformation. We then expand the system Hamiltonian and the generator formally in powers of $\epsilon$
\begin{subequations}
\begin{align}
&\hat{\mathcal{H}}_{\text{s}}=\hat{\mathcal{H}}_2-\epsilon\hat{\mathcal{H}}_4+\epsilon^2\hat{\mathcal{H}}_6+ \cdots, 
\label{eqn:Model-H_s expansion}\\
&\hat{G}=\epsilon\hat{G}_4+\epsilon^2\hat{G}_6+\cdots,
\label{eqn:Model-G expansion}
\end{align}
\end{subequations}
where $\hat{\mathcal{H}}_n$ can be found e.g. from Eq.~(\ref{eqn:Model-Hs in normpic}) for the model discussed here. The conditions for successive removal of the nonlinearity in the Hamiltonian yield, as shown in Sec.~\ref{Sec:HierEqs}, hierarchical operator equations for $\hat{G}_n$ which can be solved through computer algebra. 

An important feature that emerges in this framework is that such a unitary transformation can only remove the non-secular (number non-conserving) terms, while the secular terms are left behind contributing to an effective Hamiltonian that is diagonal in the Fock space. 
The lowest order effective Hamiltonian is the two-mode Kerr (multimode Kerr if more resonator modes are retained), identical to the one obtained when the number non-conserving terms are neglected from the outset in Eq.~(\ref{eqn:Model-Hs 1}) as implemented in Ref.~[\onlinecite{Nigg_BlackBox_2012}]. 

The role of non-secular terms is revealed when accounting for the system-bath coupling under such transformation. The effect of the removal of the non-secular terms reappears in the action of the transformation on the system quadratures that couple to the bath, i.e. $\hat{Y}_{\text{a,c}}$, in turn giving rise to corrections to the decay rates. To be more accurate, the corrections appear as operator-valued renormalizations of the corresponding collapse operators. These corrections to collapse operators can be recast into effective master equations, whose parameters display a nonlinear dependence on the excitation level of the system. These corrections are then organized into a perturbative expansion in the small parameter $\epsilon$ that describes the weak anharmonicity of the qubit. The excitation level is set here by the initial conditions. In Part~II, we show that, for systems driven with coherent microwave signals, the appropriate excitation level to consider is set by the amplitude and the frequency of the drives. 

The rest of the paper presents the implementation of the above approach to lowest order in $\epsilon$ for two cases, a weakly anharmonic qubit coupled to (i) a generic bath [Eq.~(\ref{eqn:EffME-DissQu-Lind Eq1})], (ii) an open single-mode resonator [Eq.~(\ref{eqn:EffME-Qu1Mode-PT Lind Eq})]. To succinctly explain the concept of an effective master equation as used here, we discuss the final results for case (i) and defer the presentation of the results for case (ii) to Sec.~\ref{SubSec:EffME-Qu1Mode}:
\begin{align}
\begin{split}
\dot{\hat{\rho}}_{\text{a}}(t)&=-i\left[\hat{\mathcal{H}}_{\text{a,eff}},\hat{\rho}_{\text{a}}(t)\right]\\
&+2\kappa_{\text{a}} \mathcal{D}\left[\left(1+\frac{\epsilon}{8}(1+\hat{n}_{\text{a}})\right)\hat{a}\right] \hat{\rho}_{\text{a}}(t)\\
&+2\kappa_{\text{a}3}\mathcal{D}\left[\frac{\epsilon}{48}\hat{a}^3\right]\hat{\rho}_{\text{a}}(t),
\end{split}
\label{Eq:Model-DissQu-Lind Eq}
\end{align}
where $\hat{\mathcal{H}}_\text{a,eff}\equiv \left(1-\frac{\epsilon}{8}\right)\omega_{\text{a}} \hat{n}_{\text{a}}-\frac{\epsilon}{8}\omega_{\text{a}}\hat{n}_{\text{a}}^2+O(\epsilon^2)$ and $\hat{n}_{\text{a}}=\hat{a}^{\dag}\hat{a}$ is the qubit number operator. Moreover, the effective dynamics to lowest order contains a one-photon loss term at the rate $\kappa_{\text{a}}\equiv S_{YY}(\omega_{\text{a}})$, and a three-photon loss term at a rate $\kappa_{\text{a}3}\equiv S_{YY}(3\omega_{\text{a}})$.

\section{Hierarchical Equations for Generators}
\label{Sec:HierEqs}

In this section, we discuss a procedure to find the unitary transformation that can effectively account for the nonlinearity of a Josephson junction artificial atom embedded in a general electromagnetic environment. When the Josephson nonlinearity is weak, as in the case of a transmon, a perturbative expansion can be found for the generator $\hat{G}$ of this unitary transformation. Here, we will not make any specific assumptions about the electromagnetic environment to which the qubit is coupled, merely considering a generic situation where the system Hamiltonian can be expanded in a small parameter $\epsilon$ as is the case of Eq.~(\ref{eqn:Model-Hs in normpic}):
\begin{align}
\hat{\mathcal{H}}_{\text{s}}=\hat{\mathcal{H}}_2-\epsilon\hat{\mathcal{H}}_4+\epsilon^2\hat{\mathcal{H}}_6+ \cdots,
\label{eqn:HierEqs-2nd Hs exp}
\end{align}
Note that the linear part of Hamiltonian~(\ref{eqn:HierEqs-2nd Hs exp}) (referred to as $\hat{\mathcal{H}}_2$ in this section) shall always be expressed in terms of the normal mode coordinates (of the original linearized circuit) and that $\hat{\mathcal{H}}_n$ contains polynomials of degree $n$ in the bosonic creation and annihilation operators corresponding to the normal modes. We will seek a unitary transformation
\begin{align}
\hat{\mathcal{H}}_{\text{s,eff}}\equiv e^{-\hat{G}} \hat{\mathcal{H}}_{\text{s}} e^{+\hat{G}},
\label{eqn:HierEqs:Def of PT trans}
\end{align}
that will remove all the non-secular terms at any arbitrary order $\epsilon^n$. To solve for the generator $\hat{G}$, we consider the following Ansatz, written as a series expansion in the small parameter $\epsilon$ as
\begin{align}
\hat{G}=\epsilon\hat{G}_4+\epsilon^2\hat{G}_6+\cdots.
\label{eqn:HierEqs-2nd G exp}
\end{align}

Let us look into the condition for the removal of the non-secular terms at order $\epsilon$. Using the Baker\cite{Baker_Alternants_1905}-Campbell\cite{Campbell_Law_1896}-Hausdorff\cite{Hausdorff_Symbolische_1906} (BCH) formula
\begin{align}
e^{-\hat{A}}\hat{B}e^{\hat{A}}=\hat{B}+[\hat{B},\hat{A}]+\frac{1}{2!}[[\hat{B},\hat{A}],\hat{A}]+\ldots,
\label{eqn:HierEqs:BCH lemma}
\end{align}
we obtain the lowest order expansion as
\begin{align}
e^{-\hat{G}}\hat{\mathcal{H}}_{\text{s}} e^{+\hat{G}}=\hat{\mathcal{H}}_2+\epsilon\left\{-\hat{\mathcal{H}}_4+[\hat{\mathcal{H}}_2,\hat{G}_4]\right\}+O(\epsilon^2).
\label{eqn:HierEqs:PT trans}
\end{align}
We then determine $\hat{G}_4$ in order to simplify the effective Hamiltonian~(\ref{eqn:HierEqs:PT trans}). Importantly, we observe that there is no $\hat{G}_4$ such that $[\hat{\mathcal{H}}_2,\hat{G}_4]$ cancels secular contributions in $\hat{\mathcal{H}}_4$. The reason is that the commutator of the harmonic Hamiltonian $\hat{\mathcal{H}}_2$ with any non-secular term remains non-secular, while with any secular term it is zero. On the other hand, all non-secular terms can be in principle canceled through this procedure. These commutator relations can be reduced to the following three rules: 
\begin{subequations}
\begin{align}
&[\text{sec},\text{sec}]=0,
\label{eqn:HierEqs-[S,S]=0}\\
&[\text{sec},\text{non-sec}]=\text{non-sec},
\label{eqn:HierEqs-[S,N]=N}\\
&[\text{non-sec},\text{non-sec}]=\text{sec}+ \text{non-sec}.
\label{eqn:HierEqs-[N,N]=S+N}
\end{align} 
\end{subequations}
We therefore write $\hat{\mathcal{H}}_4=\hat{\mathcal{S}}_4+\hat{\mathcal{N}}_4$, where $\hat{\mathcal{S}}$ stands for secular and $\hat{\mathcal{N}}$ for non-secular terms, and construct the generator $\hat{G}_4$ such that it satisfies
\begin{align}
[\hat{\mathcal{H}}_2,\hat{G}_4]-\hat{\mathcal{N}_4}=0.
\label{eqn:HierEqs-Cond G4}
\end{align}
Consequently, the system Hamiltonian is renormalized by the remaining secular terms as
\begin{align}
\hat{\mathcal{H}}_{\text{s,eff}}= \hat{\mathcal{H}}_2-\epsilon \hat{\mathcal{S}}_4 +O(\epsilon^2).
\label{eqn:HierEqs-1st Hs}
\end{align}
Equation~(\ref{eqn:HierEqs-1st Hs}) contains, up to lowest order, only the number conserving terms that contribute to transition frequency renormalization, in agreement with with the common rotating wave approximation (RWA) that leads to the Kerr theory.

Next, we briefly discuss how we can solve for the generator $\hat{G}_4$ based on Eq.~(\ref{eqn:HierEqs-Cond G4}). The key to a systematic and practical construction of $\hat{G}_4$ is the normal-ordered form of $\hat{\mathcal{N}}_4$.
To see this explicitly, consider the simple case of a  harmonic oscillator with $\hat{H}_2\equiv\hat{b}^{\dag}\hat{b}+1/2$. The commutator of the normal-ordered quadratic Hamiltonian with any normal-ordered operator monomial, i.e. a term of the form $(\hat{b}^{\dag})^m\hat{b}^{n}$, is proportional to that monomial:
\begin{align}
\left[\hat{H}_2,(\hat{b}^{\dag})^m\hat{b}^{n}\right]=(m-n)(\hat{b}^{\dag})^m\hat{b}^{n}.
\label{eqn:HierEqs-Qu-[Hq,bdag^m b^n]}
\end{align}

Therefore, based on Eqs.~(\ref{eqn:HierEqs-Cond G4}) and~(\ref{eqn:HierEqs-Qu-[Hq,bdag^m b^n]}), we conclude that $\hat{G}_4$ should include all of those monomials contained in $\hat{\mathcal{N}}_4$, but with modified coefficients. Note that in contrast to the secular terms, which can always be written in a compact form in terms of the quadratic Hamiltonian, there is in general a large number of non-secular terms and bookkeeping might seem challenging at first glance. However, the term-by-term calculation that becomes possible based on identity~(\ref{eqn:HierEqs-Qu-[Hq,bdag^m b^n]}) allows us to solve for the corresponding $\hat{G}_4$, regardless of the number of non-secular terms, given that we have access to sufficient and fast symbolic computing power. We have developed a computer algebra code in Mathematica to solve for the generator of the transformation as well as the resulting renormalization of any system operator \footnote{This package is available for public use. Please contact the authors to request a copy. It is based on the SNEG library made available by the university of Ljublijana \cite{Zitko_Sneg_2011}}. Importantly, the term-by-term computation based on identity~(\ref{eqn:HierEqs-Qu-[Hq,bdag^m b^n]}) allows us to solve for and categorize the terms in $\hat{G}_4$ that contribute to a particular relaxation process (See Table~\ref{tab:EffME-Qu1Mode}). Even though the calculation presented in this article is only for a single cavity mode and up to lowest order in $\epsilon$, this procedure could in principle be generalized to any order and any number of modes.

The discussion for the lowest order corrections so far can be generalized to include the $\epsilon^n$-contributions in the expansion of Eq.~(\ref{eqn:HierEqs:PT trans}). A similar equation can be found for $\hat{G}_6$ (See App.~\ref{App:2ndPT}):
\begin{align}
[\hat{\mathcal{H}}_2,\hat{G}_6]+\hat{\mathcal{N}}_6-[\hat{\mathcal{S}}_4,\hat{G}_4]-\frac{1}{2}\mathcal{N}\left([\hat{\mathcal{N}}_4,\hat{G}_4]\right)=0,
\label{eqn:HierEqs-Cond G6}
\end{align}
where we use $\mathcal{S}(\bullet)$ and $\mathcal{N}(\bullet)$ to refer to the secular and non-secular parts of a contribution. This equation as well as the equations for higher order $\hat{G}_n$ are hierarchical equations that depend on the previous lower-order generators, and therefore can be solved in a recursive way. Note that the structure of Eq.~(\ref{eqn:HierEqs-Cond G6}) for $\hat{G}_6$ is exactly the same as that of Eq.~(\ref{eqn:HierEqs-Cond G4}) for $\hat{G}_4$. In both cases, the \textit{unknown} generator appears inside a commutator with the quadratic Hamiltonian $\hat{\mathcal{H}}_2$, plus a collection of \textit{known} non-secular terms. Employing the identity~(\ref{eqn:HierEqs-Qu-[Hq,bdag^m b^n]}) and the discussion after it, one can determine $\hat{G}_6$ term by term such that it cancels the corresponding  monomials. Moreover, the corresponding Hamiltonian up to second order can be obtained as 
\begin{align}
\hat{\mathcal{H}}_{\text{s,eff}}=\hat{\mathcal{H}}_2-\epsilon\hat{\mathcal{S}}_4+\epsilon^2\left[\hat{\mathcal{S}}_6-\frac{1}{2}\mathcal{S}\left([\hat{\mathcal{N}}_4,\hat{G}_4]\right)\right]+O(\epsilon^3).
\label{eqn:HierEqs-2nd Hs}
\end{align}

To summarize the main results of this section, Eqs.~(\ref{eqn:HierEqs-Cond G4}) and~(\ref{eqn:HierEqs-Cond G6}) provide the conditions to determine the generator $\hat{G}$, up to first and second order in $\epsilon$, respectively. These equations can be be solved for using computer algebra. Furthermore, the effective system Hamiltonian is determined by a collection of secular terms as given by Eqs.~(\ref{eqn:HierEqs-1st Hs}) and~(\ref{eqn:HierEqs-2nd Hs}) up to the first and second orders in $\epsilon$, correspondingly.
\section{Effective Master Equations}
\label{Sec:EffME}

The method of unitary transformations discussed above can eliminate the number non-conserving terms in the Josephson nonlinearity to a given order in $\epsilon$. The resulting system Hamiltonian is then diagonal in the Fock state representation. At order $\epsilon$, this is the multimode Kerr Hamiltonian that would have been obtained had the number non-conserving terms in the Josephson nonlinearity been neglected from the start, e.g. in Eq.~(\ref{eqn:Model-Hs in normpic}) for a single-mode resonator. At higher order, the effective Hamiltonian includes more information [See last term of Eq.~(\ref{eqn:HierEqs-2nd Hs})] than the Hamiltonian obtained by simply dropping the number non-conserving terms from the Josephson potential.

\begin{table}
\begin{tabular}{|c|c|c|c|}
\hline
\textbf{Calculation} & \textbf{RWA?} & \textbf{Corrects Freq.?} & \textbf{Corrects Diss.?}\\
\hline\hline
\textbf{\textcolor{blue}{Eff ME}} & No & Yes & Yes \\
\hline
\textbf{\textcolor{red}{Kerr ME}} & Yes & Yes & No\\ 
\hline
\textbf{Linear ME} & N/A & No & No\\
\hline
\end{tabular}
\caption{Comparison of the different master equations up to lowest order in $\epsilon$. The color used to denote each theory is consistent with the corresponding numerical plots in the remainder of this article. The Kerr master equation only contains the corrections from the secular terms $\hat{\mathcal{S}}_4$, while neglects the non-secular terms $\hat{\mathcal{N}}_4$. The linear theory refers to the case of $\epsilon=0$, for which the RWA is not applicable.} 
\label{tab:EffME-Comparison}
\end{table}

At first sight, it may appear that the non-secular terms do not have an impact other than providing a Kerr-type effective Hamiltonian (at order $\epsilon$). Their impact is however revealed in two places: 1) coupling to the bath, and 2) initial density matrix. According to Eq.~(\ref{eqn:HierEqs-Cond G4}), the generator is merely determined by a subset of the terms (non-secular) in the system Hamiltonian, and hence the resulting unitary transformation commutes with any bath operator. Therefore, to obtain corrections to the relaxation rates, we only need to transform the system part of the system-bath Hamiltonian up to lowest order as
\begin{align}
\hat{Y}_{\text{a,c}} \rightarrow \hat{Y}_{\text{a,c}}+\epsilon[ \hat{Y}_{\text{a,c}},\hat{G}_4]+O(\epsilon^2).
\label{eqn:EffME-1st Hsb1}
\end{align}
Consequently, when the bath degrees of freedom are integrated out to obtain a properly secularized master equation, the effective collapse operators contain now nonlinear terms in powers of $\epsilon$. The form of the resulting effective master equation depends on the particular model considered. Finally, in order to be consistent, the initial density matrix also needs to be mapped into the new frame under the same unitary transformation:
\begin{align}
\hat{\rho}_{\text{s}}(0)\rightarrow \hat{\rho}_{\text{s}}(0)+\epsilon[\hat{\rho}_{\text{s}}(0),\hat{G}_4]+O(\epsilon^2).
\label{eqn:EffME-Def of Rhosw}
\end{align}

In the next two subsections, we derive effective master equations (``EME'') up to order $\epsilon$ for a qubit coupled (i) to a generic bath coupling to the $X$ - quadrature, and (ii) to an open single-mode resonator, respectively. We then compare the results from EME to both linear and the commonly used Kerr master equations. A summary of the expected renormalizations is presented in Table~\ref{tab:EffME-Comparison}. More details on derivation of the EME can be found in App.~\ref{App:EMEDer}.  
\subsection{Case (i): Pure qubit relaxation}
\label{SubSec:EffME-Qu}
In this section, we study the renormalization of pure relaxation that originates as the interplay between qubit flux noise and its anharmonicity. This case is also a pedagogical simplification of the model introduced in Sec.~\ref{Sec:Model}. The derivation and the final structure of the effective master equation however contain the key elements of the general argument more transparently. 

The derivation here will focus on an effective master equation at the lowest order in $\epsilon$. As discussed in Section~\ref{Sec:HierEqs}, at this order it is sufficient to retain only the quartic term in the Josephson potential. We start with the Hamiltonian of the form
\begin{align}
  \hat{\mathcal{H}}&=\hat{\mathcal{H}}_{\text{a}} +  \hat{\mathcal{H}}_{\text{sb}} + \hat{\mathcal{H}}_{\text{b}},
\label{Eq:H}
\end{align}
where the system Hamiltonian is given by 
\begin{align}
\hat{\mathcal{H}}_{\text{a}}=\omega_{\text{a}}\left(\hat{a}^{\dag}\hat{a}+\frac{1}{2}\right)-\frac{\epsilon}{48}\omega_{\text{a}}\left(\hat{a}+\hat{a}^{\dag}\right)^4.
\label{eqn:EffME-Qu-Def of Hq}
\end{align}
We note that, in the absence of a cavity mode, there is no hybridization and hence we denote all quantities without resorting to a bar for simplicity. We consider a bath Hamiltonian $\hat{\mathcal{H}}_\text{b}  = \sum_{k} \omega_{k} \hat{B}_{k}^\dagger \hat{B}_{k}$ for the qubit with the system-bath coupling $\hat{\mathcal{H}}_\text{sb} = \hat{X}_{\text{a}}  \sum_{k} g_{k} \hat{X}_{k}$. This would for instance represent flux noise acting on a transmon qubit, the dominant source of decoherence for recent weakly anharmonic tunable qubit designs \cite{Koch_Charge_2007}.
\color{black}
\begin{table}[t!]
  \begin{tabular}{|c|c|}
    \hline
    Operator  &  Coefficient \\
    \hline\hline
    $\hat{a}+\text{H.c.}$ & $\frac{1}{8}$\\ 
    \hline
    $\hat{a}^{\dag} \hat{a} \hat{a}+\text{H.c.}$& $\frac{1}{8}$ \\
    \hline
    $\hat{a}^3+\text{H.c.}$ & $-\frac{1}{48}$\\
    \hline
  \end{tabular}
  \caption{\label{tab:EffME-Qu} The contributions in $\left[ \hat{X}_{\text{a}}, \hat{G}_4\right]$, which provides the lowest order renormalization of the system quadrature. The left column shows each operator entering the sum, and the right column shows its coefficient.} 
\end{table}

For simplicity, we denote the unitless quadratic and quartic parts of $\hat{\mathcal{H}}_{\text{a}}$ as
\begin{align}
\hat{H}_{\text{a}}\equiv \hat{n}_{\text{a}}+\frac{1}{2}, \quad 
\hat{H}_4\equiv \left(\hat{a}+\hat{a}^{\dag}\right)^4,
\label{eqn:EffME-Qu-Def of Hq and H4}
\end{align}
where $\hat{n}_{\text{a}}$ is the number operator. The first step is to separate the quartic anharmonicity in terms of secular and non-secular parts as $\hat{H}_4=\hat{S}_4+\hat{N}_4$. In this rather simple case, it is possible to categorize all the terms in $\hat{S}_4$ and $\hat{N}_4$. There are six distinct secular terms that can be expressed as a polynomial of $\hat{H}_{\text{a}}$ as (See App.~\ref{SubApp:1stPT-DissQu}) 
\begin{align}
\begin{split}
\hat{S}_4=6\hat{H}_{\text{a}}^2+\frac{3}{2}=6\hat{n}_{\text{a}}^2+6\hat{n}_{\text{a}}+3.
\end{split}
\label{eqn:EffME-Qu-Def of S4}
\end{align}
We showed in Eq.~(\ref{eqn:HierEqs-1st Hs}) that the Hamiltonian is renormalized only by the secular terms up to lowest order. Therefore, we obtain the lowest order correction to the effective Hamiltonian
\begin{align}
\begin{split}
\hat{\mathcal{H}}_{\text{a,eff}} & = \omega_{\text{a}} \hat{H}_{\text{a}}-\frac{\epsilon\omega_{\text{a}}}{8}\hat{H}_{\text{a}}^2+O(\epsilon^2) \\
& = \left(1-\frac{\epsilon}{8}\right)\omega_{\text{a}} \hat{n}_{\text{a}}-\frac{\epsilon}{8}\omega_{\text{a}}\hat{n}_{\text{a}}^2+O(\epsilon^2),
\end{split}
\label{eqn:EffME-Qu-H_PT}
\end{align} 
where the second line explicitly shows the quadratic as well as the quartic self-Kerr corrections to the transition frequency of the qubit. 

Next, we focus on the non-secular contributions. The remaining ten non-secular terms can be expressed in normal-ordered form as
\begin{align}
\begin{split}
\hat{N}_4=\hat{a}^4+\left(\hat{a}^{\dag}\right)^4+4\left[\hat{a}^{\dag}\hat{a}^3+\left(\hat{a}^{\dag}\right)^3\hat{a}\right]+6\left[\hat{a}^2+\left(\hat{a}^{\dag}\right)^2\right].
\end{split}
\label{Eq:1stSWPT-DissQu-Expan of N4}
\end{align}
Then, we construct the $\epsilon$-order generator such that it cancels the non-secular contributions, which results in
\begin{align}
[\hat{H}_{\text{a}},\hat{G}_4]-\frac{1}{48}\hat{N}_4=0.
\label{eqn:EffME-Qu-Cond for G}
\end{align}
Employing the identity~(\ref{eqn:HierEqs-Qu-[Hq,bdag^m b^n]}), we are able to build the generator $\hat{G}_4$ term by term. The result is
\begin{align}
\begin{split}
\hat{G}_4&=\frac{1}{192}\left[\left(\hat{a}^{\dag}\right)^4-\hat{a}^4\right]+\frac{1}{24}\left[\left(\hat{a}^{\dag}\right)^3\hat{a}-\hat{a}^{\dag}\hat{a}^3\right]\\
&+\frac{1}{16}\left[\left(\hat{a}^{\dag}\right)^2-\hat{a}^2\right].
\end{split}
\label{eqn:EffME-DissQu-Sol G4}
\end{align}
Even though the non-secular terms are completely removed from the system Hamiltonian, their effects are at the end translated to modifications to the relaxation rates after the transformation is applied to system-bath coupling. For the model system considered here, the qubit couples through the quadrature $\hat{X}_{\text{a}}$. The transformation of this quadrature produces a variety of multi-photon transition processes, which up to $\epsilon$-order can be written as (See also Table~\ref{tab:EffME-Qu})
\begin{align}
\begin{split}
e^{- \hat{G}} \hat{X}_{\text{a}} e^{+\hat{G}} = \left[ 1 + \frac{\epsilon}{8} \left( 1 + \hat{n}_{\text{a}} \right) \right]  \hat{a}  -\frac{\epsilon}{48} \hat{a}^3  + \text{H.c.} +  O(\epsilon^2).
\end{split}
\label{eqn:EffME-DissQu-Xq Trans}
\end{align}

The resulting effective master equation to order $\epsilon$ is obtained
\begin{align}
\begin{split}
\dot{\hat{\rho}}_{\text{a}}(t)& = -i\left[\hat{\mathcal{H}}_{\text{a,eff}},\hat{\rho}_{\text{a}}(t)\right]\\
&+2\kappa_{\text{a}} \mathcal{D}\left[\left(1+\frac{\epsilon}{8}(1+\hat{n}_{\text{a}})\right)\hat{a}\right] \hat{\rho}_{\text{a}}(t)\\
&+2\kappa_{\text{a}3}\mathcal{D}\left[\frac{\epsilon}{16}\hat{a}^3\right]\hat{\rho}_{\text{a}}(t),
\end{split}
\label{eqn:EffME-DissQu-Lind Eq1}
\end{align}
where $2\kappa_{\text{a}}\equiv S_{XX}(\omega_{\text{a}})$, $2\kappa_{\text{a}3}\equiv S_{XX}(3\omega_{\text{a}})$. It is important to notice the operator nature of the relaxation renormalization, which becomes
manifest with the appearance of a nonlinear collapse operator
correction at order $\epsilon$. In arriving at the EME~\ref{eqn:EffME-DissQu-Lind Eq1}, we have assumed that the bath spectral function is insensitive within an $O(\epsilon)$ window around the qubit frequency. A more exact representation can be achieved by projecting onto the qubit eigenbasis in which the Hamiltonian is diagonal. The derivation and connection between these two representations is discussed in App.~\ref{App:EMEDer}.

\begin{figure}[t!]
\centering
\subfloat[\label{subfig:SWCorr-Qu-PhSpace1}]{%
\includegraphics[scale=0.39]{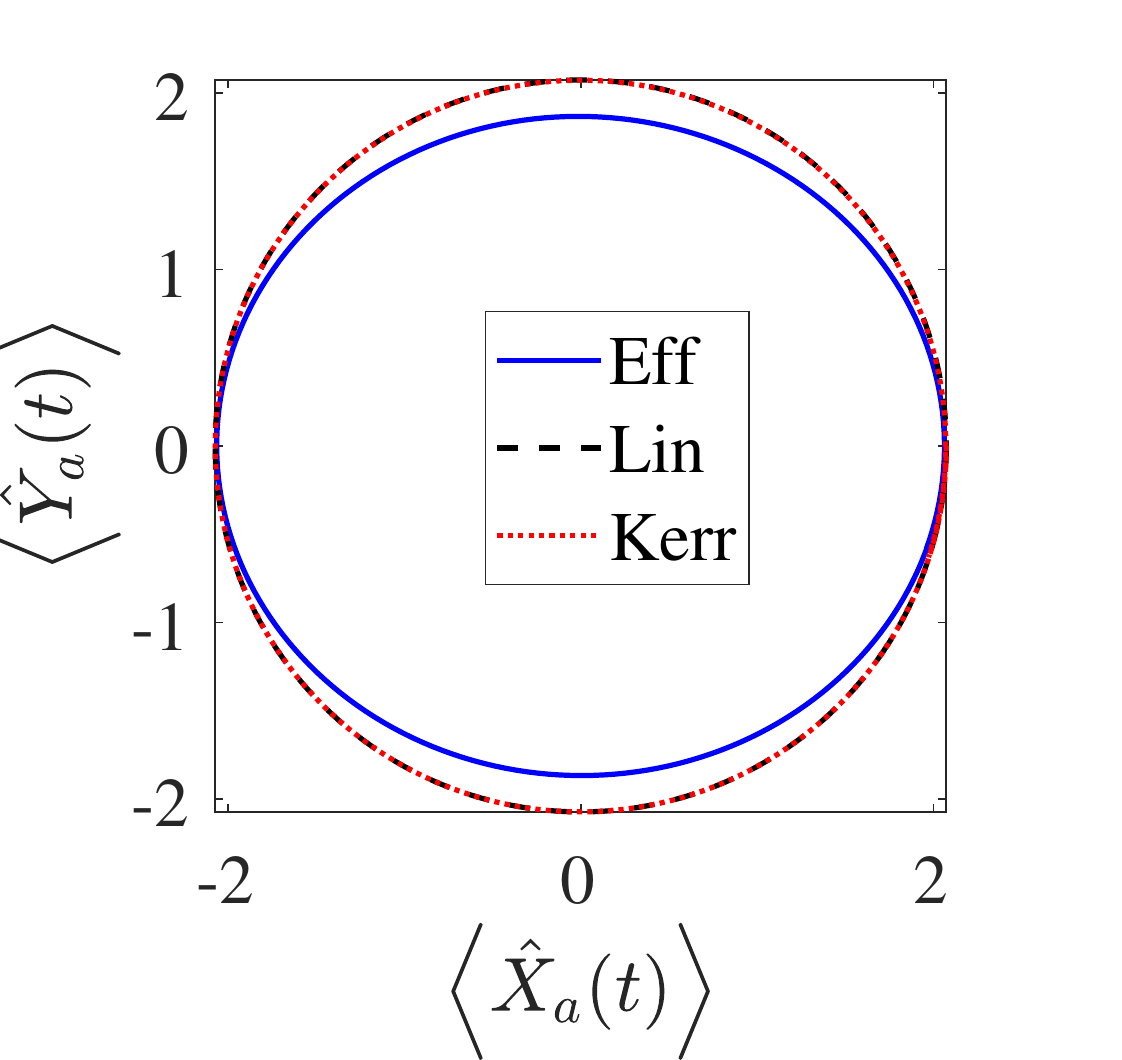}%
}
\subfloat[\label{subfig:SWCorr-Qu-PhSpace2}]{%
\includegraphics[scale=0.39]{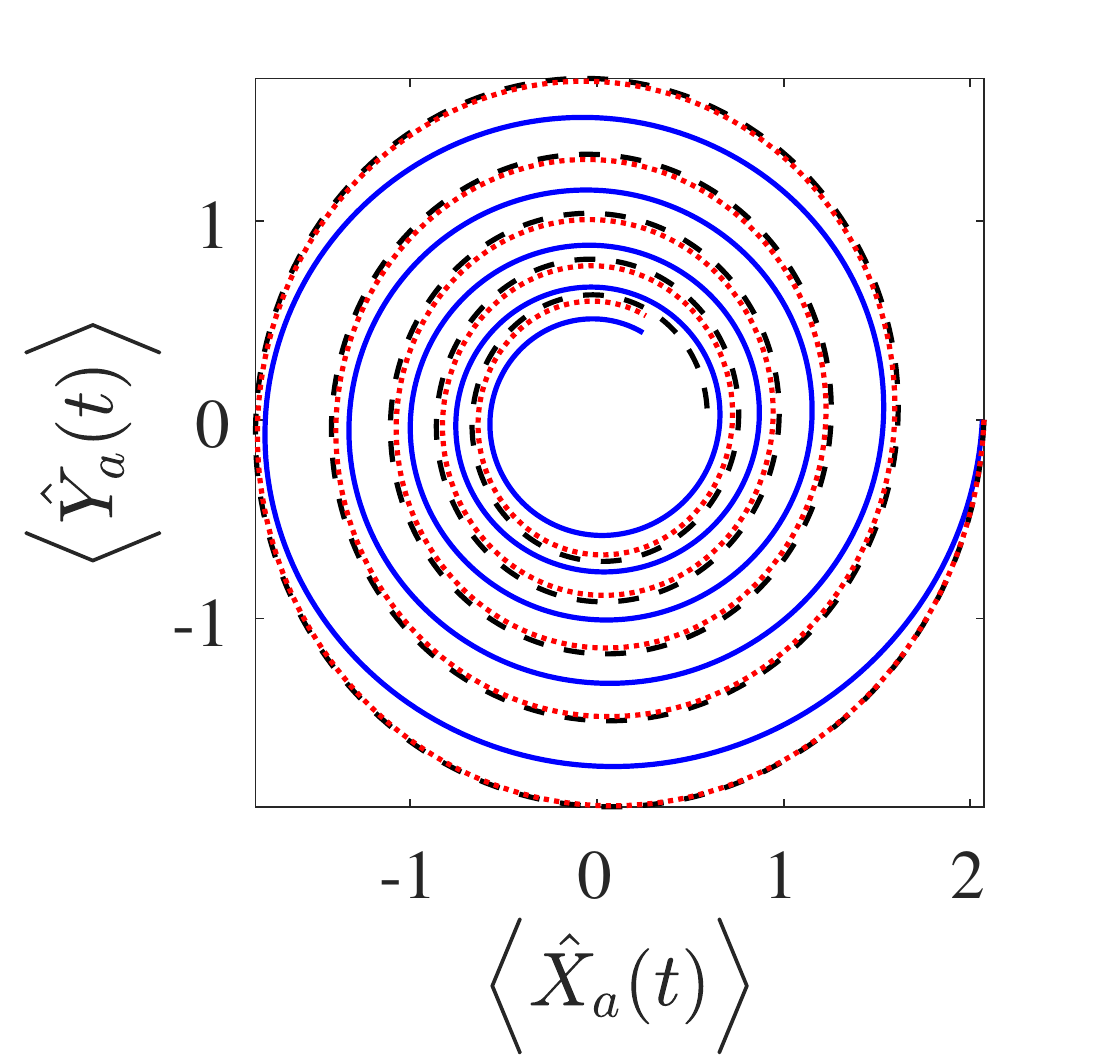}%
}\\
\subfloat[\label{subfig:SWCorr-Qu-ada}]{%
\includegraphics[scale=0.275]{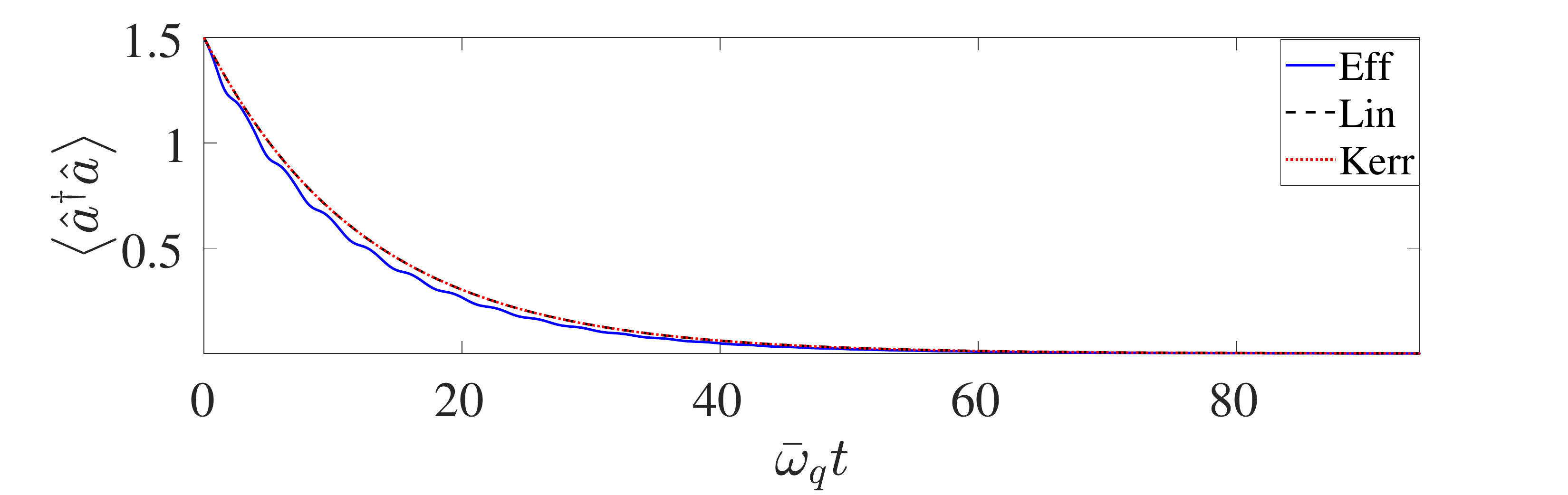}%
}
\caption{Comparison between different theories mentioned in Table~\ref{tab:EffME-Comparison} for $E_{\text{j}}=50E_{\text{c}}$ $(\epsilon=0.2)$ and qubit initial condition $\ket{\Psi_{\text{a}}(0)}=\frac{1}{\sqrt{4}}\sum_{m=0}^{3}\ket{m}$. This initial condition is adopted to have non-zero matrix elements for both single- and three-photon relaxation processes. a) Phase space during the first period for $\kappa_{\text{a}}=\kappa_{\text{a3}}=0$, b) Phase space and c) Qubit occupation number for $\kappa_{\text{a}}=\kappa_{\text{a3}}=\omega_{\text{a}}/25$. The bath spectral function $S_{YY}(\omega)$ is assumed to be flat for simplicity.}
\color{black}
\label{Fig:SWCorr-Qu-PhSpace}
\end{figure}

Next, we compare the numerical predictions from the effective theory to the Kerr and linear theories as introduced in Table~\ref{tab:EffME-Comparison}. The corrections in the effective theory, compared to Kerr, can be summarized in terms of two separate effects. First, note that the starting order-$\epsilon$ qubit Hamiltonian~(\ref{eqn:EffME-Qu-Def of Hq}) does not commute with and hence does not conserve the qubit number operator $\hat{n}_{\text{a}}$ as opposed to the Kerr Hamiltonian~(\ref{eqn:EffME-Qu-H_PT}). This suggests that the Kerr theory predicts \textit{circular} constant energy contours, while the true constant energy contours are non-circular and obey $X_{\text{a}}^2+Y_{\text{a}}^2-\epsilon/12 X_{\text{a}}^4=\text{C}$. Within the context of our method, the information about number non-conserving terms are implicitly encoded in the transformation $\exp(-\epsilon\hat{G}_4)$, which needs to be consistently applied to the density matrix at all times $t\geq 0$ when calculating the expectation values [See e.g. the discussion under Eq.~(21)]. Second, on top of this non-circular time evolution, the EME provides renormalized dissipators, that cause an increase in both the effective single-photon and three-photon relaxation rates.

The first effect can be clearly observed by turning off the dissipation and plotting the phase space of the qubit for the first oscillation period (Fig.~\ref{subfig:SWCorr-Qu-PhSpace1}). The Kerr theory predicts a circular orbit and lies on top of the linear theory ($\epsilon=0$). This is expected since the Kerr theory is diagonal in the number basis and only corrects the transition frequencies, hence in phase space the oscillator only rotates with a slower angular frequency. On the other hand, the effective theory accounts for the effect of non-secular terms which renormalize the constant energy contours. Next, we turn on dissipation and compare the dynamics for the phase space as well as the qubit ocupation number in Figs.~\ref{subfig:SWCorr-Qu-PhSpace2}-\ref{subfig:SWCorr-Qu-ada}, respectively. On top of the non-circular phase space evolution, that leads to a non-exponential decay in the occupation number dynamics, we observe that the effective theory predicts a faster rate compared to the Kerr theory which is more or less the same as that of the linear theory.
\color{black}
\subsection{Case (ii): Purcell physics}
\label{SubSec:EffME-Qu1Mode}

\begin{table}
  \begin{tabular}{|c|c|}
    \hline  
    Operator & Coefficient \\
    \hline\hline
     $\hat{a}$ & $\frac{i}{8}\frac{\bar{\omega}_{\text{a}}}{\omega_{\text{a}}}u_{\text{aa}}^2\left(u_{\text{aa}}^2+u_{\text{ac}}^2\right)$\\   
    \hline
    $\hat{a}^{\dag}$ & $\text{c.c.}$\\   
    \hline
     $\hat{a}^\dagger \hat{a} \hat{a}$ & $\frac{i}{8}\frac{\bar{\omega}_{\text{a}}}{\omega_{\text{a}}}u_{\text{aa}}^4$ \\
 	\hline
    $\hat{a}^\dagger \hat{a}^{\dag} \hat{a}$ & $\text{c.c.}$ \\
 	\hline
     $\hat{a}^3$ & $\frac{i}{16}\frac{\bar{\omega}_{\text{a}}}{\omega_{\text{a}}}u_{\text{aa}}^4$ \\ 
     \hline
     $\left(\hat{a}^{\dag}\right)^3$ & $\text{c.c.}$ \\ 
    \hline\hline
      $\hat{c}$   & $\frac{i}{4}\left(\frac{\bar{\omega}_{\text{a}}}{\omega_{\text{a}}+\omega_{\text{c}}}-\frac{\bar{\omega}_{\text{a}}}{\omega_{\text{a}}-\omega_{\text{c}}}\right)u_{\text{aa}}u_{\text{ac}}\left(u_{\text{aa}}^2+u_{\text{ac}}^2\right)$\\
    \hline
      $\hat{c}^{\dag}$   & $\text{c.c.}$\\
    \hline
     $\hat{c}^\dagger \hat{c} \hat{c}$ & $\frac{i}{4}\left(\frac{\bar{\omega}_{\text{a}}}{\omega_{\text{a}}+\omega_{\text{c}}}-\frac{\bar{\omega}_{\text{a}}}{\omega_{\text{a}}-\omega_{\text{c}}}\right)u_{\text{aa}}u_{\text{ac}}^3$ \\
    \hline        
     $\hat{c}^\dagger \hat{c} ^{\dag}\hat{c}$ & $\text{c.c.}$ \\
    \hline        
     $\hat{c}^3$ & $\frac{i}{12}\left(\frac{\bar{\omega}_{\text{a}}}{\omega_{\text{a}}+3\omega_{\text{c}}}-\frac{\bar{\omega}_{\text{a}}}{\omega_{\text{a}}-3\omega_{\text{c}}}\right)u_{\text{aa}}u_{\text{ac}}^3$ \\
     \hline
     $\left(\hat{c}^{\dag}\right)^3$ & $\text{c.c.}$\\
    \hline\hline
      $\hat{c}^\dagger \hat{c}\hat{a}$  &  $\frac{i}{4}\frac{\bar{\omega}_{\text{a}}}{\omega_{\text{a}}} u_{\text{aa}}^2u_{\text{ac}}^2$\\
    \hline
    $\hat{a}^{\dag} \hat{c}^{\dag}\hat{c}$  &  $\text{c.c.}$\\
    \hline
   $\hat{a}^\dagger \hat{a}\hat{c}$  &  $\frac{i}{2}\left(\frac{\bar{\omega}_{\text{a}}}{\omega_{\text{a}}+\omega_{\text{c}}}-\frac{\bar{\omega}_{\text{a}}}{\omega_{\text{a}}-\omega_{\text{c}}}\right)u_{\text{aa}}^3u_{\text{ac}}$\\
    \hline
     $\hat{c}^{\dag} \hat{a}^{\dag}\hat{a}$  &  $\text{c.c.}$\\
    \hline
   $\hat{c}^2\hat{a}$   &  $\frac{i}{8}\left(\frac{\bar{\omega}_{\text{a}}}{\omega_{\text{c}}+\omega_{\text{a}}}+\frac{\bar{\omega}_{\text{a}}}{\omega_{\text{c}}}\right)u_{\text{aa}}^2 u_{\text{ac}}^2$ \\
    \hline
  $\hat{a}^{\dag}\left(\hat{c}^{\dag}\right)^2$   &  $\text{c.c.}$ \\
    \hline
    $\left(\hat{c}^\dagger\right)^2\hat{a} $   & $\frac{i}{8}\left(\frac{\bar{\omega}_{\text{a}}}{\omega_{\text{a}}-\omega_{\text{c}}}-\frac{\bar{\omega}_{\text{a}}}{\omega_{\text{c}}}\right)u_{\text{ac}}^2 u_{\text{aa}}^2$\\    
    \hline
        $\hat{a}^{\dag}\hat{c}^2$   & $c.c.$\\    
    \hline
    $\hat{a}^2 \hat{c}$   &  $\frac{i}{4}\left(\frac{\bar{\omega}_{\text{a}}}{\omega_{\text{c}}+3 \omega_{\text{a}}}+\frac{\bar{\omega}_{\text{a}}}{\omega_{\text{c}}+\omega_{\text{a}}}\right)u_{\text{ac}} u_{\text{aa}}^3$\\    
    \hline
    $\hat{c}^{\dag} \left(\hat{a}^{\dag}\right)^2$ & $\text{c.c.}$\\    
    \hline
    $\left(\hat{a}^{\dag}\right)^2 \hat{c}$   &  $\frac{i}{4}\left(\frac{\bar{\omega}_{\text{a}}}{\omega_{\text{c}}-\omega_{\text{a}}}+\frac{\bar{\omega}_{\text{a}}}{\omega_{\text{c}}-3 \omega_{\text{a}}}\right)u_{\text{ac}} u_{\text{aa}}^3$\\    
    \hline
    $\hat{c}^{\dag} \hat{a}^2$   &  $\text{c.c.}$\\    
    \hline
  \end{tabular}
\caption{\label{tab:EffME-Qu1Mode}The contributions in $\left[ \hat{Y}_{\text{a}}, \hat{G}_4\right]$, which provide the lowest order renormalization of the qubit-like quadrature. The left column shows each operator entering the sum, and the right column shows its coefficient. The double horizontal lines separate the contributions into three distinct categories from top to bottom: 1) self, 2) cross and 3) mixed. The result for cavity-like quadrature, i.e. $[\hat{Y}_{\text{c}},\hat{G}_4]$, can be immediately found from this table by the simultaneous replacements $u_{\text{aa}}\leftrightarrow u_{\text{ac}}$, $\omega_{\text{a}}\leftrightarrow \omega_{\text{c}}$, $\hat{a}\leftrightarrow \hat{c}$, while the bare qubit frequency $\bar{\omega}_{\text{a}}$ remains intact.}
\end{table}
\color{black}

This section is devoted to the case of a weakly anharmonic qubit coupled to a single open resonator mode as a typical setup for studying the Purcell effect \cite{Purcell_Resonance_1946, Purcell_Spontaneous_1995}. This is the model introduced in Section~\ref{Sec:Model}. With respect to the case treated in the previous section, here the focus will be on the physics of mode mixing and its implications on both frequency and decay rate renormalization. We derive an effective master equation that accounts for this renormalization at order $\epsilon$, for which it is sufficient to retain only the quartic terms in the Josephson potential. Additional details for the calculations presented in this section can be found in App.~\ref{SubApp:1stSWPT-Qu1Mode}. 

In the normal mode basis, the system Hamiltonian is  
\begin{align}
\begin{split}
\hat{\mathcal{H}}_{\text{s}}&\equiv\omega_{\text{a}}\left(\hat{a}^{\dag}\hat{a}+\frac{1}{2}\right)+\omega_{\text{c}}\left(\hat{c}^{\dag}\hat{c}+\frac{1}{2}\right)\\
&-\frac{\epsilon\bar{\omega}_{\text{a}}}{48}\left[u_{\text{aa}}\left(\hat{a}+\hat{a}^{\dag}\right)+u_{\text{ac}}\left(\hat{c}+\hat{c}^{\dag}\right)\right]^4,
\end{split}
\label{eqn:Model-Quartic Hs in normpic}
\end{align}
up to lowest order in the anharmonicity. In the following, we apply a unitary transformation such that the effect of the weak quartic anharmonicity in $\hat{\mathcal{H}}_{\text{s}}$ is explicitly accounted for both the Hamiltonian and relaxation rates up to lowest order. 

We start by decomposing the quartic anharmonicity given in Eq.~(\ref{eqn:Model-Quartic Hs in normpic}),  $\hat{H}_4 = \left[u_{\text{aa}}\left(\hat{a}+\hat{a}^{\dag}\right)+u_{\text{ac}}\left(\hat{c}+\hat{c}^{\dag}\right)\right]^4$ into secular and non-secular terms as $\hat{H}_4\equiv \hat{S}_4+\hat{N}_4$. When expanded in terms of qubit-like and cavity-like bosonic operators, there are a total of $256$ distinct monomials in $\hat{H}_4$. The secular terms $\hat{S}_4$ can be expressed in terms of the number operators $\hat{n}_{\text{a,c}}$ as    
\begin{align}
\begin{split}
\hat{S}_4&=6\left(u_{\text{aa}}^4+2u_{\text{aa}}^2u_{\text{ac}}^2\right)\hat{n}_{\text{a}}+6\left(u_{\text{ac}}^4+2u_{\text{aa}}^2u_{\text{ac}}^2\right)\hat{n}_{\text{c}}\\
&+6u_{\text{aa}}^4\hat{n}_{\text{a}}^2+6u_{\text{ac}}^4\hat{n}_{\text{c}}^2+24u_{\text{aa}}^2u_{\text{ac}}^2\hat{n}_{\text{a}}\hat{n}_{\text{c}}.
\end{split}
\label{eqn:EffME-Qu1Mode-S4}
\end{align}
Following our previous discussion in Sec.~\ref{Sec:HierEqs}, only non-secular terms can be removed by a unitary transformation, and hence the secular terms $\hat{S}_4$ provide the lowest order correction to the Hamiltonian as (See App.~\ref{SubApp:1stSWPT-Qu1Mode} for details)
\begin{align}
\begin{split}
\hat{\mathcal{H}}_{\text{s,eff}}&=\left[\omega_{\text{a}}-\frac{\epsilon\bar{\omega}_{\text{a}}}{8}\left(u_{\text{aa}}^4+2u_{\text{aa}}^2u_{\text{ac}}^2\right)\right]\hat{n}_{\text{a}}\\
&+\left[\omega_{\text{c}}-\frac{\epsilon\bar{\omega}_{\text{a}}}{8}\left(u_{\text{ac}}^4+2u_{\text{ac}}^2u_{\text{aa}}^2\right)\right]\hat{n}_{\text{c}}\\
&-\frac{\epsilon\bar{\omega}_{\text{a}}}{8}\left(u_{\text{aa}}^4\hat{n}_{\text{a}}^2+u_{\text{ac}}^4\hat{n}_{\text{c}}^2+4u_{\text{aa}}^2u_{\text{ac}}^2\hat{n}_{\text{a}}\hat{n}_{\text{c}}\right).
\end{split}
\label{eqn:EffME-Qu1Mode-Hsw}
\end{align}
Equation~(\ref{eqn:EffME-Qu1Mode-Hsw}) describes the normal mode oscillations of a qubit-resonator system renormalized by self-Kerr and cross-Kerr contributions, whose strength is determined by the hybridization coefficients. This result is consistent with the common Kerr theory, which is derived by applying RWA to the original model Eq.(~\ref{eqn:Model-Hs 1}).

The generator $\hat{G}_4$ that removes the non-secular terms in $\hat{H}_4$ can be found by solving 
\begin{align}
[\omega_{\text{a}}\hat{H}_{\text{a}}+\omega_{\text{c}}\hat{H}_{\text{c}},\hat{G}_4]-\frac{\bar{\omega}_{\text{a}}}{48}\hat{N}_4=0,
\label{eqn:EffME-Qu1Mode-Eq for N4}
\end{align}
where we have replaced $\hat{\mathcal{H}}_2\equiv \omega_{\text{a}}\hat{H}_{\text{a}}+\omega_{\text{c}}\hat{H}_{\text{c}}$ and $\hat{\mathcal{N}}_4\equiv \bar{\omega}_{\text{a}}/48\hat{N}_4$ in the generic condition~(\ref{eqn:HierEqs-Cond G4}). Due to the large number of distinct monomials in $\hat{N}_4$, it is not straightforward to bookkeep them manually. The resulting correction to the qubit- and cavity-like quadratures are presented in Table~\ref{tab:EffME-Qu1Mode}, which accounts for all the processes involving single- and three-photon nonlinear interaction with the bath.   

Which of the single or the three-photon interactions are dominant in the qubit dynamics depends on the system parameters (e.g. the relative normal mode frequencies) as well as the initial conditions. For example, assuming that the qubit is initially prepared in the linear combination of the ground and the first excited states, then the three-photon processes play little role in the dynamics. With this assumption, keeping only the renormalization originating from the single-photon system-bath interactions, we obtain from Table~\ref{tab:EffME-Qu1Mode}
\begin{subequations}
\begin{align}
\begin{split}
&e^{-\hat{G}}\hat{Y}_{\text{a}} e^{\hat{G}} = -i \hat{a}\\
&+i \frac{\epsilon}{8}\frac{\bar{\omega}_{\text{a}}}{\omega_{\text{a}}}u_{\text{aa}}^2\left(u_{\text{aa}}^2+u_{\text{ac}}^2+u_{\text{aa}}^2\hat{n}_{\text{a}}+2u_{\text{ac}}^2\hat{n}_{\text{c}}\right)\hat{a}\\
&+i\frac{\epsilon}{2}\frac{\bar{\omega}_{\text{a}}\omega_{\text{c}}}{\omega_{\text{c}}^2-\omega_{\text{a}}^2}u_{\text{aa}}u_{\text{ac}}\left(u_{\text{aa}}^2+u_{\text{ac}}^2+u_{\text{ac}}^2\hat{n}_{\text{c}}+2u_{\text{aa}}^2\hat{n}_{\text{a}}\right)\hat{c},\\
&+\text{H.c.}+O(\epsilon^2),
\end{split}
\label{eqn:EffME-Qu1Mode-X_q,pt-Y_q}\\
\begin{split}
&e^{-\hat{G}}\hat{Y}_\text{c} e^{\hat{G}} = -i\hat{c}\\
&+i\frac{\epsilon}{8}\frac{\bar{\omega}_{\text{a}}}{\omega_{\text{c}}}u_{\text{ac}}^2\left(u_{\text{ac}}^2+u_{\text{aa}}^2+u_{\text{ac}}^2\hat{n}_{\text{c}}+2u_{\text{aa}}^2\hat{n}_{\text{a}}\right)\hat{c}\\
&+i\frac{\epsilon}{2}\frac{\bar{\omega}_{\text{a}}\omega_{\text{a}}}{\omega_{\text{a}}^2-\omega_{\text{c}}^2}u_{\text{ac}}u_{\text{aa}}\left(u_{\text{ac}}^2+u_{\text{aa}}^2+u_{\text{aa}}^2\hat{n}_{\text{a}}+2u_{\text{ac}}^2\hat{n}_{\text{c}}\right)\hat{a},\\
&+\text{H.c.}+O(\epsilon^2).
\end{split}
\label{eqn:EffME-Qu1Mode-X_c,pt-Y_c}
\end{align}
\end{subequations}
\color{black}
According to Eqs.~(\ref{eqn:EffME-Qu1Mode-X_q,pt-Y_q}-\ref{eqn:EffME-Qu1Mode-X_c,pt-Y_c}), we find  that the interaction of each normal mode with the bath obtains corrections that are proportional to both itself and the other normal mode. At zero coupling, i.e. where $g=0$ and hence $u_{\text{aa}}=1$ and $u_{\text{ac}}=0$, we recover the linear correction for case (i) in Eq.~(\ref{eqn:EffME-DissQu-Xq Trans}). Moreover, we need to recall that the bare cavity quadrature coupling to the bath translates to $\hat{\bar{Y}}_{\text{c}}=v_{\text{cc}}\hat{Y}_{\text{c}}+v_{\text{ca}}\hat{Y}_{\text{a}}$ in terms of the normal modes. Combining the linear and nonlinear renormalizations, we can obtain an effective $\epsilon$-order Lindblad equation as
\begin{align}
\begin{split}
\dot{\hat{\rho}}_{\text{s}}(t)=-i\left[\hat{\mathcal{H}}_{\text{s,eff}},\hat{\rho}_{\text{s}}\right]&+2\kappa_{\text{c}}\mathcal{D}[\hat{C}_{\text{c,eff}}]\hat{\rho}_{\text{s}}(t)
\\
&+2\kappa_{\text{a}}\mathcal{D}[\hat{C}_{\text{a,eff}}]\hat{\rho}_{\text{s}}(t).
\end{split}
\label{eqn:EffME-Qu1Mode-PT Lind Eq}
\end{align}
where $2\kappa_{\text{a},\text{c}}\equiv S_{YY}(\omega_{\text{a},\text{c}})$. Moreover, the effective qubit- and cavity-like single-photon collapse operators read
\begin{subequations}
\begin{align}
\begin{split}
&\hat{C}_{\text{a,eff}}=\left[v_{\text{ca}}-\frac{\epsilon}{8}\frac{\bar{\omega}_{\text{a}}}{\omega_{\text{a}}}v_{\text{ca}}u_{\text{aa}}^2\left(u_{\text{aa}}^2+u_{\text{ac}}^2+u_{\text{aa}}^2\hat{n}_{\text{a}}+2u_{\text{ac}}^2\hat{n}_{\text{c}}\right)\right.\\
&-\left.\frac{\epsilon}{2}\frac{\bar{\omega}_{\text{a}}\omega_{\text{a}}}{\omega_{\text{a}}^2-\omega_{\text{c}}^2}v_{\text{cc}}u_{\text{ac}}u_{\text{aa}}\left(u_{\text{ac}}^2+u_{\text{aa}}^2+u_{\text{aa}}^2\hat{n}_{\text{a}}+2u_{\text{ac}}^2\hat{n}_{\text{c}}\right)\right]\hat{a},
\end{split}
\label{eqn:EffME-Qu1Mode-Def of c_{q,PT}}
\end{align}
\begin{align}
\begin{split}
&\hat{C}_{\text{c,eff}}=\left[v_{\text{cc}}-\frac{\epsilon}{8}\frac{\bar{\omega}_{\text{a}}}{\omega_{\text{c}}}v_{\text{cc}}u_{\text{ac}}^2\left(u_{\text{ac}}^2+u_{\text{aa}}^2+u_{\text{ac}}^2\hat{n}_{\text{c}}+2u_{\text{aa}}^2\hat{n}_{\text{a}}\right)\right.\\
&-\left.\frac{\epsilon}{2}\frac{\bar{\omega}_{\text{a}}\omega_{\text{c}}}{\omega_{\text{c}}^2-\omega_{\text{a}}^2}v_{\text{ca}}u_{\text{aa}}u_{\text{ac}}\left(u_{\text{aa}}^2+u_{\text{ac}}^2+u_{\text{ac}}^2\hat{n}_{\text{c}}+2u_{\text{aa}}^2\hat{n}_{\text{a}}\right)\right]\hat{c}.
\end{split}
\label{eqn:EffME-Qu1Mode-Def of c_{c,PT}}
\end{align}
\end{subequations}
\color{black}
\begin{figure}[t!]
\centering
\subfloat[\label{subfig:HybCoeffsForXWq08WcFuncOfg}]{%
\includegraphics[scale=0.375]{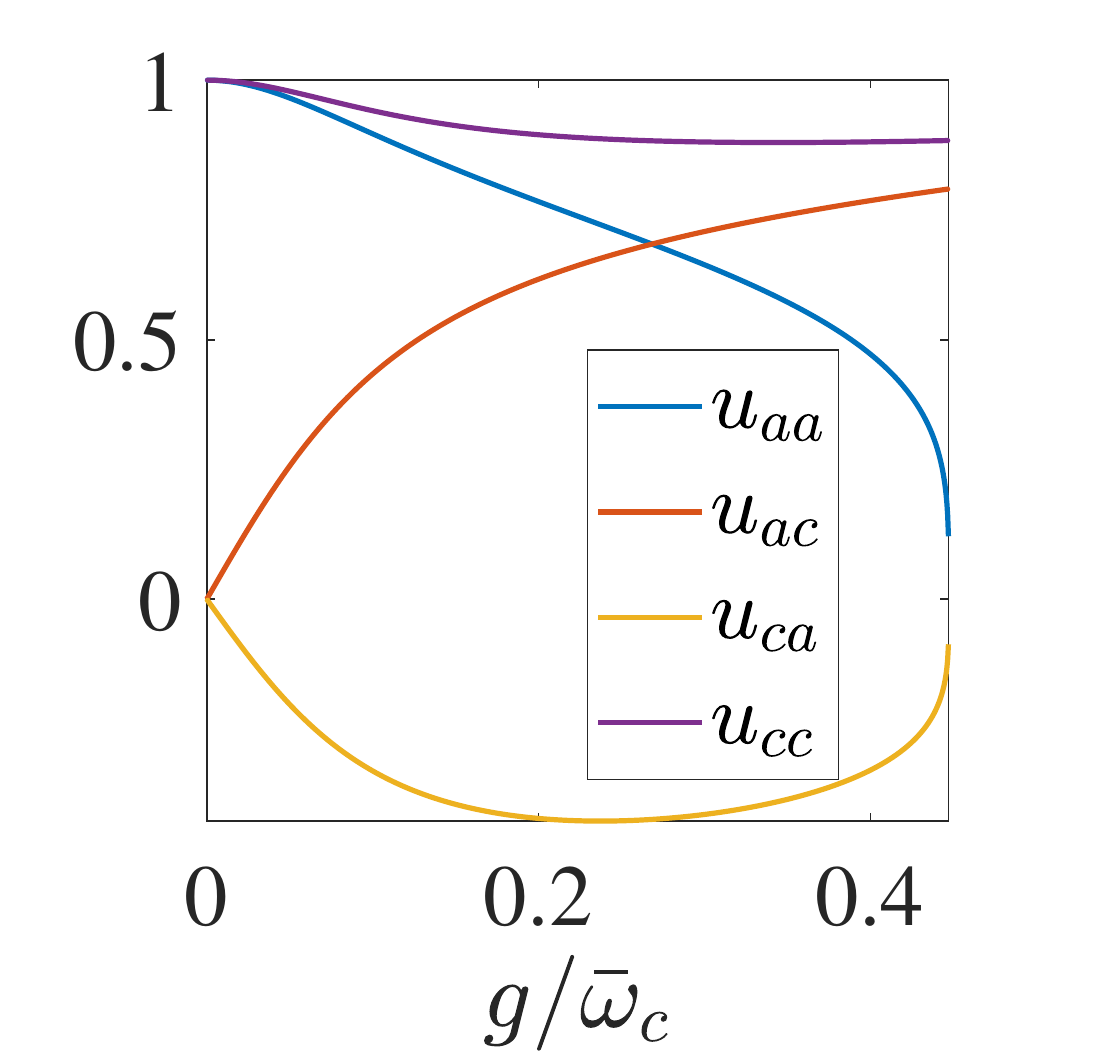}%
}
\subfloat[\label{subfig:HybCoeffsForYWq08WcFuncOfg}]{%
\includegraphics[scale=0.375]{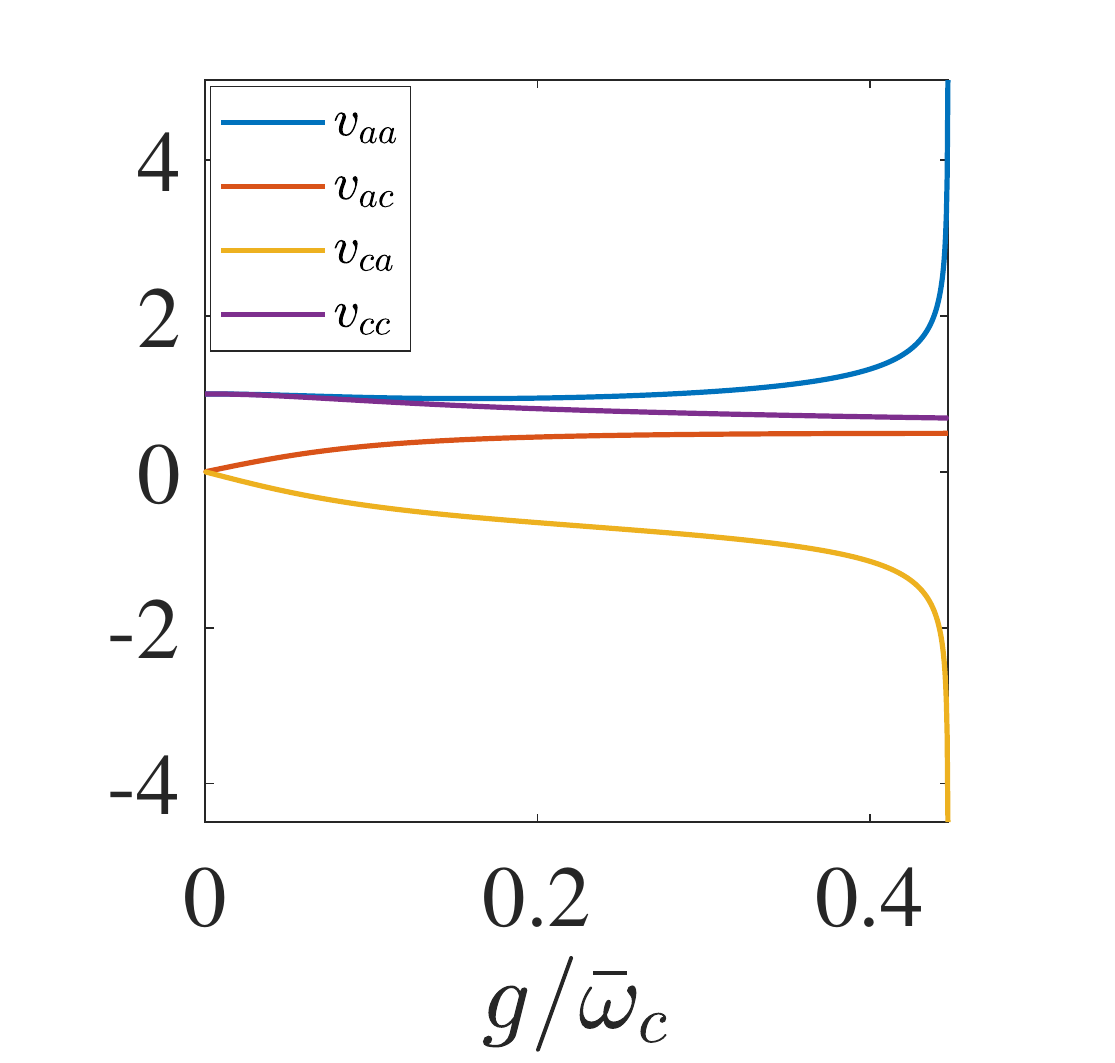}%
}\\
\subfloat[\label{subfig:NormFreqsWq08WcFuncOfg}]{%
\includegraphics[scale=0.375]{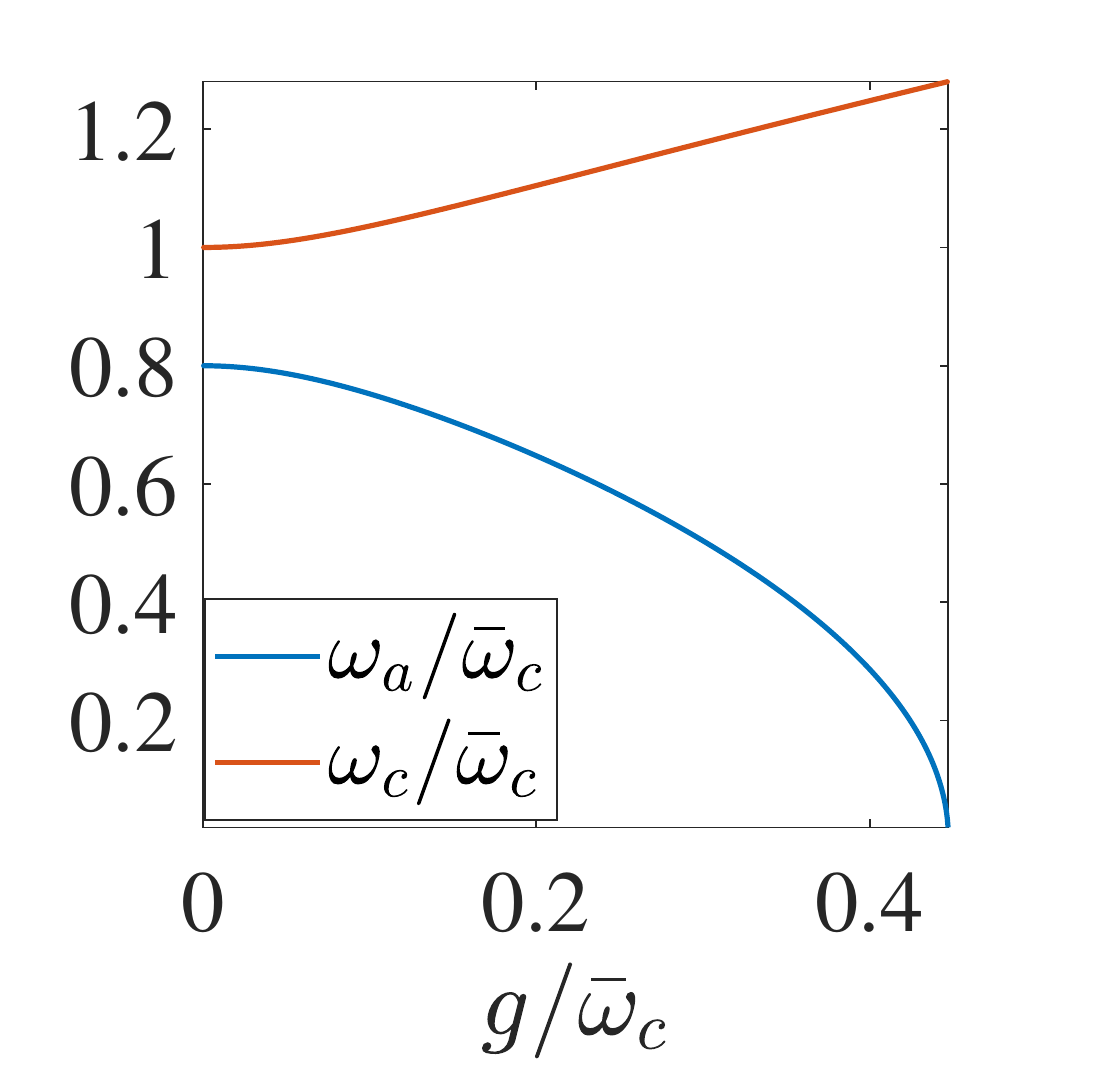}%
}
\subfloat[\label{subfig:DissCorrectionWq08WcFuncOfg}]{%
\includegraphics[scale=0.375]{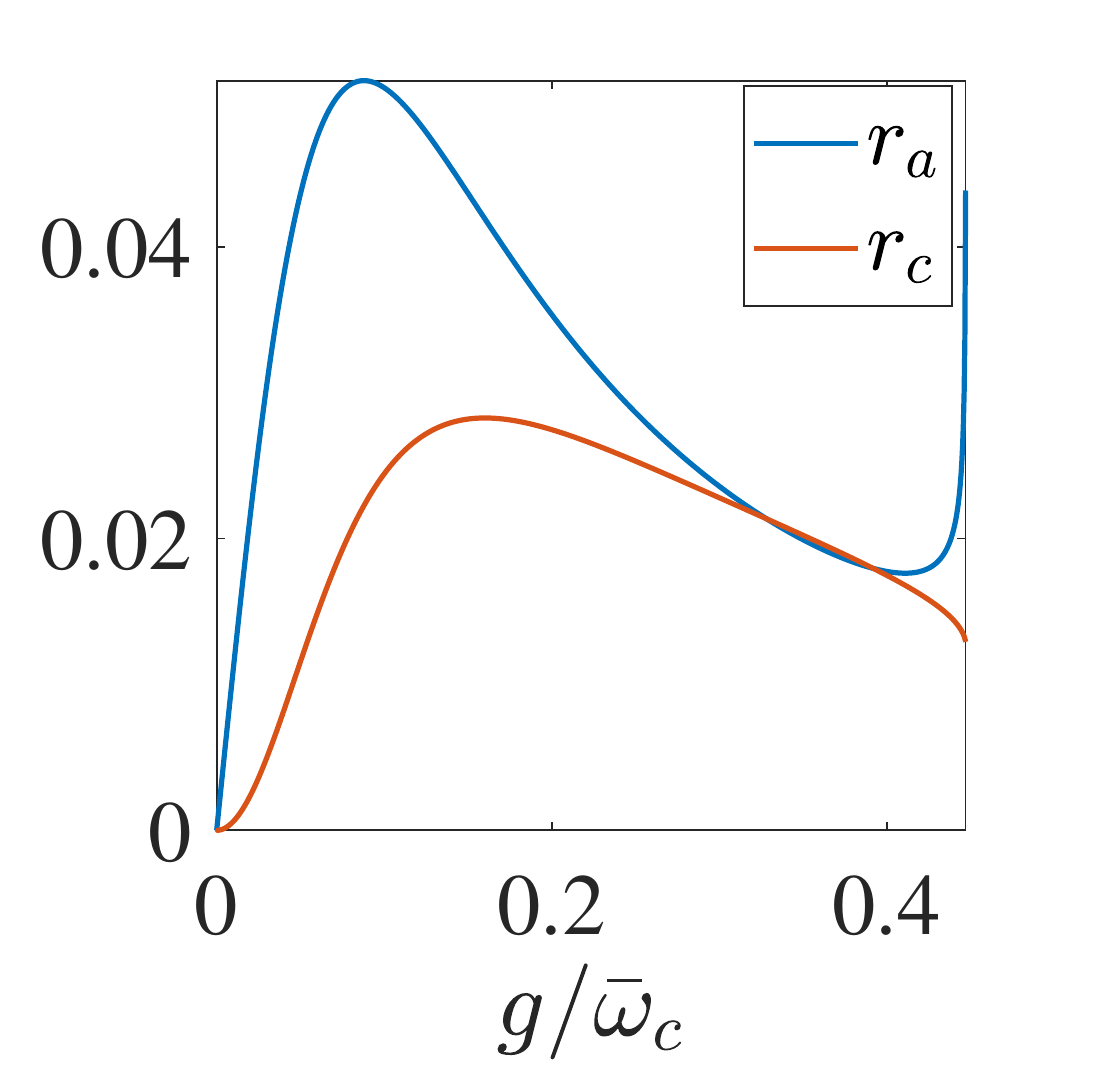}%
}
\caption{Hybridization as a function of $g$ for $\bar{\omega}_{\text{a}}=0.8\bar{\omega}_{\text{c}}$ ($\bar{\omega}_{\text{a}}<\bar{\omega}_{\text{c}}$). a,b) show hybridization coefficients obtained from Eqs.~(\ref{Eq:1stPT-Qu1Mode-Def of HybCoefForX}-\ref{Eq:1stPT-Qu1Mode-Def of HybCoefForY}),  c) show the normal mode frequencies obtained from Eqs.~(\ref{Eq:1stPT-Qu1Mode-Def of bar(v)_q}-\ref{Eq:1stPT-Qu1Mode-Def of bar(v)_c}), and d) show the order $\epsilon$ correction to the normal mode dissipators according to Eqs.~(\ref{eqn:EffMe-Qu1Mode-Def of r_q}-\ref{eqn:EffMe-Qu1Mode-Def of r_c}). The last value of $g$ is chosen in each case such that the lower normal mode frequency hits $0^+$.\color{black}}
\label{Fig:NormFreqs&HybWq08Wc}
\end{figure}
\begin{figure}
\subfloat[\label{subfig:HybCoeffsForXWc08WqFuncOfg}]{%
\includegraphics[scale=0.375]{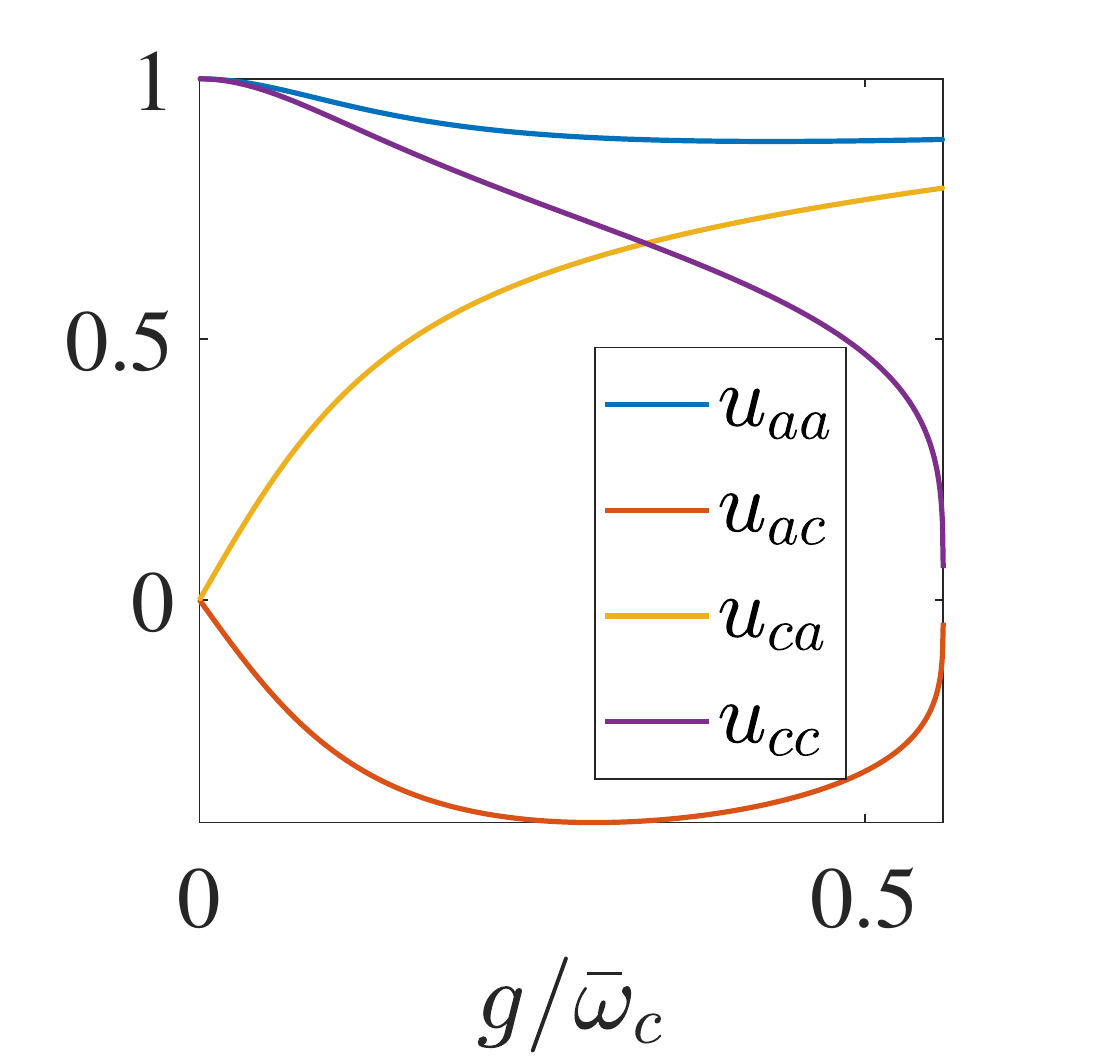}%
}
\subfloat[\label{subfig:HybCoeffsForYWc08WqFuncOfg}]{%
\includegraphics[scale=0.375]{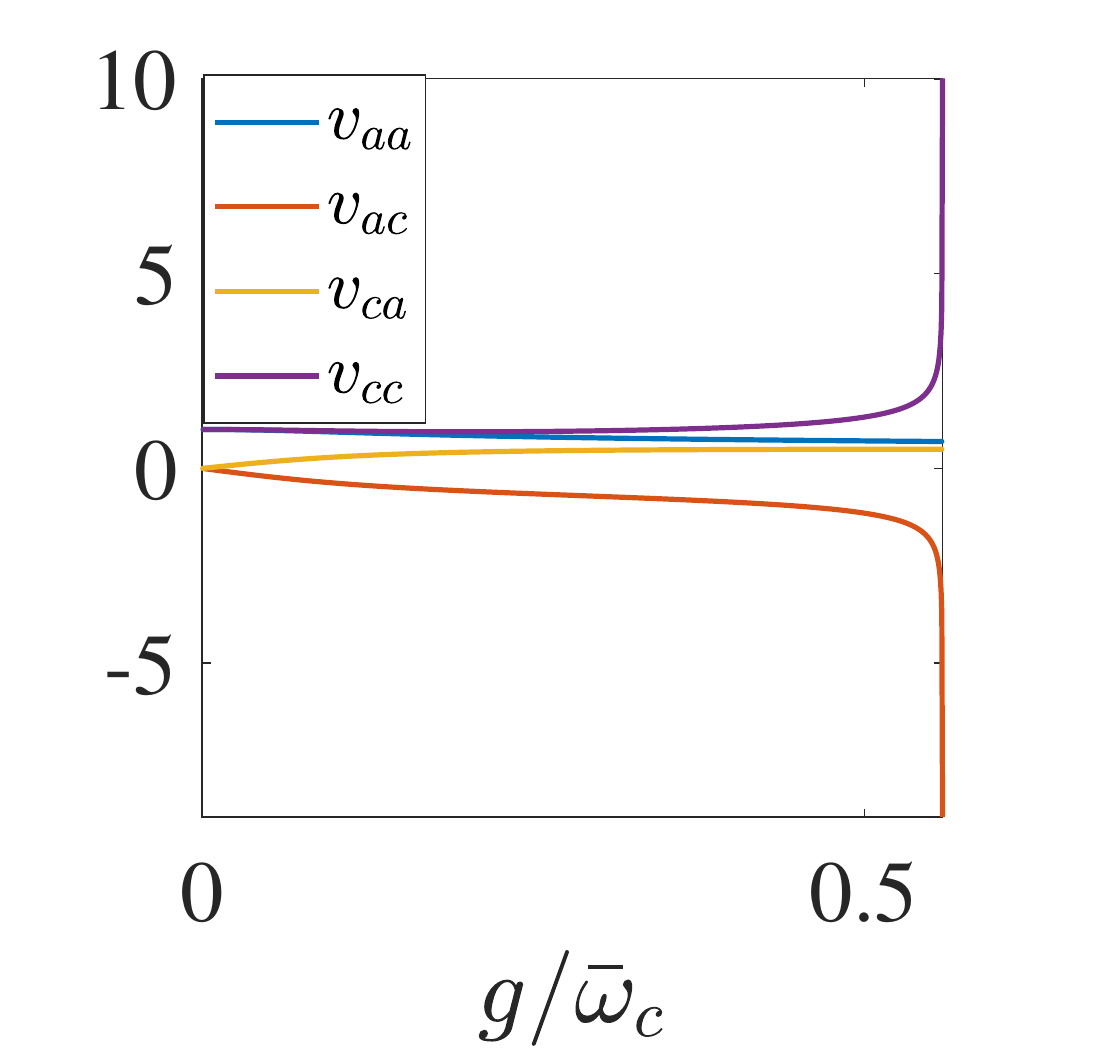}%
}\\
\subfloat[\label{subfig:NormFreqsWc08WqFuncOfg}]{%
\includegraphics[scale=0.375]{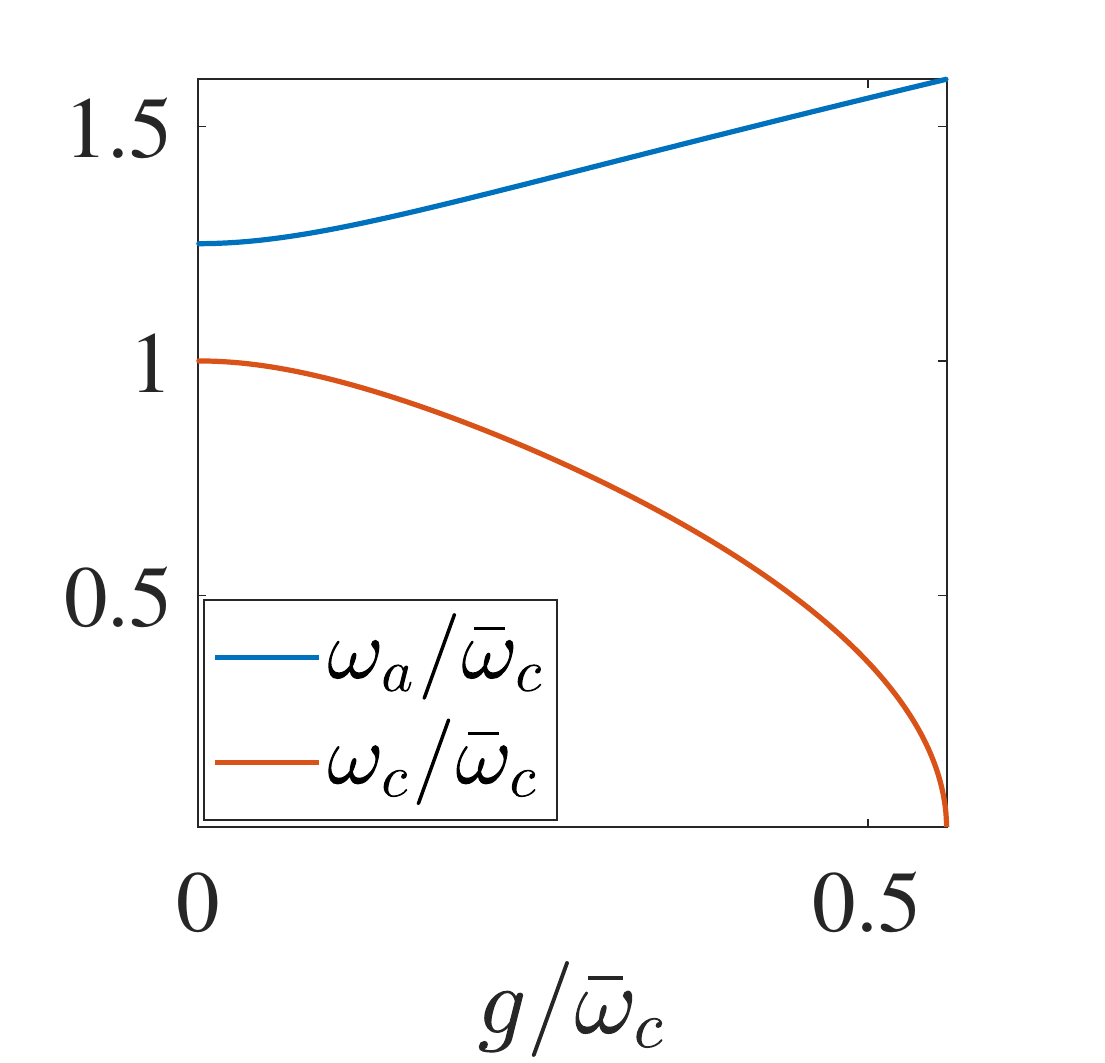}%
}
\subfloat[\label{subfig:DissCorrectionWc08WqFuncOfg}]{%
\includegraphics[scale=0.375]{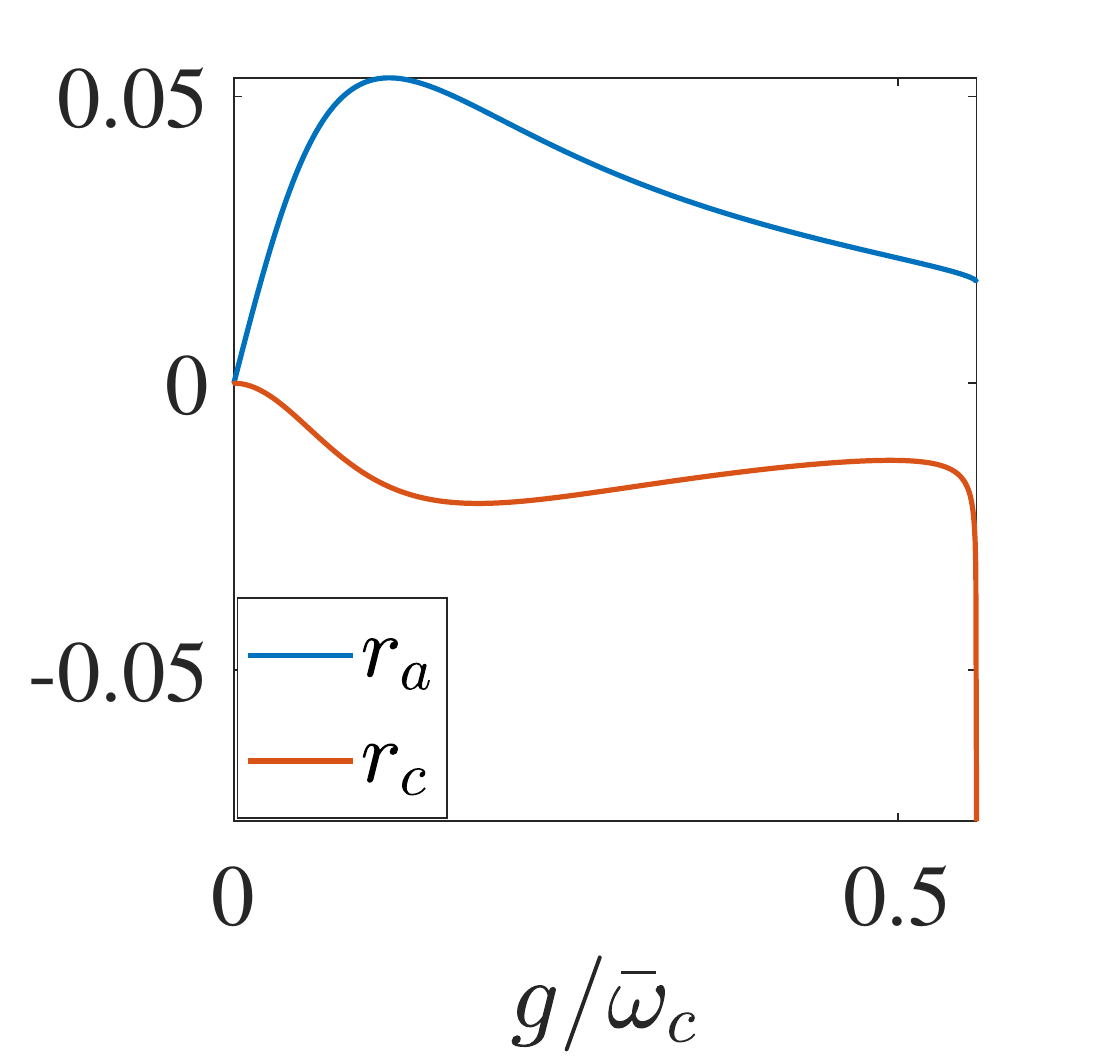}%
}
\caption{Hybridization as a function of $g$ for $\bar{\omega}_c=0.8\bar{\omega}_{\text{a}}$ ($\bar{\omega}_{\text{a}}>\bar{\omega}_c$). a,b) show hybridization coefficients obtained from Eqs.~(\ref{Eq:1stPT-Qu1Mode-Def of HybCoefForX}-\ref{Eq:1stPT-Qu1Mode-Def of HybCoefForY}),  c) show the normal mode frequencies obtained from Eqs.~(\ref{Eq:1stPT-Qu1Mode-Def of bar(v)_q}-\ref{Eq:1stPT-Qu1Mode-Def of bar(v)_c}), and d) show the order $\epsilon$ correction to the normal mode dissipators according to Eqs.~(\ref{eqn:EffMe-Qu1Mode-Def of r_q}-\ref{eqn:EffMe-Qu1Mode-Def of r_c}). The last value of $g$ is chosen in each case such that the lower normal mode frequency hits $0^+$.\color{black}}
\label{Fig:NormFreqs&HybWc08Wq}
\end{figure}

Next, we examine the dissipator renormalizations~(\ref{eqn:EffME-Qu1Mode-Def of c_{q,PT}}-\ref{eqn:EffME-Qu1Mode-Def of c_{c,PT}}) of each normal mode in more detail. We observe that the dissipator renormalizations depend on the hybridization coefficients as well as on the relative position of the qubit-like and cavity-like frequencies. While for a qubit directly coupled to a generic bath [case (i)], the relaxation rate to lowest order can only increase when increasing the anharmonicity parameter $\epsilon$, the additional dependences here suggest a richer possibility for corrections. 

To this end, let us consider first the case where the qubit is detuned below the cavity mode. We study the hybridization coefficients as well as the sign of relaxation renormalization (Fig.~\ref{Fig:NormFreqs&HybWq08Wc}). This choice of parameters leads to a non-trivial hybridization of each normal mode coupling to the bath with opposite signs, i.e. $v_{\text{cc}}>0$ and $v_{\text{ca}}<0$ (Fig.~\ref{subfig:HybCoeffsForXWq08WcFuncOfg}). To assess the overall sign of relaxation correction, we need to compare the sign of $O(\epsilon)$ corrections in Eqs.~(\ref{eqn:EffME-Qu1Mode-Def of c_{q,PT}}-\ref{eqn:EffME-Qu1Mode-Def of c_{c,PT}}) to the $O(1)$ values of $u_{\text{cc}}$ and $u_{\text{ca}}$. Although the $O(\epsilon)$ corrections inside the dissipators are operator-valued, we can assess the sign in terms of the following quantities
\begin{subequations}
\begin{align}
r_{\text{a}}\equiv - \left(\frac{\epsilon}{8}\frac{\bar{\omega}_{\text{a}}}{\omega_{\text{a}}}v_{\text{ca}}u_{\text{aa}}^2+\frac{\epsilon}{2}\frac{\bar{\omega}_{\text{a}}\omega_{\text{a}}}{\omega_{\text{a}}^2-\omega_{\text{c}}^2}v_{\text{cc}}u_{\text{ac}}u_{\text{aa}}\right)\left(u_{\text{aa}}^2+u_{\text{ac}}^2\right),
\label{eqn:EffMe-Qu1Mode-Def of r_q}\\
r_c\equiv - \left(\frac{\epsilon}{8}\frac{\bar{\omega}_{\text{a}}}{\omega_{\text{c}}}v_{\text{cc}}u_{\text{ac}}^2+\frac{\epsilon}{2}\frac{\bar{\omega}_{\text{a}}\omega_{\text{c}}}{\omega_{\text{c}}^2-\omega_{\text{a}}^2}v_{\text{ca}}u_{\text{aa}}u_{\text{ac}}\right)\left(u_{\text{aa}}^2+u_{\text{ac}}^2\right),
\label{eqn:EffMe-Qu1Mode-Def of r_c}
\end{align}
\end{subequations}
that are obtained by setting $n_{\text{a,c}}=0$ in these corrections. We find that both $r_{\text{a,c}}$ are \textit{positive} for $\omega_{\text{a}}<\omega_c$ and for all values of $g$ as seen in Fig.~\ref{subfig:DissCorrectionWq08WcFuncOfg}. Given the fact that any dissipator of the form $\mathcal{D}[\hat{C}]$ is quadratic in terms of its collapse operator $\hat{C}$ and that $u_{\text{ca}}$ is negative for $\omega_{\text{a}}<\omega_c$, we find that the single-photon relaxation is \textit{enhanced (suppressed)} for the normal \textit{cavity (qubit)} mode. We note that setting $\omega_{\text{a}}>\omega_c$ will reverse the result, in which the single-photon relaxation is \textit{enhanced (suppressed)} for the normal \textit{qubit (cavity)} mode (Fig.~\ref{Fig:NormFreqs&HybWc08Wq}). A summary of the sign of single-photon relaxation renormalizations is given in Table~\ref{tab:SignOfCorrections}. 
\begin{figure}[h!]
\centering
\subfloat[\label{subfig:SWCorr-Qu1Mod-RelFreqs}]{%
\includegraphics[scale=0.135]{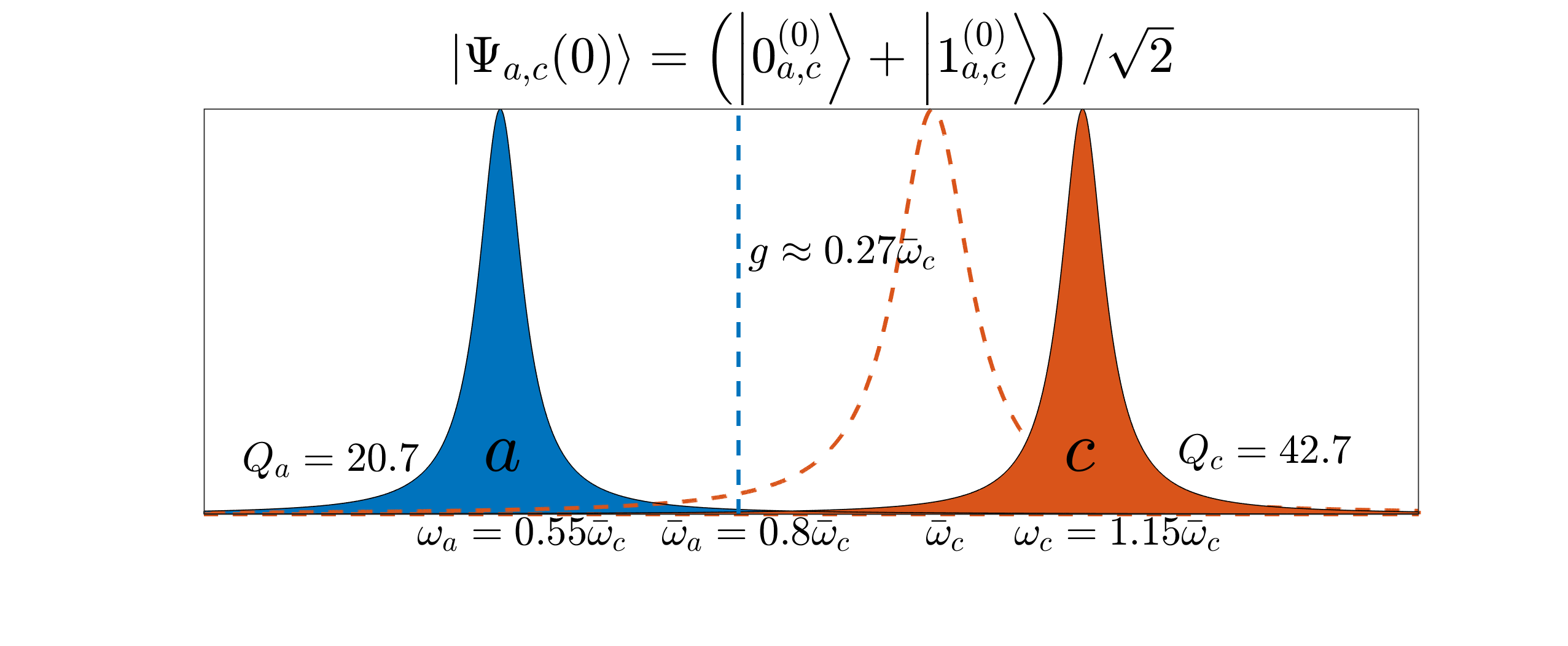}%
}\\
\subfloat[\label{subfig:SWCorr-Qu1Mod-Ada}]{%
\includegraphics[scale=0.275]{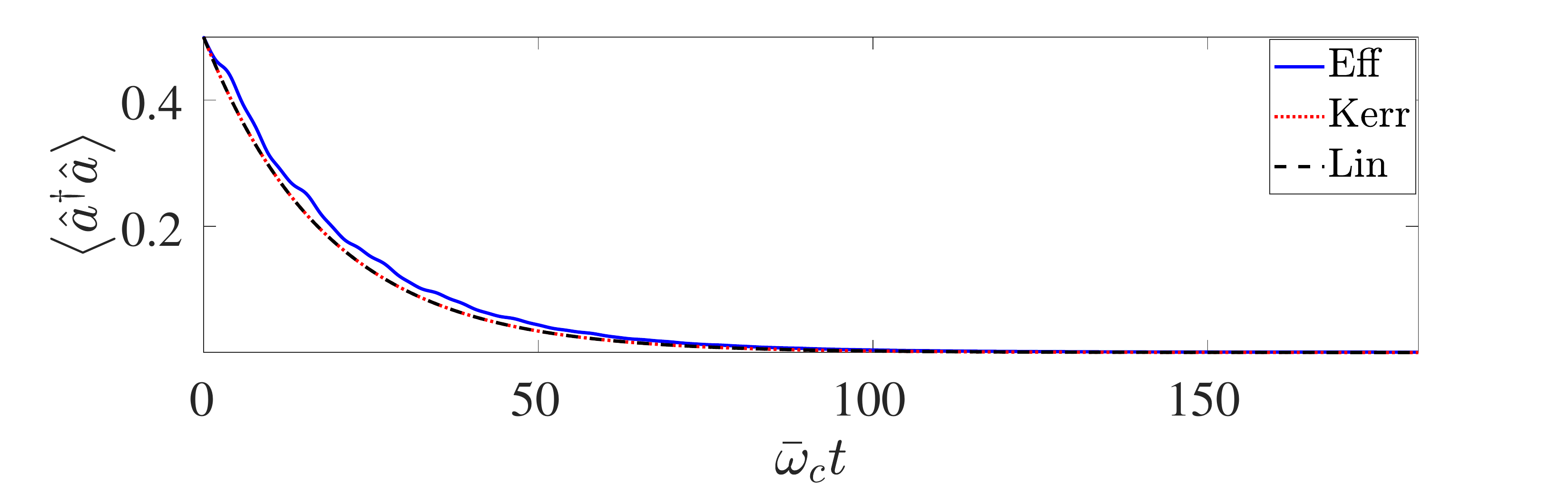}%
}\\
\subfloat[\label{subfig:SWCorr-Qu1Mod-Cdc}]{%
\includegraphics[scale=0.275]{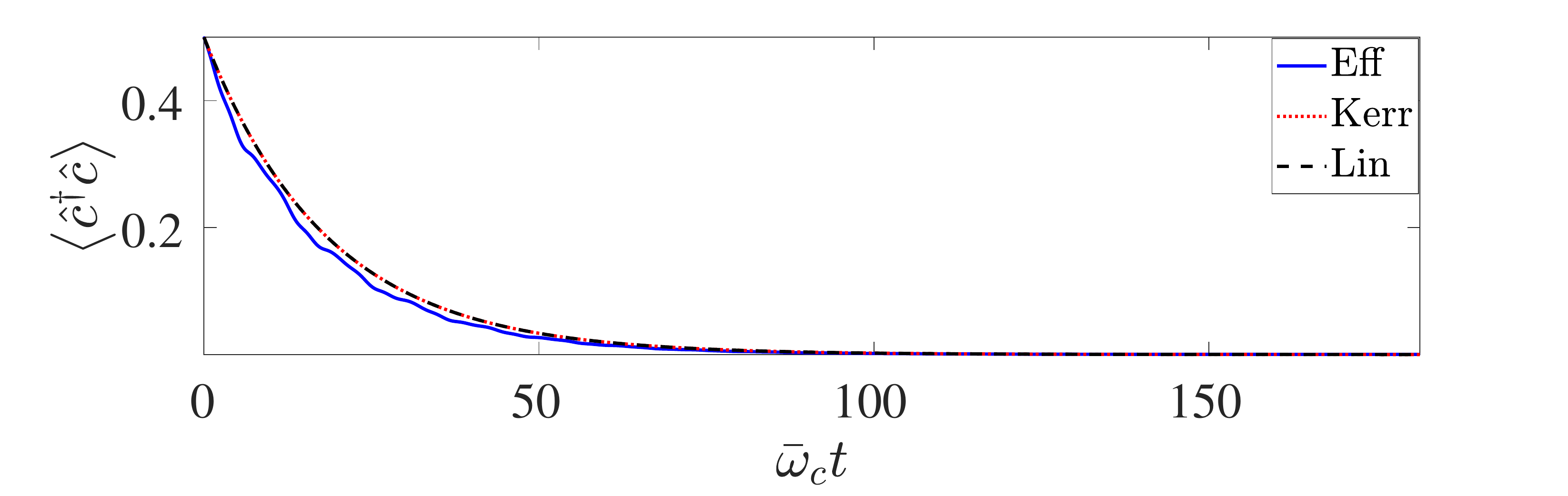}%
}
\caption{Comparison of different theories given in Table~\ref{tab:EffME-Comparison} for $E_{\text{j}}=50E_{\text{c}}$ $(\epsilon=0.2)$, $\omega_{\text{a}}=0.8\bar{\omega}_c$ and $g\approx 0.27\bar{\omega}_c$, resulting in hybridized quantities $\omega_{\text{a}}\approx 0.55\omega_{\text{c}}$, $\omega_{\text{c}}\approx 1.15\omega_{\text{c}}$, $u_{\text{aa}}=u_{\text{ac}}\approx 0.69$ and $v_{\text{cc}}=v_{\text{ca}}\approx 0.76$. We assume a flat bath spectral function, $S_{YY}(\omega_{\text{a}})=S_{YY}(\omega_c)$, such that $Q_{\text{a}}\equiv\omega_{\text{a}}/[v_{\text{ca}}^2S_{YY}(\omega_{\text{a}})]\approx 20.7$ and $Q_{\text{c}}\equiv\omega_{c}/[v_{\text{cc}}^2S_{YY}(\omega_c)]\approx 42.7$. Low Q values are chosen for simplicity and are irrelevant under the flat-bath assumption. The qubit and cavity initial conditions are set as $\ket{\Psi_{\text{a}/\text{c}}(0)}=(\ket{0_{\text{a}/\text{c}}}+\ket{1_{\text{a}/\text{c}}})/\sqrt{2}$. a) Schematic of the Harmonic parameters, b-c) Qubit and cavity occupation numbers. We note that the parameters chosen here correspond to a sizable hybridization $g\approx 0.27\bar{\omega}_{\text{c}}$ (sometimes referred to as ultrastrong coupling \cite{Bourassa_Ultrastrong_2009, Niemczyk_Circuit_2010, Forn_Observation_2010}).}
\color{black}
\label{Fig:SWCorr-Qu1Mode}
\end{figure}
\begin{table}[t!]
  \begin{tabular}{|c|c|c|}
    \hline
    parameters & $\bar{\omega}_{\text{a}}<\bar{\omega}_c$  & $\bar{\omega}_{\text{a}}>\bar{\omega}_c$\\
    \hline\hline
	$u_{\text{aa}}$, $v_{\text{qq}}$ & $+$ & $+$ \\ 
    \hline
	$u_{\text{ac}}$, $v_{\text{ac}}$ & $+$ & $-$ \\
	\hline
	$u_{\text{ca}}$, $v_{\text{ca}}$ & $-$ & $+$ \\    
	\hline
	$u_{\text{cc}}$, $v_{\text{cc}}$ & $+$ & $+$ \\    
	\hline
	$r_{\text{a}}$ & $+$ & $+$ \\    
	\hline
	$r_c$ & $+$ & $-$ \\    
	\hline
    $\mathcal{D}[\hat{C}_{\text{a,eff}}]$ & $\downarrow$ & $\uparrow$\\    
    \hline
    $\mathcal{D}[\hat{C}_{\text{c,eff}}]$ & $\uparrow$ & $\downarrow$\\
    \hline 
  \end{tabular}
  \caption{\label{tab:SignOfCorrections} Summary of the renormalization of the single-photon relaxation rates. The labels $+$ and $-$ show the sign of a quantity, while the arrows $\uparrow$ and $\downarrow$ show whether a quantity is increased or decreased, respectively.} 
\end{table}
\color{black}

Next, we compare the dynamics predicted by the EME~(\ref{eqn:EffME-Qu1Mode-PT Lind Eq}) to the corresponding Kerr and linear master equations. In order to show the possibility of a qualitatively different behavior compared to the direct bath coupling discussed in case (i), we consider the scenario where the qubit is detuned below the cavity as in Fig.~\ref{Fig:NormFreqs&HybWq08Wc}. Moreover, we fix the light-matter coupling $g$ such that the anharmonicity is almost equally shared between the two normal modes (i.e. $u_{\text{aa}}\approx u_{\text{ac}}>0$), hence the cross mode correction in the renormalizations~(\ref{eqn:EffME-Qu1Mode-Def of c_{q,PT}}-\ref{eqn:EffME-Qu1Mode-Def of c_{c,PT}}) become significant. This choice of parameters is demonstrated schematically in Fig.~\ref{subfig:SWCorr-Qu1Mod-RelFreqs}. The occupation number plots~\ref{subfig:SWCorr-Qu1Mod-Ada}-\ref{subfig:SWCorr-Qu1Mod-Cdc} reveal that the Kerr theory barely provides any renormalization with respect to the linear exponential decay. On the other hand, the EME show an increase (decrease) in the relaxation of the normal cavity (qubit) mode with respect to the linear Purcell rates.

\color{black} 
\section{Conclusion}

We presented a computational framework to derive an effective master equation for the dynamics of a weakly anharmonic superconducting qubit (e.g. a transmon) embedded in a given electromagnetic environment. An effective master equation was presented for two different cases of the electromagnetic environment: (i) a flux bath, (ii) a single-mode resonator coupled to an open resonator. The procedure based on unitary transformations yields in each case an effective master equation whose parameters (frequencies, self- and cross-Kerr terms, relaxation rates) depend nonlinearly on the initial excitation level in a systematic expansion in the small parameter characterizing the weak anharmonicity, $\epsilon = \sqrt{2 E_{\text{c}} / E_{\text{j}}}$.

Our findings in case (i) show that the relaxation rate of the qubit \textit{increases} with the strength of the anharmonicity $\epsilon$ and with the initial excitation. The presented approach explicitly shows that the renormalization of the relaxation rate originates from the number non-conserving terms in the nonlinearity of the qubit. Findings in case (ii) demonstrate the complex dependence of the renormalization of the relaxation rates on the hybridization of the qubit with its electromagnetic environment, allowing for the distinct situations where the qubit relaxation rate may increase or decrease.   

We note that for an anharmonicity that corresponds to the typical experiments with transmon qubits ($E_{\text{j}}/E_{\text{c}} \approx 50$ corresponding to $\epsilon \approx 0.2$), and initialization in the first excited state of the transmon, the transient dynamics as captured by the effective master equation is not substantially different from either the (hybridized) linear theory or the Kerr result. The differences may not be observable in an experiment. Nonetheless these results have an important implication. When either the electromagnetic environment is highly excited or the Josephson junction is initialized at a higher excitation level, the Kerr theory (as the linear hybridized theory) will display discernible deviations from the exact transient dynamics. In the effective master equation these differences will be captured by the occupation dependent relaxation rates as well as three-photon loss terms that get activated at higher excitation. The most dramatic appearance of these renormalization effects will be when the resonator-qubit system is driven by a coherent microwave tone, as in a typical quantum non-demolition readout setup. The mathematical procedure to extract an effective master equation in that case involves additional techniques, and will be discussed in Part II. 

The methodology for the derivation of an effective master equation discussed here is broadly applicable to multi-oscillator superconducting circuit devices. In combination with an accurate computational technique for modeling complex electromagnetic environments presented in Refs.~[\onlinecite{Malekakhlagh_NonMarkovian_2016}] and~[\onlinecite{Malekakhlagh_Cutoff-Free_2017}], the approach presented here provides a compelling theoretical framework for studying the quantum dynamics of large integrated quantum circuits in a way that is accurate and resource-efficient. 

\section{Acknowledgements}
We appreciate insightful discussions with Alexandre Blais, Michel Devoret, Ioan Pop, Shantanu Mundhada and Shyam Shankar. This research was supported by the US Department of Energy, Office of Basic Energy Sciences, Division of Materials Sciences and Engineering under Award No. DE-SC0016011.
\appendix
\section{First order perturbation theory}
\label{App:1stSWPT}
In this appendix, we discuss the details of the first order perturbation theory in order to calculate the effect of the weak anharmonicity of a qubit on both transition frequencies and relaxation rates of the system. The main results are presented as effective Lindblad equations with renormalized Hamiltonian and dissipators. In Sec.~\ref{SubApp:1stPT-DissQu}, we consider the case of a weakly anharmonic qubit coupled to bath through its flux quadrature. Next, in Sec.~\ref{SubApp:1stSWPT-Qu1Mode}, we discuss the case of a weakly anharmonic qubit coupled to an open resonator. 
\subsection{Qubit coupled to a bath}
\label{SubApp:1stPT-DissQu}

We model this system by the overall Hamiltonian
\begin{align}
  \hat{\mathcal{H}}&=\hat{\mathcal{H}}_{\text{a}} +  \hat{\mathcal{H}}_{\text{sb}} + \hat{\mathcal{H}}_{\text{b}},
\label{Eq:1stPT-DissQu-H}
\end{align}
where the system Hamiltonian is given by 
\begin{align}
\begin{split}
\hat{\mathcal{H}}_{\text{a}}&\equiv\frac{\omega_{\text{a}}}{4}\left(\hat{X}_{\text{a}}^2+\hat{Y}_{\text{a}}^2-\frac{\epsilon}{12}\hat{X}_{\text{a}}^4\right)+O(\epsilon^2)\\
&=\omega_{\text{a}}\left(\hat{a}^{\dag}\hat{a}+\frac{1}{2}\right)-\frac{\epsilon}{48}\omega_{\text{a}}\left(\hat{a}+\hat{a}^{\dag}\right)^4+O(\epsilon^2).
\end{split}
\label{Eq:1stPT-DissQu-Def of Hq}
\end{align}
We describe the bath by a continuum set of modes $\hat{\mathcal{H}}_\text{b}  = \sum_{k} \omega_{k} \hat{B}_{k}^\dagger \hat{B}_{k}$ that models the flux noise of the qubit through the system-bath coupling $\hat{\mathcal{H}}_\text{sb} = \hat{X}_{\text{a}}  \sum_{k} g_{k} (\hat{B}_{k} + \hat{B}_{k}^\dagger)$. For notation simplicity, we define unitless quadratic and quartic operators as
\begin{subequations}  
\begin{align}
&\hat{H}_{\text{a}}\equiv \frac{1}{2}\left(\hat{a}^{\dag}\hat{a}+\hat{a}\hat{a}^{\dag}\right)=\hat{a}^{\dag}\hat{a}+\frac{1}{2},
\label{Eq:1stPT-DissQu-Def of Hb}\\
&\hat{H}_4\equiv \left(\hat{a}+\hat{a}^{\dag}\right)^4.
\label{Eq:1stPT-DissQu-Def of H4}
\end{align}
\end{subequations}

Our analysis begins by applying a unitary transformation to the overall Hamiltonian~(\ref{Eq:1stPT-DissQu-H}) as 
\begin{align}
\hat{\mathcal{H}}_{\text{eff}}\equiv e^{-\hat{G}}\hat{\mathcal{H}}e^{+\hat{G}},
\label{Eq:1stPT-DissQu-PT Trans}
\end{align}
where $\hat{G}$ is an anti-Hermitian operator and the generator of the transformation. We seek an order-by-order solution for this generator in such a fashion that the system Hamiltonian becomes simpler as we see in the following. Up to lowest order in $\epsilon$ we can write
\begin{align}
\hat{G}=\epsilon\hat{G}_4+O(\epsilon^2),
\label{Eq:1stPT-DissQu-G Exp}
\end{align}
where the subscript ``$4$'' is chosen to match the lowest order nonlinear expansion of the Hamiltonian that is quartic. Upon inserting Eqs.~(\ref{Eq:1stPT-DissQu-Def of Hq}) and~(\ref{Eq:1stPT-DissQu-G Exp}) into Eq.~(\ref{Eq:1stPT-DissQu-PT Trans}), we obtain the lowest order transformation of $\hat{\mathcal{H}}_{\text{a,eff}}$ as
\begin{align}
\begin{split}
\hat{\mathcal{H}}_{\text{a,eff}}=\omega_{\text{a}}\left(\hat{H}_{\text{a}}-\frac{\epsilon}{48}\hat{H}_4\right)+\epsilon\omega_{\text{a}}[\hat{H}_{\text{a}},\hat{G}_4]+O(\epsilon^2).
\end{split}
\end{align}
Then, $\hat{G}_4$ needs to be determined such that the transformed lowest order Hamiltonian, i.e. $\epsilon\omega_{\text{a}}\left([\hat{H}_{\text{a}},\hat{G}_4]-\frac{1}{48}\hat{H}_4\right)$, gets simplified. After we obtain the desired operator $\hat{G}_4$, the overall Hamiltonian~(\ref{Eq:1stPT-DissQu-H}), and in particular the system-bath coupling, also need to be transformed accordingly.

It is important to note that any higher order anharmonicity can be partitioned into secular and non-secular contributions. In particular, the quartic anharmonicity $\hat{H}_4$ consists of six secular and ten non-secular terms such that we can write 
\begin{align}
\hat{H}_4\equiv\hat{S}_4+\hat{N}_4.
\label{Eq:1stPT-DissQu-Def of S4&N4}
\end{align}
Moreover, the secular terms can be written in terms of the harmonic Hamiltonian $\hat{H}_{\text{a}}$ as
\begin{align}
\begin{split}
\hat{S}_4&=\hat{a}^{\dag}\hat{a}^{\dag}\hat{a}\hat{a}+\hat{a}^{\dag}\hat{a}\hat{a}^{\dag}\hat{a}+\hat{a}^{\dag}\hat{a}\hat{a}\hat{a}^{\dag}\\
&+\hat{a}\hat{a}\hat{a}^{\dag}\hat{a}^{\dag}+\hat{a}\hat{a}^{\dag}\hat{a}\hat{a}^{\dag}+\hat{a}\hat{a}^{\dag}\hat{a}^{\dag}\hat{a}\\
&=6\hat{H}_{\text{a}}^2+\frac{3}{2}=6\hat{n}_{\text{a}}^2+6\hat{n}_{\text{a}}+3.
\end{split}
\label{Eq:1stPT-DissQu-S4 in terms of H0}
\end{align}
From the algebra of the bosonic operators, we find that there does not exist any operator $\hat{G}_4$ such that $[\hat{H}_{\text{a}},\hat{G}_4]$ could cancel any of the secular contributions in $\hat{S}_4$. The reason is that the commutator of the harmonic Hamiltonian $\hat{H}_{\text{a}}$ with any non-secular term remains non-secular, while with any secular term is zero. The discussion can be summarized by the following commutator rules:
\begin{subequations}
\begin{align}
&[\text{sec},\text{sec}]=0,
\label{Eq:1stPT-[S,S]=0}\\
&[\text{sec},\text{non-sec}]=\text{non-sec},
\label{Eq:1stPT-[S,N]=N}\\
&[\text{non-sec},\text{non-sec}]=\text{sec}+\text{non-sec}.
\label{Eq:1stPT-[N,N]=S+N}
\end{align} 
\end{subequations}
As a result, all the non-secular terms in the system Hamiltonian can be in principle canceled through this procedure. Therefore, we are looking for an operator $\hat{G}_4$ such that
\begin{align}
[\hat{H}_{\text{a}},\hat{G}_4]-\frac{1}{48}\hat{N}_4=0,
\label{Eq:1stPT-DissQu-Cond to find X}
\end{align}
and the lowest order effective Hamiltonian then becomes
\begin{align}
\begin{split}
\hat{\mathcal{H}}_{\text{a,eff}} & = \omega_{\text{a}} \hat{H}_{\text{a}}-\frac{\epsilon\omega_{\text{a}}}{8}\hat{H}_{\text{a}}^2+O(\epsilon^2) \\
& = \left(1-\frac{\epsilon}{8}\right)\omega_{\text{a}} \hat{n}_{\text{a}}-\frac{\epsilon}{8}\omega_{\text{a}}\hat{n}_{\text{a}}^2+O(\epsilon^2),
\end{split}
\label{Eq:1stPT-DissQu-H_q,eff}
\end{align} 
where we have replaced $\hat{S}_4$ from Eq.~(\ref{Eq:1stPT-DissQu-S4 in terms of H0}). Note that the new effective Hamiltonian~(\ref{Eq:1stPT-DissQu-H_q,eff}) is diagonal in the original number basis of the harmonic Hamiltonian and agrees with the common Kerr theory that could alternatively be obtained by applying the RWA on Hamiltonian~(\ref{Eq:1stPT-DissQu-Def of Hq}) from the outset. As we will see in App.~\ref{App:2ndPT} for the second order perturbation, the correspondence to the Kerr theory is coincidental for the lowest order, while additional corrections appear in the effective Hamiltonian that a simple RWA can not recover.     

The first step towards finding $\hat{G}_4$ is to obtain different contributions in $\hat{N}_4$ and write them in normal ordering as
\begin{align}
\begin{split}
\hat{N}_4=\hat{a}^4+\left(\hat{a}^{\dag}\right)^4+4\left[\hat{a}^{\dag}\hat{a}^3+\left(\hat{a}^{\dag}\right)^3\hat{a}\right]+6\left[\hat{a}^2+\left(\hat{a}^{\dag}\right)^2\right].
\end{split}
\label{Eq:1stPT-DissQu-Expan of N4}
\end{align}
Next, we use the fact that the commutator of $\hat{H}_{\text{a}}$ with any non-secular monomial is proportional to that monomial as
\begin{align}
\left[\hat{H}_{\text{a}},(\hat{a}^{\dag})^m\hat{a}^{n}\right]=(m-n)(\hat{a}^{\dag})^m\hat{a}^{n}.
\label{Eq:1stPT-DissQu-[H2,bdag^m b^n]}
\end{align}
To see this explicitly, consider the commutator of $\hat{H}_{\text{a}}$ with each individual term in Eq.~(\ref{Eq:1stPT-DissQu-Expan of N4}):
\begin{subequations}
\begin{align}
&\left[\hat{H}_{\text{a}},\hat{a}^4\right]=-4\hat{a}^4, 
\label{Eq:1stPT-DissQu-[H0,N4] Term1}\\ 
&\left[\hat{H}_{\text{a}},\left(\hat{a}^{\dag}\right)^4\right]=+4\left(\hat{a}^{\dag}\right)^4,
\label{Eq:1stPT-DissQu-[H0,N4] Term2}\\
&\left[\hat{H}_{\text{a}},\hat{a}^{\dag}\hat{a}^3\right]=-2\hat{a}^{\dag}\hat{a}^3,
\label{Eq:1stPT-DissQu-[H0,N4] Term3}\\
&\left[\hat{H}_{\text{a}},\left(\hat{a}^{\dag}\right)^3\hat{a}\right]=+2\left(\hat{a}^{\dag}\right)^3\hat{a},
\label{Eq:1stPT-DissQu-[H0,N4] Term4}\\
&\left[\hat{H}_{\text{a}},\hat{a}^2\right]=-2\hat{a}^2, 
\label{Eq:1stPT-DissQu-[H0,N4] Term5}\\ 
&\left[\hat{H}_{\text{a}},\left(\hat{a}^{\dag}\right)^2\right]=+2\left(\hat{a}^{\dag}\right)^2.
\label{Eq:1stPT-DissQu-[H0,N4] Term6}
\end{align}
\end{subequations}
From Eq.~(\ref{Eq:1stPT-DissQu-[H2,bdag^m b^n]}), we understand that the generator $\hat{G}_4$ that obeys the condition~(\ref{Eq:1stPT-DissQu-Cond to find X}) will contain the same set of monomials as $\hat{N}_4$, but only with different coefficients. Therefore, we directly construct the operator $\hat{G}_4$ in terms of $\hat{N}_4$ as
\begin{align}
\begin{split}
\hat{G}_4&=\frac{1}{192}\left[\left(\hat{a}^{\dag}\right)^4-\hat{a}^4\right]+\frac{1}{24}\left[\left(\hat{a}^{\dag}\right)^3\hat{a}-\hat{a}^{\dag}\hat{a}^3\right]\\
&+\frac{1}{16}\left[\left(\hat{a}^{\dag}\right)^2-\hat{a}^2\right].
\end{split}
\label{Eq:1stPT-DissQu-Sol G4}
\end{align}

Next, we calculate the effect of this transformation on the full Hamiltonian. Since the generator $\hat{G}_4$ was only determined in terms of system operators, the resulting transformation only acts on the system quadrature of the system-bath Hamiltonian. Therefore, up to lowest order in $\epsilon$, we need to calculate
\begin{align}
e^{- \hat{G}}\hat{X}_{\text{a}}e^{+\hat{G}}=\hat{X}_{\text{a}}+\epsilon\left[\hat{X}_{\text{a}},\hat{G}_4\right]+O(\epsilon^2).
\label{Eq:1stPT-DissQu-TransOf b+b^dag 1}
\end{align}
Using expression~(\ref{Eq:1stPT-DissQu-Sol G4}) for  $\hat{G}_4$ we obtain
\begin{subequations}
\begin{align}
[\hat{a},\hat{G}_4]&=\frac{1}{48}\left(\hat{a}^{\dag}\right)^3+\frac{1}{8}\left(\hat{a}^{\dag}\right)^2\hat{a}-\frac{1}{24}\hat{a}^3+\frac{1}{8}\hat{a}^{\dag},
\label{Eq:1stPT-DissQu-[b,G]}\\
[\hat{a}^{\dag},\hat{G}_4]&=\frac{1}{48}\hat{a}^3+\frac{1}{8}\hat{a}^{\dag}\hat{a}^2-\frac{1}{24}\left(\hat{a}^{\dag}\right)^3+\frac{1}{8}\hat{a},
\label{Eq:1stPT-DissQu-[b^dag,G]}
\end{align}
and by adding them we find
\begin{align}
\begin{split}
[\hat{X}_{\text{a}},\hat{G}_4]&=\frac{1}{8}\left(\hat{a}+\hat{a}^{\dag}\right)\\
&+\frac{1}{8}\left[\left(\hat{a}^{\dag}\right)^2\hat{a}+\hat{a}^{\dag}\hat{a}^2\right]\\
&-\frac{1}{48}\left[\hat{a}^3+\left(\hat{a}^{\dag}\right)^3\right].
\end{split}
\label{Eq:1stPT-DissQu-[b+bdag,G]}
\end{align}
\end{subequations}
We observe from Eq.~(\ref{Eq:1stPT-DissQu-[b+bdag,G]}) that a variety of multi-photon couplings appear up to $\epsilon$-order.  For instance, the first and second line of Eq.~(\ref{Eq:1stPT-DissQu-[b+bdag,G]}) produce transitions between successive energy levels of the oscillator, while the third line cause transitions between every third energy levels. In particular, the single-photon terms can be reexpressed more compactly in terms of $\hat{n}_{\text{a}}$ as
\begin{align}
\begin{split}
\hat{a}+\hat{a}^{\dag}+\left(\hat{a}^{\dag}\right)^2\hat{a}+\hat{a}^{\dag}\hat{a}^2 = \left( 1 + \hat{n}_{\text{a}} \right) \hat{a}+\text{H.c.},
\end{split}
\label{Eq:1stPT-DissQu-Identity {H0,b+bdag}}
\end{align}
Using identity~(\ref{Eq:1stPT-DissQu-Identity {H0,b+bdag}}), the transformation~(\ref{Eq:1stPT-DissQu-TransOf b+b^dag 1}) can be written in the following compact form
\begin{align}
\begin{split}
e^{- \hat{G}} \hat{X}_\text{a} e^{+\hat{G}} & = \left[ 1 + \frac{\epsilon}{8} \left(1+\hat{n}_{\text{a}}\right) \right] \hat{a}\\
&-\frac{\epsilon}{48}\hat{a}^3 + \text{H.c.} +  O(\epsilon^2).
\end{split}
\label{Eq:1stPT-DissQu-TransOf b+b^dag 2}
\end{align}

Following the common derivation of the Lindblad master equation, one obtains the $\epsilon$-order effective master equation as
\begin{align}
\begin{split}
\dot{\hat{\rho}}_{\text{a}}(t)& = -i\left[\hat{\mathcal{H}}_{\text{a,eff}},\hat{\rho}_{\text{a}}(t)\right]\\
&+2\kappa_{\text{a}} \mathcal{D}\left[\left(1+\frac{\epsilon}{8}(1+\hat{n}_{\text{a}})\right)\hat{a}\right] \hat{\rho}_{\text{a}}(t)\\
&+2\kappa_{\text{a}3}\mathcal{D}\left[\frac{\epsilon}{16}\hat{a}^3\right]\hat{\rho}_{\text{a}}(t),
\end{split}
\label{Eq:1stPT-DissQu-Lind Eq1}
\end{align}
where $\kappa_{\text{a}}\equiv S_{XX}(\omega_{\text{a}})$, $\kappa_{\text{a}3}\equiv S_{XX}(3\omega_{\text{a}})$ are the single-photon and three-photon relaxation rates. Moreover, $S_{YY} (\omega) = \int_{-\infty}^\infty d\tau \, \text{e}^{-i\omega \tau} \text{tr} \left[(1/Z_{\text{b}})\text{e}^{- \hat{\mathcal{H}}_{\text{b}}/ k_\text{B} T} \hat{X}_{\text{b}}(\tau) \, \hat{X}_{\text{b}}(0) \right]$ is the bath spectral function with $\hat{X}_{\text{b}}\equiv\sum_{k} g_{k}(\hat{B}_{k}+\hat{B}_{k}^{\dag})$ being the bath flux quadrature that couples to the qubit. Note that the cross terms (mixtures of single- and three-photon couplings) are canceled out due to the Markov approximation and the resulting secular condition. It is important to notice the operator nature of the relaxation renormalization, which is manifest as a nonlinear collapse operator.
\color{black}

The order $\epsilon$ effective Lindblad Eq.~(\ref{Eq:1stPT-DissQu-Lind Eq1}) along with the corresponding renormalized Hamiltonian~(\ref{Eq:1stPT-DissQu-H_q,eff}) dissipators are the main results of this appendix, and are employed in Sec.~\ref{SubSec:EffME-Qu} of the main body of the paper. 
\subsection{Qubit coupled to a single-mode open resonator}
\label{SubApp:1stSWPT-Qu1Mode}
Here, we return to the problem of a weakly anharmonic qubit coupled to an open resonator. For simplicity, we consider a single mode, while our results can be trivially generalized to a multimode scenario. The system Hamiltonian up to lowest order in $\epsilon$ reads
\begin{align}
\begin{split}
\hat{\mathcal{H}}_{\text{s}}\equiv\frac{\bar{\omega}_{\text{a}}}{4}\left(\hat{\bar{X}}_{\text{a}}^2+\hat{\bar{Y}}_{\text{a}}^2-\frac{\epsilon}{12}\hat{\bar{X}}_{\text{a}}^4\right)+\frac{\bar{\omega}_{\text{c}}}{4}\left(\hat{\bar{X}}_{\text{c}}^2+\hat{\bar{Y}}_{\text{c}}^2\right)+g\hat{\bar{Y}}_{\text{c}}\hat{\bar{Y}}_{\text{a}}
\end{split},
\label{Eq:1stPT-Qu1Mode-Def of Hs}
\end{align}
with $\bar{\omega}_{\text{a}}$ and $\bar{\omega}_{\text{c}}$ being the qubit and the cavity bare frequencies and $g$ the coupling strength. Furthermore, we consider a resonator-bath coupling with the bath Hamiltonian $\hat{\mathcal{H}}_\text{b}  = \sum_{k} \omega_{k} \hat{B}_{k}^\dagger \hat{B}_{k}$ and the coupling Hamiltonian $\hat{\mathcal{H}}_\text{sb} = \hat{\bar{Y}}_{\text{c}}  \sum_{k} g_{k} \hat{Y}_{k}$. 

To simplify the perturbative calculation, we work in the normal mode basis, in which the quadratic part of the Hamiltonian~(\ref{Eq:1stPT-Qu1Mode-Def of Hs}) becomes diagonal. The desired transformation can be obtained by the successive applications of non-uniform scaling and rotation \cite{Jellal_Two_2005} as shown in the following.

With this aim, we first introduce scaled sets of canonical cavity/qubit operators 
\begin{subequations}
\begin{align}
\hat{\bar{X}}_{\text{a}}\equiv s_1\hat{X}_{\text{a}}', \quad \hat{\bar{Y}}_{\text{a}}\equiv s_1^{-1}\hat{Y}_{\text{a}}',
\label{Eq:1stPT-Qu1Mode-Def of X'_q/Y'_q}\\
\hat{\bar{X}}_{\text{c}}\equiv s_1^{-1}\hat{X}_{\text{c}}', \quad \hat{\bar{Y}}_{\text{c}}\equiv s_1\hat{Y}_{\text{c}}',
\label{Eq:1stPT-Qu1Mode-Def of X'_c/Y'_c}
\end{align}
\end{subequations}
where $s_1\equiv (\bar{\omega}_{\text{c}}/\bar{\omega}_{\text{a}})^{1/4}$. In terms of the prime canonical quadratures, the quadratic part of Hamiltonian~(\ref{Eq:1stPT-Qu1Mode-Def of Hs}) becomes
\begin{align}
\hat{\mathcal{H}}_2\equiv\frac{\omega'_{\text{ac}}}{4}\left(\hat{X}_{\text{a}}^{'2}+\hat{X}_{\text{c}}^{'2}\right)+\frac{\omega'_{\text{a}}}{4}\hat{Y}_{\text{a}}^{'2}+\frac{\omega'_{\text{c}}}{4}\hat{Y}_{\text{c}}^{'2}+g\hat{Y}'_{\text{c}}\hat{Y}'_{\text{a}},
\label{Eq:1stPT-Qu1Mode-Def of H 2}
\end{align}
where $\omega'_{\text{ac}}\equiv (\bar{\omega}_{\text{a}}\bar{\omega}_{\text{c}})^{1/2}$, $\omega'_{\text{a}}\equiv(\bar{\omega}_{\text{a}}^3/\bar{\omega}_{\text{c}})^{1/2}$ and $\omega'_{\text{c}}\equiv(\bar{\omega}_{\text{c}}^3/\bar{\omega}_{\text{a}})^{1/2}$. 

Second, we introduce the following unitary rotations
\begin{subequations}
\begin{align}
\begin{bmatrix}
\hat{X}'_{\text{a}}\\
\hat{X}'_{\text{c}}
\end{bmatrix}=
\begin{bmatrix}
\cos(\theta) & \sin(\theta)\\
-\sin(\theta) & \cos(\theta)
\end{bmatrix}
\begin{bmatrix}
\hat{X}''_{\text{a}}\\
\hat{X}''_{\text{c}}
\end{bmatrix},
\label{Eq:1stPT-Qu1Mode-Def of Rot 1}\\
\begin{bmatrix}
\hat{Y}'_{\text{a}}\\
\hat{Y}'_{\text{c}}
\end{bmatrix}=
\begin{bmatrix}
\cos(\theta) & \sin(\theta)\\
-\sin(\theta) & \cos(\theta)
\end{bmatrix}
\begin{bmatrix}
\hat{Y}''_{\text{a}}\\
\hat{Y}''_{\text{c}}
\end{bmatrix},
\label{Eq:1stPT-Qu1Mode-Def of Rot 2}
\end{align}
\end{subequations}
in terms of the new double-prime set of canonical operators. The rotation angle $\theta$ that removes the off-diagonal terms in Hamiltonian~(\ref{Eq:1stPT-Qu1Mode-Def of H 2}), i.e. $g\hat{Y}_{\text{c}}'\hat{Y}_{\text{a}}'$, is then found from the condition
\begin{align}
\tan(2\theta)=\frac{4g}{\omega'_{\text{c}}-\omega'_{\text{a}}}=\frac{4g\sqrt{\bar{\omega}_{\text{a}}\bar{\omega}_{\text{c}}}}{\bar{\omega}_{\text{c}}^2-\bar{\omega}_{\text{a}}^2},
\label{Eq:1stPT-Qu1Mode-Cond for Theta}
\end{align}
for which Hamiltonian~(\ref{Eq:1stPT-Qu1Mode-Def of H 2}) becomes
\begin{align}
\begin{split}
\hat{\mathcal{H}}_2&\equiv\frac{\omega'_{\text{ac}}}{4}\hat{X}_{\text{a}}^{''2}+\frac{\omega'_{\text{ac}}}{4}\hat{X}_{\text{c}}^{''2}\\
&+\frac{\omega'_{\text{a}}\cos^2(\theta)+\omega'_{\text{c}}\sin^2(\theta)-2g\sin(2\theta)}{4}\hat{Y}_{\text{a}}^{''2}\\
&+\frac{\omega'_{\text{c}}\cos^2(\theta)+\omega'_{\text{a}}\sin^2(\theta)+2g\sin(2\theta)}{4}\hat{Y}_{\text{c}}^{''2}.
\end{split}
\label{Eq:1stPT-Qu1Mode-Def of H 3}
\end{align}

Third, we need to introduce another non-uniform scaling transformation into the final normal modes (denoted by bar) as
\begin{subequations}
\begin{align}
\hat{X}''_{\text{a}}\equiv s_2\hat{X}_{\text{a}}, \quad \hat{Y}''_{\text{a}}\equiv s_2^{-1}\hat{Y}_{\text{a}},
\label{Eq:1stPT-Qu1Mode-Def of bar(X/Y)_q}\\
\hat{X}''_{\text{c}}\equiv s_3\hat{X}_{\text{c}}, \quad \hat{Y}''_{\text{c}}\equiv s_3^{-1}\hat{Y}_{\text{c}},
\label{Eq:1stPT-Qu1Mode-Def of bar(X/Y)_c}
\end{align}
\end{subequations}
The scales $s_2$ and $s_3$ in Eqs.~(\ref{Eq:1stPT-Qu1Mode-Def of bar(X/Y)_q}-\ref{Eq:1stPT-Qu1Mode-Def of bar(X/Y)_c}) are evaluated as 
\begin{subequations}
\begin{align}
&s_2\equiv\left[\frac{\bar{\omega}_{\text{a}}^2\cos^2(\theta)+\bar{\omega}_{\text{c}}^2\sin^2(\theta)-2g\sqrt{\bar{\omega}_{\text{a}}\bar{\omega}_{\text{c}}}\sin(2\theta)}{\bar{\omega}_{\text{a}}\bar{\omega}_{\text{c}}}\right]^{1/4},
\label{Eq:1stPT-Qu1Mode-Def of s2}\\
&s_3\equiv\left[\frac{\bar{\omega}_{\text{c}}^2\cos^2(\theta)+\bar{\omega}_{\text{a}}^2\sin^2(\theta)+2g\sqrt{\bar{\omega}_{\text{a}}\bar{\omega}_{\text{c}}}\sin(2\theta)}{\bar{\omega}_{\text{a}}\bar{\omega}_{\text{c}}}\right]^{1/4},
\label{Eq:1stPT-Qu1Mode-Def of s3}
\end{align}
\end{subequations}
such that Hamiltonian~(\ref{Eq:1stPT-Qu1Mode-Def of H 3}) becomes diagonal as
\begin{align}
\hat{\mathcal{H}}_2=\frac{\omega_{\text{a}}}{4}\left(\hat{X}_{\text{a}}^2+\hat{Y}_{\text{a}}^2\right)+\frac{\omega_{\text{c}}}{4}\left(\hat{X}_{\text{c}}^2+\hat{Y}_{\text{c}}^2\right).
\label{Eq:1stPT-Qu1Mode-H2 in NorMod}
\end{align}
The qubit-like and cavity-like normal mode frequencies read
\begin{subequations}
\begin{align}
\omega_{\text{a}}\equiv\left[\bar{\omega}_{\text{a}}^2\cos^2(\theta)+\bar{\omega}_{\text{c}}^2\sin^2(\theta)-2g\sqrt{\bar{\omega}_{\text{a}}\bar{\omega}_{\text{c}}}\sin(2\theta)\right]^{1/2},
\label{Eq:1stPT-Qu1Mode-Def of bar(v)_q}\\
\omega_{\text{c}}\equiv\left[\bar{\omega}_{\text{c}}^2\cos^2(\theta)+\bar{\omega}_{\text{a}}^2\sin^2(\theta)+2g\sqrt{\bar{\omega}_{\text{a}}\bar{\omega}_{\text{c}}}\sin(2\theta)\right]^{1/2}.
\label{Eq:1stPT-Qu1Mode-Def of bar(v)_c}
\end{align}
\end{subequations}
Putting the result of the three transformations~(\ref{Eq:1stPT-Qu1Mode-Def of X'_q/Y'_q}-\ref{Eq:1stPT-Qu1Mode-Def of X'_c/Y'_c}),~(\ref{Eq:1stPT-Qu1Mode-Def of Rot 1}-\ref{Eq:1stPT-Qu1Mode-Def of Rot 2}) and~(\ref{Eq:1stPT-Qu1Mode-Def of bar(X/Y)_q}-\ref{Eq:1stPT-Qu1Mode-Def of bar(X/Y)_c}) together, one can relate the initial and normal mode quadratures via a set of hybridization coefficients 
\begin{subequations}
\begin{align}
\begin{bmatrix}
\hat{\bar{X}}_{\text{a}}\\
\hat{\bar{X}}_{\text{c}}
\end{bmatrix}=
\begin{bmatrix}
u_{\text{aa}} & u_{\text{ac}}\\
u_{\text{ca}} & u_{\text{cc}}
\end{bmatrix}
\begin{bmatrix}
\hat{X}_{\text{a}}\\
\hat{X}_{\text{c}}
\end{bmatrix},
\label{Eq:1stPT-Qu1Mode-LinHybTransForX}
\end{align}
\begin{align}
\begin{bmatrix}
\hat{\bar{Y}}_{\text{a}}\\
\hat{\bar{Y}}_{\text{c}}
\end{bmatrix}=
\begin{bmatrix}
v_{\text{qq}} & v_{\text{ac}}\\
v_{\text{ca}} & v_{\text{cc}}
\end{bmatrix}
\begin{bmatrix}
\hat{Y}_{\text{a}}\\
\hat{Y}_{\text{c}}
\end{bmatrix},
\label{Eq:1stPT-Qu1Mode-LinHybTransForY}
\end{align}
\end{subequations}
that are obtained as
\begin{subequations}
\begin{align}
\begin{split}
\begin{bmatrix}
u_{\text{aa}} & u_{\text{ac}}\\
u_{\text{ca}} & u_{\text{cc}}
\end{bmatrix}&=
\begin{bmatrix}
s_1 & 0 \\
0 & s_1^{-1}
\end{bmatrix}
\begin{bmatrix}
\cos(\theta) & \sin(\theta)\\
-\sin(\theta) & \cos(\theta)
\end{bmatrix}
\begin{bmatrix}
s_2 & 0 \\
0 & s_3
\end{bmatrix}\\
&=
\begin{bmatrix}
s_1s_2\cos(\theta) & s_1s_3 \sin(\theta)\\
-s_1^{-1}s_2\sin(\theta) & s_1^{-1}s_3 \cos(\theta)
\end{bmatrix}.
\end{split}
\label{Eq:1stPT-Qu1Mode-Def of HybCoefForX}
\end{align}
\begin{align}
\begin{split}
\begin{bmatrix}
v_{\text{qq}} & v_{\text{ac}}\\
v_{\text{ca}} & v_{\text{cc}}
\end{bmatrix}&=
\begin{bmatrix}
s_1^{-1} & 0 \\
0 & s_1
\end{bmatrix}
\begin{bmatrix}
\cos(\theta) & \sin(\theta)\\
-\sin(\theta) & \cos(\theta)
\end{bmatrix}
\begin{bmatrix}
s_2^{-1} & 0 \\
0 & s_3^{-1}
\end{bmatrix}\\
&=
\begin{bmatrix}
s_1^{-1}s_2^{-1}\cos(\theta) & s_1^{-1}s_3^{-1} \sin(\theta)\\
-s_1 s_2^{-1}\sin(\theta) & s_1 s_3^{-1} \cos(\theta)
\end{bmatrix}.
\end{split}
\label{Eq:1stPT-Qu1Mode-Def of HybCoefForY}
\end{align}
\end{subequations}
Examples of the dependence of the normal mode frequencies and hybridization coefficients on coupling $g$ is studied in Figs.~\ref{Fig:NormFreqs&HybWq08Wc} and~\ref{Fig:NormFreqs&HybWc08Wq}.
\color{black}

We can then rewrite the system Hamiltonian~(\ref{Eq:1stPT-Qu1Mode-Def of Hs}) in the normal mode picture as
\begin{align}
\begin{split}
\hat{\mathcal{H}}_{\text{s}}&\equiv\omega_{\text{a}}\left(\hat{a}^{\dag}\hat{a}+\frac{1}{2}\right)+\omega_{\text{c}}\left(\hat{c}^{\dag}\hat{c}+\frac{1}{2}\right)\\
&-\frac{\epsilon\bar{\omega}_{\text{a}}}{48}\left[u_{\text{aa}}\left(\hat{a}+\hat{a}^{\dag}\right)+u_{\text{ac}}\left(\hat{c}+\hat{c}^{\dag}\right)\right]^4.
\end{split}
\label{Eq:1stPT-Qu1Mode-Hs in NorMod}
\end{align}
Note that the quartic anharmonicity induces nonlinear mixing between the normal modes, whose intensity is given by the hybridization coefficients $u_{\text{aa}}$ and $u_{\text{ac}}$. Moreover, due to hybridization, the original system-bath coupling now acts on both normal modes as
\begin{align}
\hat{\mathcal{H}}_{\text{sb}}=(v_{\text{cc}}\hat{Y}_{\text{c}}+v_{\text{ca}}\hat{Y}_{\text{a}})\sum\limits_{k} g_{k} \hat{Y}_k.
\label{Eq:1stPT-Hsb in normpic}
\end{align} 

In the following, we apply a unitary transformation to the overall Hamiltonian to obtain corrections to both oscillation frequency and relaxation rates in orders of weak anharmonicity measure $\epsilon$. Based on Hamiltonian~(\ref{Eq:1stPT-Qu1Mode-Hs in NorMod}), we introduce the following unitless operators 
\begin{subequations}
\begin{align}
&\hat{H}_{\text{a}}\equiv \hat{a}^{\dag}\hat{a}+\frac{1}{2}, 
\label{Eq:1stPT-Qu1Mode-Def of Ha}\\
&\hat{H}_{\text{c}}\equiv \hat{c}^{\dag}\hat{c}+\frac{1}{2}, 
\label{Eq:1stPT-Qu1Mode-Def of Hc}\\
&\hat{H}_4\equiv \left[u_{\text{aa}}\left(\hat{a}+\hat{a}^{\dag}\right)+u_{\text{ac}}\left(\hat{c}+\hat{c}^{\dag}\right)\right]^4,
\label{Eq:1stPT-Qu1Mode-Def of H4}
\end{align}
\end{subequations}
to simplify our calculation. Expanding the generator of the transformation up to lowest order in $\epsilon$ we can write
\begin{align}
\begin{split}
\hat{\mathcal{H}}_{\text{s,eff}}&\equiv e^{-\hat{G}}\hat{\mathcal{H}}_{\text{s}} e^{+\hat{G}}=\omega_{\text{a}}\hat{H}_{\text{a}}+\omega_{\text{c}}\hat{H}_{\text{c}}\\
&+\epsilon\left\{[\omega_{\text{a}}\hat{H}_{\text{a}}+\omega_{\text{c}}\hat{H}_{\text{c}},\hat{G}_4]-\frac{\bar{\omega}_{\text{a}}}{48}\hat{H}_4\right\}+O(\epsilon^2),
\end{split}
\label{Eq:1stPT-Qu1Mode-Def of Hsw}
\end{align}
where in the last step we used Eqs.~(\ref{Eq:1stPT-Qu1Mode-Def of Ha}-\ref{Eq:1stPT-Qu1Mode-Def of H4}). The generator $\hat{G}_4$ is then determined such that the renormalized Hamiltonian up to lowest order becomes diagonal in the number basis. We then decompose the quartic Hamiltonian $\hat{H}_4$ into secular and non-secular contributions as
\begin{align}
\hat{H}_4=\hat{S}_4+\hat{N}_4.
\label{Eq:1stPT-Qu1Mode-H4=S4+N4}
\end{align}
Following our discussion in Sec.~\ref{SubApp:1stPT-DissQu}, we know that it is only possible to remove the non-secular contributions $\hat{N}_4$, i.e. the generator $\hat{G}_4$ is determined via 
\begin{align}
[\omega_{\text{a}}\hat{H}_{\text{a}}+\omega_{\text{c}}\hat{H}_{\text{c}},\hat{G}_4]-\frac{\bar{\omega}_{\text{a}}}{48}\hat{N}_4=0.
\label{Eq:1stPT-Qu1Mode-Eq for N4}
\end{align}
On the other hand, the secular contributions $\hat{S}_4$ provides the lowest order correction to the Hamiltonian.

The secular terms $\hat{S}_4$ can always be expressed in terms of the quadratic Hamiltonians. For the current system we find    
\begin{align}
\begin{split}
\hat{S}_4&=6u_{\text{aa}}^4\hat{H}_{\text{a}}^2+6u_{\text{ac}}^4\hat{H}_{\text{c}}^2+24u_{\text{aa}}^2u_{\text{ac}}^2\hat{H}_{\text{a}}\hat{H}_{\text{c}}\\
&=6\left(u_{\text{aa}}^4+2u_{\text{aa}}^2u_{\text{ac}}^2\right)\hat{n}_{\text{a}}+6\left(u_{\text{ac}}^4+2u_{\text{aa}}^2u_{\text{ac}}^2\right)\hat{n}_{\text{c}}\\
&+6u_{\text{aa}}^4\hat{n}_{\text{a}}^2+6u_{\text{ac}}^4\hat{n}_{\text{c}}^2+24u_{\text{aa}}^2u_{\text{ac}}^2\hat{n}_{\text{a}}\hat{n}_{\text{c}}.
\end{split}
\label{Eq:1stPT-Qu1Mode-S4 Sol}
\end{align}
Therefore, up to lowest order, we obtain the effective system Hamiltonian as
\begin{align}
\begin{split}
\hat{\mathcal{H}}_{\text{s,eff}}&=\left[\omega_{\text{a}}-\frac{\epsilon\bar{\omega}_{\text{a}}}{8}\left(u_{\text{aa}}^4+2u_{\text{aa}}^2u_{\text{ac}}^2\right)\right]\hat{n}_{\text{a}}\\
&+\left[\omega_{\text{c}}-\frac{\epsilon\bar{\omega}_{\text{a}}}{8}\left(u_{\text{ac}}^4+2u_{\text{ac}}^2u_{\text{aa}}^2\right)\right]\hat{n}_{\text{c}}\\
&-\frac{\epsilon\bar{\omega}_{\text{a}}}{8}\left(u_{\text{aa}}^4\hat{n}_{\text{a}}^2+u_{\text{ac}}^4\hat{n}_{\text{c}}^2+4u_{\text{aa}}^2u_{\text{ac}}^2\hat{n}_{\text{a}}\hat{n}_{\text{c}}\right).
\end{split}
\label{Eq:1stPT-Qu1Mode-Hsw Sol}
\end{align}
According to Eq.~(\ref{Eq:1stPT-Qu1Mode-Hsw Sol}), the Hamiltonian for each normal mode is renormalized due to two contributions, self-Kerr and cross-Kerr, whose strength is determined by the hybridization coefficients.

Next, we solve for the generator $\hat{G}_4$ from Eq.~(\ref{Eq:1stPT-Qu1Mode-Eq for N4}). For this matter, we use the fact that the commutator of the quadratic Hamiltonian with any monomial of creation and annihilation operators is proportional to that monomial as discussed in Eq.~(\ref{Eq:1stPT-DissQu-[H2,bdag^m b^n]}). The generalization for the two bosonic mode case is found as
\begin{align}
\begin{split}
&\left[\hat{\mathcal{H}}_2,(\hat{a}^{\dag})^m\hat{a}^{n}(\hat{c}^{\dag})^l\hat{c}^{p}\right]=\left[\omega_{\text{a}}\hat{H}_{\text{a}}+\omega_{\text{c}}\hat{H}_{\text{c}},(\hat{a}^{\dag})^m\hat{a}^{n}(\hat{c}^{\dag})^l\hat{c}^{p}\right]\\
&=\left[(m-n)\omega_{\text{a}}+(l-p)\omega_{\text{c}}\right](\hat{a}^{\dag})^m\hat{a}^{n}(\hat{c}^{\dag})^l\hat{c}^{p}.\end{split}
\label{Eq:1stPT-Duff1Mode-[H2,bdag^m b^n]}
\end{align}
As a result, if there is a monomial of the form $N(\hat{a}^{\dag})^m\hat{a}^{n}(\hat{c}^{\dag})^l\hat{c}^{p}$ in $\hat{N}_4$, we require $\hat{G}_4$ to contain the same monomial but with modified coefficients determined by Eq.~(\ref{Eq:1stPT-Qu1Mode-Eq for N4}) and identity~(\ref{Eq:1stPT-Duff1Mode-[H2,bdag^m b^n]}) as
\begin{align}
N=\frac{\bar{\omega}_{\text{a}}}{48\left[(m-n)\omega_{\text{a}}+(l-p)\omega_{\text{c}}\right]}G.
\label{Eq:1stPT-Duff1Mode-N4 ITO G4}
\end{align}
 
Note that there are $4^4=256$ distinct terms in $\hat{H}_4$, among which $36$ are secular and $220$ non-secular. Therefore, due to high number of non-secular terms, the bookkeeping can not be done manually. Note that due to the term-by-term calculation that is possible based on solution~(\ref{Eq:1stPT-Duff1Mode-N4 ITO G4}), we can categorize all the terms that contribute to a particular multi-photon process.
 
Up to here, we have found the required transformation to remove the non-secular terms from the system Hamiltonian. We need to consistently apply this transformation to obtain the renormalization to the system-bath Hamiltonian as well. The system quadratures $\hat{Y}_{\text{a,c}}$ are transformed up to lowest order as
\begin{align}
\begin{split}
e^{-\hat{G}} \hat{Y}_{\text{a,c}} e^{+\hat{G}}=\hat{Y}_{\text{a,c}}+\epsilon\left[\hat{Y}_{\text{a,c}},\hat{G}_4\right]+O(\epsilon^2).
\end{split}
\label{Eq:1stPT-Qu1Mode-b_(q,c)+b_(q,c)^dag trans}
\end{align}
The $\epsilon$-order correction to the qubit- and cavity-like quadratures, i.e. $[\hat{Y}_{\text{a},\text{c}},\hat{G}_4]$, are categorized in Table~\ref{tab:EffME-Qu1Mode} including a multitude of single- and three-photon nonlinear interaction with the bath.  

In particular, the single-photon contributions can be added together to give the renormalizations
\begin{subequations}
\begin{align}
\begin{split}
&e^{-\hat{G}}\hat{Y}_{\text{a}} e^{\hat{G}} = -i \hat{a}\\
&+i \frac{\epsilon}{8}\frac{\bar{\omega}_{\text{a}}}{\omega_{\text{a}}}u_{\text{aa}}^2\left(u_{\text{aa}}^2+u_{\text{ac}}^2+u_{\text{aa}}^2\hat{n}_{\text{a}}+2u_{\text{ac}}^2\hat{n}_{\text{c}}\right)\hat{a}\\
&+i\frac{\epsilon}{2}\frac{\bar{\omega}_{\text{a}}\omega_{\text{c}}}{\omega_{\text{c}}^2-\omega_{\text{a}}^2}u_{\text{aa}}u_{\text{ac}}\left(u_{\text{aa}}^2+u_{\text{ac}}^2+u_{\text{ac}}^2\hat{n}_{\text{c}}+2u_{\text{aa}}^2\hat{n}_{\text{a}}\right)\hat{c},\\
&+\text{H.c.}+O(\epsilon^2),
\end{split}
\label{Eq:1stPT-Qu1Mode-X_q,pt-X_q}\\
\begin{split}
&e^{-\hat{G}}\hat{Y}_\text{c} e^{\hat{G}} = -i\hat{c}\\
&+i\frac{\epsilon}{8}\frac{\bar{\omega}_{\text{a}}}{\omega_{\text{c}}}u_{\text{ac}}^2\left(u_{\text{ac}}^2+u_{\text{aa}}^2+u_{\text{ac}}^2\hat{n}_{\text{c}}+2u_{\text{aa}}^2\hat{n}_{\text{a}}\right)\hat{c}\\
&+i\frac{\epsilon}{2}\frac{\bar{\omega}_{\text{a}}\omega_{\text{a}}}{\omega_{\text{a}}^2-\omega_{\text{c}}^2}u_{\text{ac}}u_{\text{aa}}\left(u_{\text{ac}}^2+u_{\text{aa}}^2+u_{\text{aa}}^2\hat{n}_{\text{a}}+2u_{\text{ac}}^2\hat{n}_{\text{c}}\right)\hat{a},\\
&+\text{H.c.}+O(\epsilon^2).
\end{split}
\label{Eq:1stPT-Qu1Mode-X_c,pt-X_c}
\end{align}
\end{subequations}
From Eq.~(\ref{Eq:1stPT-Qu1Mode-X_q,pt-X_q}-\ref{Eq:1stPT-Qu1Mode-X_c,pt-X_c}), we find that due to nonlinear mixing the quadrature of the qubit/cavity-like modes will transform into linear combinations of both normal quadratures, with coefficients that depend on both the hybridization coefficients as well as the relative position of the normal mode frequencies.

According to Eq.~(\ref{Eq:1stPT-Hsb in normpic}), the bare cavity quadrature coupling to the bath translates as $v_{\text{cc}}\hat{Y}_{\text{c}}+v_{\text{ca}}\hat{Y}_{\text{a}}$ in terms of the normal modes. Combining the linear and nonlinear renormalizations, we can obtain an effective $\epsilon$-order Lindblad equation as
\begin{align}
\begin{split}
\dot{\hat{\rho}}_{\text{s}}(t)=-i\left[\hat{\mathcal{H}}_{\text{s,eff}},\hat{\rho}_{\text{s}}\right]&+2\kappa_{\text{c}}\mathcal{D}[\hat{C}_{\text{c,eff}}]\hat{\rho}_{\text{s}}(t)
\\
&+2\kappa_{\text{a}}\mathcal{D}[\hat{C}_{\text{a,eff}}]\hat{\rho}_{\text{s}}(t).
\end{split}
\label{Eq:1stPT-Qu1Mode-PT Lind Eq}
\end{align}
where $2\kappa_{\text{a,c}}\equiv S_{YY}(\omega_{\text{a,c}})$ the effective qubit- and cavity-like single photon collapse operators read
\begin{subequations}
\begin{align}
\begin{split}
&\hat{C}_{\text{a,eff}}=\left[v_{\text{ca}}-\frac{\epsilon}{8}\frac{\bar{\omega}_{\text{a}}}{\omega_{\text{a}}}v_{\text{ca}}u_{\text{aa}}^2\left(u_{\text{aa}}^2+u_{\text{ac}}^2+u_{\text{aa}}^2\hat{n}_{\text{a}}+2u_{\text{ac}}^2\hat{n}_{\text{c}}\right)\right.\\
&-\left.\frac{\epsilon}{2}\frac{\bar{\omega}_{\text{a}}\omega_{\text{a}}}{\omega_{\text{a}}^2-\omega_{\text{c}}^2}v_{\text{cc}}u_{\text{ac}}u_{\text{aa}}\left(u_{\text{ac}}^2+u_{\text{aa}}^2+u_{\text{aa}}^2\hat{n}_{\text{a}}+2u_{\text{ac}}^2\hat{n}_{\text{c}}\right)\right]\hat{a},
\end{split}
\label{Eq:1stPT-Qu1Mode-Def of c_{q,PT}}\\
\begin{split}
&\hat{C}_{\text{c,eff}}=\left[v_{\text{cc}}-\frac{\epsilon}{8}\frac{\bar{\omega}_{\text{a}}}{\omega_{\text{c}}}v_{\text{cc}}u_{\text{ac}}^2\left(u_{\text{ac}}^2+u_{\text{aa}}^2+u_{\text{ac}}^2\hat{n}_{\text{c}}+2u_{\text{aa}}^2\hat{n}_{\text{a}}\right)\right.\\
&-\left.\frac{\epsilon}{2}\frac{\bar{\omega}_{\text{a}}\omega_{\text{c}}}{\omega_{\text{c}}^2-\omega_{\text{a}}^2}v_{\text{ca}}u_{\text{aa}}u_{\text{ac}}\left(u_{\text{aa}}^2+u_{\text{ac}}^2+u_{\text{ac}}^2\hat{n}_{\text{c}}+2u_{\text{aa}}^2\hat{n}_{\text{a}}\right)\right]\hat{c}.
\end{split}
\label{Eq:1stPT-Qu1Mode-Def of c_{c,PT}}
\end{align}
\end{subequations}
The effective Lindblad Eq.~(\ref{Eq:1stPT-Qu1Mode-PT Lind Eq}) along with the renormalized Hamiltonian~(\ref{Eq:1stPT-Qu1Mode-Hsw Sol}) and colllapse operators~(\ref{Eq:1stPT-Qu1Mode-Def of c_{q,PT}}-\ref{Eq:1stPT-Qu1Mode-Def of c_{c,PT}}) is the main result of this appendix.  
\section{Second order perturbation theory}
\label{App:2ndPT}
In this appendix, we discuss the generalization of our perturbation to second order in weak anharmonicity $\epsilon$. Contrary to App.~\ref{App:1stSWPT}, we only provide the generic conditions for frequency and lifetime renormalization. In practice, these higher order results can be applied to the specific cases studied in Apps.~\ref{SubApp:1stPT-DissQu}-\ref{SubApp:1stSWPT-Qu1Mode} only by symbolic computer algebra, due to the large number of terms that grow exponentially with the order of perturbation.

We start by a unitary transformation of the form 
\begin{align}
\hat{\mathcal{H}}_{\text{eff}}\equiv e^{-\hat{G}}\hat{\mathcal{H}}e^{+\hat{G}},
\label{Eq:2ndPT-Def of H_eff}
\end{align}
where $\hat{G}$ is the generator of this transformation and $\hat{\mathcal{H}}$ stands for the total Hamiltonian. Next, we employ the following formal expansions of the system Hamiltonian and the generator
\begin{subequations}
\begin{align}
&\hat{\mathcal{H}}_{\text{s}}=\hat{\mathcal{H}}_2-\epsilon\hat{\mathcal{H}}_4+\epsilon^2\hat{\mathcal{H}}_6+O(\epsilon^3),
\label{Eq:2ndPT-Hs exp}\\
&\hat{G}=\epsilon\hat{G}_4+\epsilon^2\hat{G}_6+O(\epsilon^3),
\label{Eq:2ndPT-G exp}
\end{align}
\end{subequations}
where the alternating sign in the system Hamiltonian expansion~(\ref{Eq:2ndPT-Hs exp}) is chosen to be consistent with the Taylor expansion of the Josephson potential.

To calculate the generator of this transformation we first focus on how the system Hamiltonian transforms. Employing the BCH formula we can expand the transformed system Hamiltonian as
\begin{align}
\hat{\mathcal{H}}_{\text{s,eff}}=e^{-\hat{G}}\hat{\mathcal{H}}_{\text{s}} e^{+\hat{G}}=\hat{\mathcal{H}}_{\text{s}}+[\hat{\mathcal{H}}_{\text{s}},\hat{G}]+\frac{1}{2!}[[\hat{\mathcal{H}}_{\text{s}},\hat{G}],\hat{G}]+\ldots .
\label{Eq:2ndPT-BCH Lemma}
\end{align}
Keeping the BCH formula up to the second order in $\hat{G}$ and plugging Eqs.~(\ref{Eq:2ndPT-Hs exp}-\ref{Eq:2ndPT-G exp}) we obtain
\begin{align}
\begin{split}
\hat{\mathcal{H}}_{\text{s,eff}}&=\hat{\mathcal{H}}_2-\epsilon\hat{\mathcal{H}}_4+\epsilon^2\hat{\mathcal{H}}_6+[\hat{\mathcal{H}}_2-\epsilon\hat{\mathcal{H}}_4,\epsilon\hat{G}_4+\epsilon^2\hat{G}_6]\\
&+\frac{1}{2}[[\hat{\mathcal{H}}_2-\epsilon\hat{\mathcal{H}}_4,\epsilon\hat{G}_4+\epsilon^2\hat{G}_6],\epsilon\hat{G}_4+\epsilon^2\hat{G}_6]+O(\epsilon^3).
\end{split}
\label{Eq:2ndPT-PT Exp 1}
\end{align}
Collecting distinct powers of $\epsilon$ in Eq.~(\ref{Eq:2ndPT-PT Exp 1}) results in
\begin{align}
\begin{split}
&\hat{\mathcal{H}}_{\text{s,eff}}=\hat{\mathcal{H}}_2+\epsilon \left\{-\hat{\mathcal{H}}_4+[\hat{\mathcal{H}}_2,\hat{G}_4]\right\}\\
&+\epsilon^2\left\{\hat{\mathcal{H}}_6+[\hat{\mathcal{H}}_2,\hat{G}_6]-[\hat{\mathcal{H}}_4,\hat{G}_4]+\frac{1}{2}[[\hat{\mathcal{H}}_2,\hat{G}_4],\hat{G}_4]\right\},
\end{split}
\label{Eq:2ndPT-PT Exp 2}
\end{align}
from which we can determine $\hat{G}_4$ and $\hat{G}_6$ order by order such that it simplifies the system Hamiltonian. 

Next, we partition both the quartic and the sextic contributions $\hat{\mathcal{H}}_{4}$ and $\hat{\mathcal{H}}_6$ into secular and non-secular parts according to
\begin{align}
\hat{\mathcal{H}}_{2n}=\hat{\mathcal{S}}_{2n}+\hat{\mathcal{N}}_{2n},
\label{Eq:2ndPT-H2n Partition}
\end{align}
and plug them into the second order expansion~(\ref{Eq:2ndPT-PT Exp 2}). As discussed in Sec.~\ref{Sec:HierEqs}, $\hat{G}_4$ is determined such that it cancels all the non-secular terms up to the first order as
\begin{align}
[\hat{\mathcal{H}}_2,\hat{G}_4]-\hat{\mathcal{N}_4}=0,
\label{Eq:2ndPT-Eq for G4}
\end{align}
leaving behind the secular terms $\hat{\mathcal{S}}_4$ to renormalize the system Hamiltonian. We next focus on the second order contributions in Eq.~(\ref{Eq:2ndPT-PT Exp 2}). Using the the first order result~(\ref{Eq:2ndPT-Eq for G4}), we are able to simplify the following terms as
\begin{align}
\begin{split}
-[\hat{\mathcal{H}}_4,\hat{G}_4]&+\frac{1}{2}[[\hat{\mathcal{H}}_2,\hat{G}_4],\hat{G}_4]\\
&=-[\hat{\mathcal{S}}_4,\hat{G}_4]-\frac{1}{2}[\hat{\mathcal{N}}_4,\hat{G}_4].
\end{split}
\label{Eq:2ndPT-Simplification}
\end{align}
To proceed, we need to categorize the remaining contributions in Eq.~(\ref{Eq:2ndPT-Simplification}) into secular and non-secular. Based on Eqs.~(\ref{Eq:1stPT-[S,S]=0}-\ref{Eq:1stPT-[N,N]=S+N}) and the fact that $\hat{G}_4$ only includes non-secular terms, we find that $[\hat{\mathcal{S}}_4,\hat{G}_4]$ only contains non-secular contributions, while $[\hat{\mathcal{N}}_4,\hat{G}_4]$ includes both types. We then use $\mathcal{S}(\bullet)$ and $\mathcal{N}(\bullet)$ to denote the secular and non-secular parts of an operator-valued expression, respectively. Using Eq.~(\ref{Eq:2ndPT-Simplification}) and the fact that only the non-secular terms in Eq.~(\ref{Eq:2ndPT-PT Exp 2}) can be removed, we obtain the condition to determine $\hat{G}_6$ as
\begin{subequations}
\begin{align}
[\hat{\mathcal{H}}_2,\hat{G}_6]+\hat{\mathcal{N}}_6-[\hat{\mathcal{S}}_4,\hat{G}_4]-\frac{1}{2}\mathcal{N}\left([\hat{\mathcal{N}}_4,\hat{G}_4]\right)=0.
\label{Eq:2ndPT-Eq for G6}
\end{align}
Moreover, we find that the Hamiltonian is renormalized due to secular contributions that originate both directly through the system Hamiltonian as well as indirectly via the remaining commutators as
\begin{align}
\hat{\mathcal{H}}_{\text{s,eff}}=\hat{\mathcal{H}}_2-\epsilon\hat{\mathcal{S}}_4+\epsilon^2\left[\hat{\mathcal{S}}_6-\frac{1}{2}\mathcal{S}\left([\hat{\mathcal{N}}_4,\hat{G}_4]\right)\right]+O(\epsilon^3).
\label{Eq:2ndPT-Hs}
\end{align}
\end{subequations}
Equations~(\ref{Eq:2ndPT-Eq for G6}) and~(\ref{Eq:2ndPT-Hs}) are the generic main results for the second order perturbation theory. From here on, one needs to apply the resulting transformation on the system-bath interaction to determine the renormalization of the dissipators.

\section{Effective Master Equation Derivation}
\label{App:EMEDer}

Here, we provide a derivation for the EMEs discussed in App.~\ref{App:1stSWPT} as well as Sec.~\ref{Sec:EffME} of the main body of the paper. The discussion here makes the distinction between the two possible representation of the EME, namely Eqs.~(\ref{eqn:EffME-DissQu-Lind Eq1}) and~(\ref{eqn:EffME-DissQu-Lind Eq2}), more clear.  

Our starting point is the von Neumann equation for the full density matrix as
\begin{align}
\dot{\hat{\rho}}(t)=-i[ \hat{\mathcal{H}}, \hat{\rho}(t)],
\label{Eq:EMEDer-vonNeumann Eq1}
\end{align}
where $\hat{\mathcal{H}}$ is the full Hamiltonian given by
\begin{align}
\hat{\mathcal{H}}= \hat{\mathcal{H}}_\text{s} + \hat{\mathcal{H}}_\text{b}+\hat{\mathcal{H}}_{\text{sb}}.
\label{Eq:EMEDer-Def of Hfull}
\end{align}
We take the following steps to derive the desired effective master equation. First, we apply a unitary transformation to the Von Neuman Eq.~(\ref{Eq:EMEDer-vonNeumann Eq1}). The generator of this transformation is solved for such that it removes the non-secular term from the system Hamiltonian $\hat{\mathcal{H}}_s$. The resulting equation for the generator has been discussed in App.~\ref{App:2ndPT} in Eqs.~(\ref{Eq:2ndPT-Eq for G4}) and~(\ref{Eq:2ndPT-Eq for G6}) up to the sfirst and second order in perturbation, respectively. Second, to be consistent, we need to apply the resulting unitary transformation on the system-bath interaction. Third, we move to the resulting interaction picture  and obtain a Redfield equation \cite{Redfield_Theory_1965, Breuer_Theory_2002}, from which a Lindblad master equation is derived.  

We apply the following unitary transformation 
\begin{align}
\hat{\rho}(t) \rightarrow e^{-\hat{G}} \hat{\rho}(t) e^{\hat{G}}.
\label{Eq:EMEDer-Def of rho'(t)}
\end{align}
where the new density matrix $\hat{\rho}'(t)$ obeys the transformed von Neumann equation:
\begin{align}
\dot{\hat{\rho}}(t)=-i\left[e^{-\hat{G}}\hat{\mathcal{H}} e^{\hat{G}},\hat{\rho}(t)\right].
\label{Eq:EMEDer-von Neumann Eq2}
\end{align}
The generator $\hat{G}=\epsilon\hat{G}_4+O(\epsilon^2)$ is determined such that it eliminates the number non-conserving terms in the system Hamiltonian, resulting in the condition
\begin{align}
[\hat{\mathcal{H}}_2,\hat{G}_4]=\hat{\mathcal{N}}_4,
\label{Eq:EMEDer-Eq for N4}
\end{align}
up to lowest order in $\epsilon$. Then, the effective von Neumann equation reads
\begin{align}
\dot{\hat{\rho}}(t)=-i\left[\hat{\mathcal{H}}_{\text{s},\text{eff}}+\hat{\mathcal{H}}_{\text{b}}+\hat{\mathcal{H}}_{\text{sb},\text{eff}},\hat{\rho}(t)\right],
\label{Eq:EMEDer-von Neumann Eq3}
\end{align}
where $\hat{\mathcal{H}}_{\text{s}}$ and $\hat{\mathcal{H}}_{\text{sb}}$ are found as
\begin{subequations}
\begin{align}
&\hat{\mathcal{H}}_{\text{s},\text{eff}}=\hat{\mathcal{H}}_2-\epsilon\hat{\mathcal{S}}_4++O(\epsilon^2),
\label{Eq:EMEDer-Def of Hs,eff}\\
&\hat{\mathcal{H}}_{\text{sb},\text{eff}}=\hat{\mathcal{H}}_{\text{sb}}+\epsilon\left[\hat{\mathcal{H}}_{\text{sb}}, \hat{G}_4\right]+O(\epsilon^2).
\label{Eq:EMEDer-Def of Hsb,eff}
\end{align}
\end{subequations}

Next, we move to an interaction picture with respect to the effective system and bath Hamiltonian $\hat{\mathcal{H}}_{\text{s},\text{eff}} +\hat{\mathcal{H}}_\text{b}$. Accounting for the new effective system-bath interaction $\hat{\mathcal{H}}_{\text{sb},\text{eff}}$ perturpatively under Born approximation, we arrive at the Redfield equation \cite{Redfield_Theory_1965, Breuer_Theory_2002}
\begin{align}
&\dot{\hat{\tilde{\rho}}}_{\text{s}}(t)=-\int_{0}^{t}dt' \text{tr}_b\left[\hat{\mathcal{\tilde{H}}}_{\text{sb,eff}}(t),[\hat{\mathcal{\tilde{H}}}_{\text{sb,eff}}(t'),\hat{\tilde{\rho}}_{\text{s}}(t')\otimes\hat{\rho}_{\text{b}}(0)]\right]
\label{Eq:EMEDer-Redfield Eq}
\end{align}
where interaction picture operators are denoted by a tilde and are defined as
\begin{align}
\hat{\tilde{\mathcal{O}}}(t)=e^{i \left(\hat{\mathcal{H}}_{\text{s,eff}} +\hat{\mathcal{H}}_\text{b} \right) t} \hat{\mathcal{O}} e^{-i \left(\hat{\mathcal{H}}_{\text{s,eff}} + \hat{\mathcal{H}}_\text{b} \right) t}.
\label{Eq:EMEDer-Def of tilde(O)}
\end{align}
The system-bath Hamiltonian in the interaction picture can be written as
\begin{align}
\begin{split}
&\hat{\mathcal{\tilde{H}}}_{\text{sb,eff}}(t) = e^{i\hat{\mathcal{H}}_\text{b} t}\left[\sum\limits_k g_k (\hat{B}_k+\hat{B}_k^{\dag})\right]e^{-i\hat{\mathcal{H}}_\text{b} t}\otimes\\ 
&e^{i \left(\hat{\mathcal{H}}_{2} - \epsilon \hat{\mathcal{S}}_{4}\right)t} \left\{ \hat{X}_\text{s} + \epsilon \left[\hat{X}_\text{s},\hat{G}_4\right]\right\}e^{-i \left(\hat{\mathcal{H}}_{2} - \epsilon \hat{\mathcal{S}}_{4}\right)t}+O(\epsilon^2). 
\end{split}
\label{Eq:EMEDer-tilde(Hsb,eff)}
\end{align}
For simplicity, we denote the transformed bath quadrature as
\begin{align}
\begin{split}
\hat{\tilde{X}}_b(t)&\equiv e^{i\hat{\mathcal{H}}_\text{b} t}\left[\sum\limits_k g_k (\hat{B}_k+\hat{B}_k^{\dag})\right]e^{-i\hat{\mathcal{H}}_\text{b} t}\\
&=\sum\limits_k g_k \left(\hat{B}_k e^{-i\omega_k t}+\hat{B}_k^{\dag} e^{i\omega_k t}\right).
\end{split}
\label{Eq:EMEDer-Def of tilde(X)_b}
\end{align}
The correction to the system quadrature, i.e. commutator $\left[  \hat{X}_\text{s},\hat{G}_4\right]$, can be assembled from the result for $\hat{G}_4$ as provided in Tables~\ref{tab:EffME-Qu} and~\ref{tab:EffME-Qu1Mode} for the cases of a qubit coupled to a flux bath and a single mode cavity, respectively.

In the following, we show the route to arrive at two possible representation of the EME, namely the number-state representation of dissipators as in Eq.~(\ref{eqn:EffME-DissQu-Lind Eq1}) and the compact operator represenatation of dissipators as in Eq.~(\ref{eqn:EffME-DissQu-Lind Eq2}) of Sec.~\ref{SubSec:EffME-Qu} for case (i). 

To arrive at the first representation, we note that the renormalized system quadrature~(\ref{Eq:EMEDer-tilde(Hsb,eff)}) can be expressed in the number-basis, that are also eigenstates of the system evolution operator $\exp[-i(\hat{\mathcal{H}}_{2} - \epsilon \hat{\mathcal{S}}_{4})t]$, as
\begin{align}
\begin{split}
& e^{i \left(\hat{\mathcal{H}}_{2} - \epsilon \hat{\mathcal{S}}_{4,i}\right)t} \left\{\hat{X}_\text{s} + \epsilon \left[ \hat{X}_\text{s}, \hat{G}_4\right] \right\} e^{-i \left(\hat{\mathcal{H}}_{2} - \epsilon \hat{\mathcal{S}}_{4}\right)t} \\ 
&= \sum_{n_{\text{a}},n_\text{s} \atop\bar{m}_{\text{a}},\bar{m}_\text{c}} \ket{n_{\text{a}},n_\text{c}} \bra{n_{\text{a}},n_\text{c}} \left\{\hat{X}_\text{s} + \epsilon \left[ \hat{X}_\text{c},+\hat{G}_4\right] \right\}\\
&\times\ket{m_{\text{a}},m_\text{c}}\bra{m_{\text{a}},m_\text{c}}  e^{i(\omega_{n_{\text{a}},n_\text{c}} - \omega_{\bar{m}_{\text{a}},\bar{m}_\text{c}})t }\equiv  \sum_j \hat{C}_{\text{eff}}(\omega_j) e^{-i \omega_j t}. 
\end{split}
\label{Eq:EMEDer-Xc FockRep}
\end{align}
In the second line of Eq.~(\ref{Eq:EMEDer-Xc FockRep}), we have considered the case of a qubit coupled to a single cavity mode, while the discussion here is generic for a multimode system. In order to go from the second to the third line of Eq.~(\ref{Eq:EMEDer-Xc FockRep}), one needs to calculate the nonzero matrix elements, and hence obtain the corresponding state-dependent frequency of each individual process. The system transition frequencies $\{\omega_{n_{\text{a}},n_\text{c}}\}$ are easily accessible due to being diagonal in the number basis as provided for example in Eq.~\ref{eqn:EffME-Qu1Mode-Hsw}. Moreover, the sum in the last step of Eq.~(\ref{Eq:EMEDer-Xc FockRep}) has, in general, less terms than the sum from which it originates: firstly, there may  be processes between distinct states that have the same frequency (as would be the case, for example, in a harmonic oscillator); secondly, certain matrix elements may be vanishing. 

Lastly, by taking the time integral in the Redfield equation~(\ref{Eq:EMEDer-Redfield Eq}), under Markov approximation, each individual processes appear as an independent relaxation channel as  
\begin{align}
\dot{\hat{\rho}}(t) = -i \left[ \hat{\mathcal{H}}_{\text{s},\text{eff}}, \hat{\rho}(t) \right] + \sum_{j}2 \kappa(\omega_j) \mathcal{D}\left[ \hat{C}_{\text{eff}}(\omega_j) \right] \hat{\rho}(t), 
\label{Eq:EMEDer-EME}
\end{align} 
where the relaxation rate $\kappa(\omega)$ is found in terms of the bath spectral function as
\begin{align}
2\kappa(\omega)=S_{YY}(\omega)\equiv\int_{-\infty}^\infty d\tau \, \text{e}^{-i\omega \tau} \text{tr}_{\text{b}} \left[ \hat{\rho}_\text{b}(0)\hat{X}_{\text{b}}(\tau) \, \hat{X}_{\text{b}}(0) \right].
\label{Eq:EMEDer-Def of S_XX}
\end{align}  
Equation~(\ref{Eq:EMEDer-EME}), together with the definition  of $\hat{C}_\text{eff}(\omega_j)$ in Eq.~(\ref{Eq:EMEDer-Xc FockRep}), forms the most general form of the EME up to lowest order in $\epsilon$. 

As an example, applying this procedure on case (i) of Sec.~\ref{SubSec:EffME-Qu}, one would arrive at
\begin{subequations}
\begin{align}
\begin{split}
\dot{\hat{\rho}}_{\text{a}}(t)=-i\left[\hat{\mathcal{H}}_{\text{eff}},\hat{\rho}_{\text{a}}(t)\right]&+\sum\limits_{n=1}^{N_{\text{c}}}2\kappa_{\text{a},n} \mathcal{D}\left[ \ket{n-1} \bra{n} \right] \hat{\rho}_{\text{a}}(t)\\
&+\sum\limits_{n=3}^{N_{\text{c}}}2\kappa_{\text{a}3,n}\mathcal{D}\left[ \ket{n-3} \bra{n} \right]\hat{\rho}_{\text{a}}(t),
\end{split}
\label{eqn:EffME-DissQu-Lind Eq2}
\end{align}  
with $\kappa_{a,n}$ and $\kappa_{q3,n}$ defined as 
\begin{align}
&\kappa_{\text{a},n}\equiv\left[1+\frac{\epsilon}{8}(1+n)\right]^2 n \, S_{XX}\left([1-\frac{\epsilon}{4}n]\omega_{\text{a}}\right), 
\label{eqn:EffME-Def of kappa_q,n}\\
&\kappa_{\text{a}3,n}\equiv\left(\frac{\epsilon}{48}\right)^2 n(n-1)(n-2) \, S_{XX}\left([1-\frac{3\epsilon}{4}(n-1)]\omega_{\text{a}}\right).
\label{eqn:EffME-Def of kappa_q3,n}
\end{align}
\end{subequations}
The above representation of the effective master equation shows that the interplay of qubit anharmonicity with the flux bath appears as relaxation rates that nonlinearly depend on the initial excitation and the weak anharmonicity measure $\epsilon$ of the qubit.

To obtain the alternative representation of the EME we take the following steps. Firstly, instead of expanding over the number basis, we can rewrite the interaction-picture system quadrature as  
\begin{align}
\begin{split}
&e^{i \left(\hat{\mathcal{H}}_{2} - \epsilon \hat{\mathcal{S}}_{4}\right)t} \left\{ \hat{X}_\text{s} + \epsilon \left[\hat{X}_\text{s},\hat{G}_4\right]\right\}e^{-i \left(\hat{\mathcal{H}}_{2} - \epsilon \hat{\mathcal{S}}_{4}\right)t}\\
&= e^{i \left(\hat{\mathcal{H}}_{2} - \epsilon \hat{\mathcal{S}}_{4}\right)t} \hat{X}_\text{s} e^{-i \left(\hat{\mathcal{H}}_{2} - \epsilon \hat{\mathcal{S}}_{4}\right)t}\\
&+ \epsilon e^{i \hat{\mathcal{H}}_{2}t} \left[\hat{X}_\text{s},\hat{G}_4\right]e^{-i \hat{\mathcal{H}}_{2} t}+O(\epsilon^2),
\label{Eq:EMEDer-Approx}
\end{split}
\end{align}
where in order to keep terms only up to lowest order in $\epsilon$, the correction to the system quadrature has to transform with the zeroth order harmonic Hamiltonian $\hat{\mathcal{H}}_2$. This makes the computation of the third line of~Eq.~(\ref{Eq:EMEDer-Approx}) rather trivial, as the result for $[\hat{X}_s,\hat{G}_4]$ (Tables~\ref{tab:EffME-Qu} and~\ref{tab:EffME-Qu1Mode} for example) can be immediately used by replacing $\hat{a}\rightarrow \hat{a}^{-i\omega_{\text{a}}t}$ and $\hat{c}\rightarrow \hat{c}^{-i\omega_{\text{c}}t}$.

Secondly, we drop the term $-\epsilon \hat{\mathcal{S}}_{4}$ in the time evolution of the bare system quadrature $\hat{X}_s$ [second line of Eq.~(\ref{Eq:EMEDer-Approx})], which amounts to corrections from the Kerr Hamiltonian to the eigenfrequencies. This will in turn change the argument of the bath spectral function by order $\epsilon$ relative to the normal mode frequencies. We assume that the bath spectral function is flat enough close to frequencies of interest such that one can write  
\begin{align}
\begin{split}
S_\text{YY}( \omega_i + \epsilon \omega_j) &= S_\text{YY}(\omega_i) +  \epsilon \omega_j S_\text{YY}'(\omega_i)  + O(\epsilon^2) \\
&\approx S_\text{YY}(\omega_i),
\end{split}
\label{Eq:EMEDer-Sxx flat1}
\end{align}
for any relevant system transition frequency corrected by the nonlinearity, expressed generically as $\omega_i + \epsilon \omega_j$. Under this approximation, one obtains the EMEs~(\ref{eqn:EffME-DissQu-Lind Eq1}) for the qubit coupled to flux bath, or~(\ref{eqn:EffME-Qu1Mode-PT Lind Eq}) for the qubit coupled to an open driven resonator.
\color{black}
\section{Equations of motion for physical observables}
\label{App:ObsEOM}
In this appendix, we discuss the derivation of equations of motion for physical observables based on the results for the lowest order effective Lindblad equation. The main motivation for this calculation is to understand how the renormalized single-photon dissipators become manifest at the level of the equations of motion for relevant physical observables such as expectation values of the qubit quadratures. 

We begin by finding the equations of motion for the expectation value of an arbitrary operator $\hat{O}$ from a generic Lindblad equation
\begin{align}
\dot{\hat{\rho}}=-i\left[\hat{\mathcal{H}},\hat{\rho}\right]+2\kappa\mathcal{D}[\hat{C}]\hat{\rho},
\label{Eq:ObsEOM-Gen Lind Eq}
\end{align}
with arbitrary Hamiltonian $\hat{\mathcal{H}}$ and collapse operator $\hat{C}$. Multiplying Eq.~(\ref{Eq:ObsEOM-Gen Lind Eq}) by $\hat{O}$ and taking the trace we obtain 
\begin{align}
\frac{d}{dt}\left<\hat{O}\right>=-i\left<[\hat{O},\hat{\mathcal{H}}]\right>+2\kappa\left<\hat{C}^{\dag}\hat{O}\hat{C}-\frac{1}{2}\left\{\hat{C}^{\dag}\hat{C},\hat{O}\right\}\right>,
\label{Eq:ObsEOM-<O> EOM 1}
\end{align}
where the expectation value is defined as
\begin{align}
\left<\hat{O}\right>\equiv \text{Tr}\left\{\hat{\rho}(t)\hat{O}\right\}.
\label{Eq:ObsEOM-Def of <O>}
\end{align}
The terms in the dissipator contribution can be rewritten in a more symmetric form as
\begin{align}
\hat{C}^{\dag}\hat{O}\hat{C}-\frac{1}{2}\left\{\hat{C}^{\dag}\hat{C},\hat{O}\right\}=\frac{1}{2}\hat{C}^{\dag}\left[\hat{O},\hat{C}\right]+\frac{1}{2}\left[\hat{C}^{\dag},\hat{O}\right]\hat{C}
\label{Eq:ObsEOM-rewrite Dissipator}
\end{align}
in terms of which Eq.~(\ref{Eq:ObsEOM-<O> EOM 1}) becomes
\begin{align}
\frac{d}{dt}\left<\hat{O}\right>=-i\left<[\hat{O},\hat{\mathcal{H}}]\right>+\kappa\left<\hat{C}^{\dag}\left[\hat{O},\hat{C}\right]+\left[\hat{C}^{\dag},\hat{O}\right]\hat{C}\right>.
\label{Eq:ObsEOM-<O> EOM 2}
\end{align}

In the following, we consider the case of a weakly anharmonic qubit coupled to a flux bath, discussed in App.~\ref{SubApp:1stPT-DissQu}, as the simplest example. We intend to obtain effective equations of motion for the quadratures, i.e. $\left<\hat{X}_{\text{a}}\right>$ and $\left<\hat{Y}_{\text{a}}\right>$ starting from Eq.~(\ref{Eq:ObsEOM-<O> EOM 2}). We showed in Eqs.~(\ref{Eq:1stPT-DissQu-H_q,eff}) and~(\ref{Eq:1stPT-DissQu-Identity {H0,b+bdag}}) that the Hamiltonian and the single-photon dissipator of the effective Lindblad equation up to lowest order read 
\begin{subequations}
\begin{align}
&\hat{\mathcal{H}}_{\text{a,eff}}=\omega_{\text{a}}\hat{H}_{\text{a}}-\frac{\epsilon}{8}\omega_{\text{a}}\hat{H}_{\text{a}}^2+O(\epsilon^2),
\label{Eq:ObsEOM-Eff PT Ham}\\
&\hat{C}_{\text{a,eff}}=\hat{a}+\frac{\epsilon}{16}\left\{\hat{H}_{\text{a}},\hat{a}\right\}+O(\epsilon^2).
\label{Eq:ObsEOM-Eff PT Diss}
\end{align} 
\end{subequations}

We first focus on the equation of motion for $\left<\hat{X}_{\text{a}}\right>$, by setting $\hat{O}=\hat{X}_{\text{a}}$ in Eq.~(\ref{Eq:ObsEOM-<O> EOM 2}). There are multiple contributions. The Hamiltonian part simplifies to
\begin{align}
[\hat{X}_{\text{a}},\hat{\mathcal{H}}_{\text{a,eff}}]=i\omega_{\text{a}}\left[\hat{Y}_{\text{a}}-\frac{\epsilon}{8}\{\hat{H}_{\text{a}},\hat{Y}_{\text{a}}\}\right].
\label{Eq:ObsEOM-[X,H_PT]}
\end{align}
We then calculate the terms originating from the dissipator in Eq.~(\ref{Eq:ObsEOM-<O> EOM 2}) one by one. The first commutator is found as
\begin{subequations}
\begin{align}
\begin{split}
\left[\hat{X}_{\text{a}},\hat{C}_{\text{a,eff}}\right]&=[\hat{X}_{\text{a}},\hat{a}]+\frac{\epsilon}{16}\left[\hat{X}_{\text{a}},\left\{\hat{H}_{\text{a}},\hat{a}\right\}\right]\\
&=-1+\frac{\epsilon}{16}\left([\hat{X}_{\text{a}},\hat{H}_{\text{a}}\hat{a}]+[\hat{X}_{\text{a}},\hat{a}\hat{H}_{\text{a}}]\right)\\
&=-1-\frac{\epsilon}{16}\left(2\hat{H}_{\text{a}}-i\{\hat{Y}_{\text{a}},\hat{a}\}\right).
\end{split}
\label{Eq:ObsEOM-[X,C_PT]}
\end{align}
Consequently, the first term in the dissipator of Eq.~(\ref{Eq:ObsEOM-<O> EOM 2}) takes the form
\begin{align}
\begin{split}
&\hat{C}_{\text{a,eff}}^{\dag}\left[\hat{X}_{\text{a}},\hat{C}_{\text{a,eff}}\right]=-\left[\hat{a}^{\dag}+\frac{\epsilon}{16}\left\{\hat{H}_{\text{a}},\hat{a}^{\dag}\right\}\right]\\
&\times\left[1+\frac{\epsilon}{16}\left(2\hat{H}_{\text{a}}-i\{\hat{Y}_{\text{a}},\hat{a}\}\right)\right].
\end{split}
\label{Eq:ObsEOM-C^dag[X,C_PT]}
\end{align}
The second commutator is obtained as
\begin{align}
\begin{split}
\left[\hat{C}_{\text{a,eff}}^{\dag},\hat{X}_{\text{a}}\right]&=\left[\hat{a}^{\dag},\hat{X}_{\text{a}}\right]+\frac{\epsilon}{16}\left[\left\{\hat{H}_{\text{a}},\hat{a}^{\dag}\right\},\hat{X}_{\text{a}}\right]\\
&=-1+\frac{\epsilon}{16}\left([\hat{H}_{\text{a}}\hat{a}^{\dag},\hat{X}_{\text{a}}]+[\hat{a}^{\dag}\hat{H}_{\text{a}},\hat{X}_{\text{a}}]\right)\\
&=-1-\frac{\epsilon}{16}\left(2\hat{H}_{\text{a}}+i\{\hat{Y}_{\text{a}},\hat{a}^{\dag}\}\right).
\end{split}
\label{Eq:ObsEOM-[C_PT^dag,X]}
\end{align}
Therefore, the second term in the commutator of Eq.~(\ref{Eq:ObsEOM-<O> EOM 2}) reads
\begin{align}
\begin{split}
\left[\hat{C}_{\text{a,eff}}^{\dag},\hat{X}_{\text{a}}\right]\hat{C}_{\text{a,eff}}&=-\left[1+\frac{\epsilon}{16}\left(2\hat{H}_{\text{a}}+i\{\hat{Y}_{\text{a}},\hat{a}^{\dag}\}\right)\right]\\
&\times \left[\hat{a}+\frac{\epsilon}{16}\left\{\hat{H}_{\text{a}},\hat{a}\right\}\right].
\end{split}
\label{Eq:ObsEOM-[C_PT^dag,X]C_PT}
\end{align}
\end{subequations}
Adding Eqs.~(\ref{Eq:ObsEOM-C^dag[X,C_PT]}) and~(\ref{Eq:ObsEOM-[C_PT^dag,X]C_PT}) and keeping the terms up to lowest order in $\epsilon$ we obtain
\begin{align}
\begin{split}
&\hat{C}_{\text{a,eff}}^{\dag}\left[\hat{X}_{\text{a}},\hat{C}_{\text{a,eff}}\right]+\left[\hat{C}_{\text{a,eff}}^{\dag},\hat{X}_{\text{a}}\right]\hat{C}_{\text{a,eff}}\\
&=-\left(\hat{X}_{\text{a}}+\frac{\epsilon}{16}\{\hat{H}_{\text{a}},\hat{X}_{\text{a}}\}\right)\\
&-\frac{\epsilon}{16}\left[\hat{a}^{\dag}\left(2\hat{H}_{\text{a}}-i\{\hat{Y}_{\text{a}},\hat{a}\}\right)+\left(2\hat{H}_{\text{a}}+i\{\hat{Y}_{\text{a}},\hat{a}^{\dag}\}\right)\hat{a}\right].
\end{split}
\label{Eq:ObsEOM-C^dag[X,C_PT]+[C_PT^dag,X]C_PT 1}
\end{align}
We then simplify the terms in the second line of Eq.~(\ref{Eq:ObsEOM-C^dag[X,C_PT]+[C_PT^dag,X]C_PT 1}) as
\begin{align}
\begin{split}
&\hat{a}^{\dag}\left(2\hat{H}_{\text{a}}-i\{\hat{Y}_{\text{a}},\hat{a}\}\right)+\left(2\hat{H}_{\text{a}}+i\{\hat{Y}_{\text{a}},\hat{a}^{\dag}\}\right)\hat{a}\\
&=(\hat{X}_{\text{a}}-i\hat{Y}_{\text{a}})\hat{H}_{\text{a}}+\hat{H}_{\text{a}}(\hat{X}_{\text{a}}+i\hat{Y}_{\text{a}})+i[\hat{Y}_{\text{a}},\hat{a}^{\dag}\hat{a}]\\
&=\{\hat{H}_{\text{a}},\hat{X}_{\text{a}}\}-i[\hat{Y}_{\text{a}},\hat{H}_{\text{a}}]+i[\hat{Y}_{\text{a}},\hat{H}_{\text{a}}]=\{\hat{H}_{\text{a}},\hat{X}_{\text{a}}\}.
\end{split}
\label{Eq:ObsEOM-simplify term in diss}
\end{align}
Plugging the result~(\ref{Eq:ObsEOM-simplify term in diss}) into Eq.~(\ref{Eq:ObsEOM-C^dag[X,C_PT]+[C_PT^dag,X]C_PT 1}) we find
\begin{align}
\begin{split}
\hat{C}_{\text{a,eff}}^{\dag}\left[\hat{X}_{\text{a}},\hat{C}_{\text{a,eff}}\right]&+\left[\hat{C}_{\text{a,eff}}^{\dag},\hat{X}_{\text{a}}\right]\hat{C}_{\text{a,eff}}\\
&=-\left(\hat{X}_{\text{a}}+\frac{\epsilon}{8}\{\hat{H}_{\text{a}},\hat{X}_{\text{a}}\}\right).
\end{split}
\label{Eq:ObsEOM-C^dag[X,C_PT]+[C_PT^dag,X]C_PT 2}
\end{align}
Finally, by inserting the Hamiltonian part~(\ref{Eq:ObsEOM-[X,H_PT]}) and the dissipator part~(\ref{Eq:ObsEOM-C^dag[X,C_PT]+[C_PT^dag,X]C_PT 2}) into the generic Eq.~(\ref{Eq:ObsEOM-<O> EOM 2}) we obtain the dynamics of $\left<\hat{X}_{\text{a}}\right>$ as
\begin{subequations}
\begin{align}
\begin{split}
\frac{d}{dt}\left<\hat{X}_{\text{a}}\right>&+\kappa_{\text{a}}\left<\left[\hat{X}_{\text{a}}+\frac{\epsilon}{8}\{\hat{H}_{\text{a}},\hat{X}_{\text{a}}\}\right]\right>\\
&-\omega_{\text{a}}\left<\left[\hat{Y}_{\text{a}}-\frac{\epsilon}{8}\{\hat{H}_{\text{a}},\hat{Y}_{\text{a}}\}\right]\right>=0.
\end{split}
\label{Eq:ObsEOM-<X> EOM}
\end{align}
Following the same procedure, we obtain an equation for $\left<\hat{Y}_{\text{a}}\right>$ as
\begin{align}
\begin{split}
\frac{d}{dt}\left<\hat{Y}_{\text{a}}\right>&+\kappa_{\text{a}}\left<\left[\hat{Y}_{\text{a}}+\frac{\epsilon}{8}\{\hat{H}_{\text{a}},\hat{Y}_{\text{a}}\}\right]\right>\\
&+\omega_{\text{a}}\left<\left[\hat{X}_{\text{a}}-\frac{\epsilon}{8}\{\hat{H}_{\text{a}},\hat{X}_{\text{a}}\}\right]\right>=0.
\end{split}
\label{Eq:ObsEOM-<Y> EOM}
\end{align}
\end{subequations}
From the lowest order results~(\ref{Eq:ObsEOM-<X> EOM}-\ref{Eq:ObsEOM-<Y> EOM}), we find that the oscillation frequency is decreased as expected due to the \textit{softening} nature of the quartic correction in the Josephson potential. More importantly, in contrast to the frequency, the decay rate increases with the same exact slope. Equations~(\ref{Eq:ObsEOM-<X> EOM}-\ref{Eq:ObsEOM-<Y> EOM}) explain the non-circular nature of the phase space orbits as shown in Fig.~(\ref{subfig:SWCorr-Qu-PhSpace2}).
\bibliographystyle{apsrev4-1}
\bibliography{SWPTBib}
\end{document}